\documentclass[a4paper, 11pt, twoside, openright]{book}

\usepackage[utf8]{inputenc}
\usepackage[T1]{fontenc}
\usepackage{lmodern}
\usepackage[UKenglish]{babel}
\usepackage[cyr]{aeguill}
\usepackage{charter}

\usepackage[top=1.1in, bottom=0.95in, outer=5cm, inner=2cm, heightrounded, marginparwidth=2.5cm, marginparsep=2cm]{geometry}

\usepackage[Sonny]{fncychap} 

\ChTitleVar{\Huge\bfseries}
\ChNameVar{\large\bfseries}
\ChNumVar{\fontsize{60}{62}\bfseries}

\usepackage{fancyhdr}
\let\headruleORIG\headrule
\renewcommand{\headrule}{\color{black} \headruleORIG}

\makeatletter
\if@twoside 
    \newlength{\textblockoffset}
\fi
\makeatother

\usepackage{pifont}
\usepackage{dsfont}
\usepackage{shorttoc}
\usepackage{eqparbox}
\usepackage{sidecap}

\usepackage{amssymb}
\usepackage{amsthm}
\usepackage{amsmath}
\usepackage{amsfonts}
\usepackage{amssymb}
\usepackage{mathtools}
\usepackage{nicefrac}

\usepackage{supertabular}
\usepackage{tabularx}
\usepackage{tabulary}
\usepackage{setspace}
\usepackage[dvips]{graphicx}
\usepackage{epsfig}
\usepackage{float}
\usepackage{psfrag}
\usepackage{epstopdf}
\usepackage{adforn}
\usepackage{array}
\usepackage{pifont}
\usepackage{textcomp}
\usepackage{multirow}

\usepackage{multirow}
\usepackage[boxed]{algorithm}
\usepackage{algorithmic}
\usepackage{rotating}
\usepackage{floatpag}

\usepackage{booktabs}

\usepackage{caption}
\usepackage{subcaption}
\usepackage[section, above, below]{placeins}

\addto\captionsenglish{}
\addto\captionsenglish{}

\usepackage{xcolor}
\usepackage{esint}

\usepackage{multicol}

\usepackage[nottoc, notlof, notlot]{tocbibind}
\usepackage[english, tight]{minitoc}

\mtcsetformat{minitoc}{tocrightmargin}{2.55em plus 1fil}
\newcommand{\multicolumnmtc}{2}
\makeatletter
\let\SV@mtc@verse\mtc@verse
\let\SV@endmtc@verse\endmtc@verse
\def\mtc@verse#1{\SV@mtc@verse#1\removelastskip%
    \begin{multicols}{\multicolumnmtc}\raggedcolumns\leavevmode\unskip
        \vskip -1.5ex \vskip -1\baselineskip}
        \def\endmtc@verse{\end{multicols}\SV@endmtc@verse}
\makeatother

\usepackage{tikz}
\usetikzlibrary{shapes,arrows}

\usepackage{pgfplots}
\pgfplotsset{compat=newest}
\pgfplotsset{plot coordinates/math parser=false}
\usetikzlibrary{plotmarks}

\usepackage{pgfplots}
\pgfplotsset{
    dirac/.style={
            mark=triangle*,
            mark options={scale=2},
            ycomb,
            scatter,
            visualization depends on={y/abs(y)-1 \as \sign},
            scatter/@pre marker code/.code={\scope[rotate=90*\sign,yshift=-2pt]}
        }
}

\theoremstyle{plain}

\theoremstyle{definition}

\theoremstyle{remark}

\usepackage{pagecolor}
\usepackage[framemethod=TikZ]{mdframed}
\usepackage{hf-tikz}
\usetikzlibrary{calc}

\usepackage[most]{tcolorbox}
\tcbuselibrary{theorems}

\newtcolorbox[auto counter, number within=section]{Thm}[2][]{%
    enhanced,
    colframe=black,
    colback=white,
    colbacktitle=white,
    coltitle=black,
    attach boxed title to top left={xshift=5mm,yshift*=-\tcboxedtitleheight/2},
    fonttitle=\bfseries,
    title=Theorem~\thetcbcounter: #2,#1}

\theoremstyle{definition}

\surroundwithmdframed[outerlinewidth=0pt, innerlinewidth=0.5pt, middlelinewidth=1pt, middlelinecolor=white]{Def}

\usepackage{thmtools, needspace}

\newcommand{\exrule}{\rule{\textwidth}{1pt}}

\declaretheoremstyle[
    spaceabove = 15pt, spacebelow=15pt,  
    headfont   = \bfseries,  
    headformat = $\blacktriangledown$ Example~\NUMBER\NOTE\vspace{-3mm},
    headpunct  = \newline\exrule\newline,    
    notefont   = \itshape,           
    notebraces = {--- }{},           
    bodyfont   = \leftskip=0em \rightskip=0em \itshape \rmfamily , 
    preheadhook   = \Needspace*{4\baselineskip},  
    postheadspace = 0pt,              
    prefoothook   = \ifhmode \hspace*{\fill}\newline \else \vspace{-5mm} \fi \exrule  
]{examplestyle}

\tcbset{
    thmbox/.style={
            enhanced,
            breakable,
            sharp corners=all,
            fonttitle=\bfseries\normalsize,
            fontupper=\normalsize\itshape,
            top=0mm,
            bottom=0mm,
            right=0mm,
            colback=white,
            colframe=white,
            colbacktitle=white,
            coltitle=black,
            attach boxed title to top left,
            boxed title style={empty, size=minimal, bottom=1.5mm},
            overlay unbroken ={
                    \draw (title.south west)--(title.south east);
                    \draw ([xshift=3.5mm]frame.north west)|-%
                    (frame.south east)--(frame.north east);},
            overlay first={
                    \draw (title.south west)--(title.south east);
                    \draw ([xshift=3.5mm]frame.north west)--([xshift=3.5mm]frame.south west);
                    \draw (frame.north east)--(frame.south east);},
            overlay middle={
                    \draw ([xshift=3.5mm]frame.north west)--([xshift=3.5mm]frame.south west);
                    \draw (frame.north east)--(frame.south east);},
            overlay last={
                    \draw ([xshift=3.5mm]frame.north west)|-%
                    (frame.south east)--(frame.north east);},
        },
    S/.style={thmbox,
            overlay unbroken ={
                    \draw (title.south west)--(title.south east);
                    \draw ([xshift=3.5mm]frame.north west)--([xshift=3.5mm]frame.south west);},
            overlay first={
                    \draw (title.south west)--(title.south east);
                    \draw ([xshift=3.5mm]frame.north west)--([xshift=3.5mm]frame.south west);},
            overlay middle={
                    \draw ([xshift=3.5mm]frame.north west)--([xshift=3.5mm]frame.south west);},
            overlay last={
                    \draw ([xshift=3.5mm]frame.north west)--([xshift=3.5mm]frame.south west);},
        },
    L/.style={thmbox,
            overlay unbroken ={
                    \draw (title.south west)--(title.south east);
                    \draw ([xshift=3.5mm]frame.north west)|-([xshift=15mm]frame.south west);},
            overlay first={
                    \draw (title.south west)--(title.south east);
                    \draw ([xshift=3.5mm]frame.north west)--([xshift=3.5mm]frame.south west);},
            overlay middle={
                    \draw ([xshift=3.5mm]frame.north west)--([xshift=3.5mm]frame.south west);},
            overlay last={
                    \draw ([xshift=3.5mm]frame.north west)|-([xshift=15mm]frame.south west);},
        },
    LQ/.style={thmbox,
            overlay unbroken ={
                    \draw (title.south west)--(title.south east);
                    \draw ([xshift=3.5mm]frame.north west)|-([xshift=10cm]frame.south west);
                    \node[anchor=east] at (frame.south east) {$\square$};},
            overlay first={
                    \draw (title.south west)--(title.south east);
                    \draw ([xshift=3.5mm]frame.north west)--([xshift=3.5mm]frame.south west);},
            overlay middle={
                    \draw ([xshift=3.5mm]frame.north west)--([xshift=3.5mm]frame.south west);},
            overlay last={
                    \draw ([xshift=3.5mm]frame.north west)|-([xshift=10cm]frame.south west);
                    \node[anchor=east] at (frame.south east) {$\square$};},
        },
}

\newtcbtheorem[]{LEM}{Lemma}{thmbox,S}{theo}
\newtcbtheorem[]{Proof}{Proof}{thmbox,LQ}{theo}

\usepackage[autostyle=tryonce,autopunct=true]{csquotes}

\usepackage{shapepar}
\usepackage{fancyvrb}
\usepackage{spot}
\usepackage{ifsym}
\usepackage{fancypar}
\usepackage{fancytabs}
\fancytabsHeight{5.25cm}
\definecolor{TestFonce}{cmyk}{1,1,1,0}
\definecolor{TestClair}{cmyk}{0.1,0.1,0.1,0}
\usepackage{textpos}
\usepackage{tikz-3dplot}
\usepackage[american]{circuitikz}
\usepackage{pgf}
\usepackage{tikz}
\usetikzlibrary{patterns}
\usetikzlibrary{shapes.geometric}

\usetikzlibrary{decorations.markings}
\usetikzlibrary{arrows}
\usetikzlibrary{fit}

\usepackage{textcomp}
\usetikzlibrary{babel}
\usepackage{schemabloc}
\usepackage{lettrine}

\usepackage{afterpage}

\usepackage[unicode=true, bookmarks=true, bookmarksnumbered=true, bookmarksopen=true, bookmarksopenlevel=1, breaklinks=true, pdfborder={0 0 0}, backref=true, colorlinks=true]{hyperref}

\definecolor{LinkCol}{rgb}{0,0,0.5}
\definecolor{CiteCol}{rgb}{0.5,0,0.1}

\usepackage{colortbl}
\arrayrulecolor{black}

\makeatletter

\def\cleardoublepage{\clearpage\if@twoside \ifodd\c@page\else%
            \hbox{}%
            \thispagestyle{empty}
            \newpage%
            \if@twocolumn\hbox{}\newpage\fi\fi\fi}

\makeatother

\usepackage[toc,page]{appendix}

\usepackage{csquotes}

\usepackage{verbatim}

\usepackage{enumitem}

\usepackage{microtype}

\newenvironment{Abstract}%
{\null\vfill\begin{center}%
        \bfseries \Large\abstractname\end{center}}%
{\vfill\null}

\usepackage[savewrites,toc, style=long,nonumberlist]{glossaries}

\newglossary[slg]{symbolslist}{syi}{syg}{Symbols}

\makeglossaries

\setacronymstyle{long-sc-short}

\newacronym{A}{***}{***} 

\newglossaryentry{S}{name=\ensuremath{***}, description={***}, type=symbolslist}

\usepackage{lipsum}

\newenvironment{eqaed}
    {\begin{equation}
    \begin{aligned}
    }
    { 
    \end{aligned}
    \end{equation}
    \ignorespacesafterend
    }

\usepackage{tikz}
\usetikzlibrary{matrix}

\tikzset{
  pretableaumatrix/.style={
    ampersand replacement=\&,
    matrix of math nodes,
    outer sep=1mm,
    inner sep=0mm,
    anchor=center,
    row sep={between borders,-\pgflinewidth},
    column sep={between borders,-\pgflinewidth},
    dottedentry/.style={densely dotted},
    dashedentry/.style={densely dashed},
    spaceentry/.style={draw=none,execute at begin node=\null},
  },
  pretableaunode/.style={
    font=\small,
    draw=gray,
    sharp corners,
    rectangle,
    anchor=base,
    text height=3.75mm,
    text depth=1.25mm,
    minimum height=5mm,
    minimum width=5mm,
    inner sep=0mm,
    outer sep=0mm,
    doublewidth/.style={minimum width=10mm},
    footnotesize/.style={font=\footnotesize},
    scriptsize/.style={font=\scriptsize},
  },
  tableaumatrix/.style={
    pretableaumatrix,
    every node/.append style={
      pretableaunode,
    },
  },
  medtableaumatrix/.style={
    pretableaumatrix,
    every node/.append style={
      pretableaunode,
      font=\footnotesize,
      text height=2.75mm,
      text depth=.75mm,
      minimum height=3.5mm,
      minimum width=3.5mm
    },
  },
  smalltableaumatrix/.style={
    pretableaumatrix,
    every node/.append style={
      pretableaunode,
      font=\scriptsize,
      text height=1.85mm,
      text depth=.15mm,
      minimum height=2.5mm,
      minimum width=2.5mm,
    },
  },
  tinytableaumatrix/.style={
    pretableaumatrix,
    every node/.append style={
      pretableaunode,
      font=\tiny,
      text height=1.25mm,
      text depth=.15mm,
      minimum height=1.75mm,
      minimum width=1.75mm
    },
  },
  tableau/.style={
    baseline=-1.25mm,
    every matrix/.style={tableaumatrix},
  },
  medtableau/.style={
    baseline=-1.25mm,
    every matrix/.style={medtableaumatrix},
  },
  smalltableau/.style={
    baseline=-1.25mm,
    every matrix/.style={smalltableaumatrix},
  },
  preshapetableaumatrix/.style={
    pretableaumatrix,
    execute at end cell={\strut},
    every node/.append style={
      draw=black,
      anchor=base,
      inner sep=0mm,
      outer sep=0mm,
    },
    shadedentry/.style={fill=gray},
    darkshadedentry/.style={fill=darkgray},
  },
  medshapetableaumatrix/.style={
    preshapetableaumatrix,
    every node/.append style={
      text height=2.75mm,
      text depth=.75mm,
      minimum height=3.5mm,
      minimum width=3.5mm
    },
  },
  shapetableaumatrix/.style={
    ampersand replacement=\&,
    matrix of math nodes,
    outer sep=0mm,
    inner sep=0mm,
    anchor=base,
    row sep={between borders,-\pgflinewidth},
    column sep={between borders,-\pgflinewidth},
    execute at begin cell={\strut},
    every node/.append style={draw,anchor=base,text height=1mm,text depth=.5mm,minimum size=1.5mm,inner sep=0mm,outer sep=0mm},
  },
  shapetableau/.style={
    every matrix/.style={shapetableaumatrix},
  },
  topalign/.style={
    every matrix/.append style={name=maintableau,anchor=maintableau-1-1.base},
    baseline,
  },
}

\newcommand*\smalltableau[2][]{\tikz[smalltableau,#1]\matrix{#2};}

\definecolor{green3}{RGB}{44,160,44}
\definecolor{red3}{RGB}{214,39,40}

\usepackage[square,numbers]{natbib}

\hypersetup{
    pdftitle={Dissertation \LaTeX\ Template},
    pdfauthor={A. Mhamdi},
    pdfsubject={Template for \LaTeX\ users},
    pdfkeywords={\LaTeX, PhD, Report.},
    colorlinks=true,
    linkcolor=red3,
    citecolor=green3,
    urlcolor=teal,
    filecolor=blue,
    pdfpagelayout=SinglePage,
    pdffitwindow=true,
    pdfnewwindow=true,
    pdfstartview=XYZ,
    pdftoolbar=true,
    pdfmenubar=true,
    plainpages=false
}

\usepackage{layout}

\begin{document}


\begin{titlepage}
\thispagestyle{empty}
\begin{center}
\begin{large}
Universit\`a degli Studi di Torino \\
{\bf Scuola di Dottorato} \\
\end{large}
\end{center}
\hrulefill

\vspace{2cm}
\begin{center}
 \includegraphics[scale=0.55]{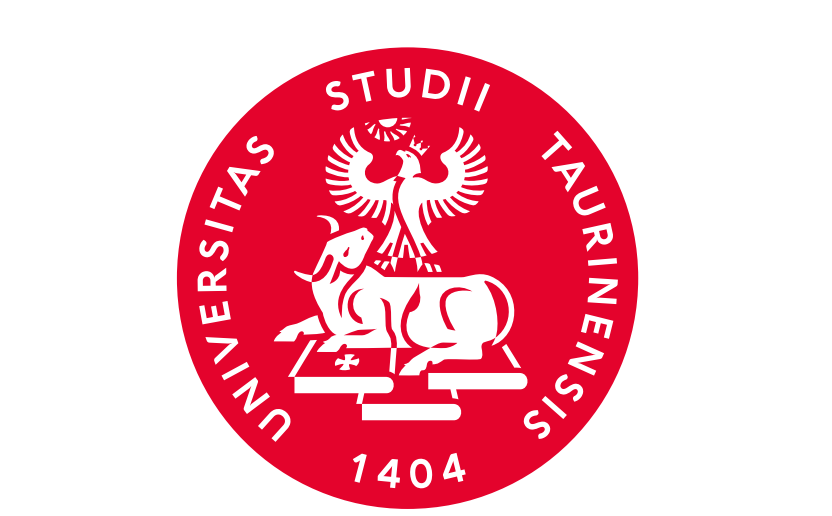}
 \end{center}

\vspace{6cm}

\Large{\bf Aspects of Stability, Rigidity and Unitarity in String Vacua}

\vspace{2cm}

\large{\bf Giorgio Leone}
\newpage
\pagenumbering{arabic}
\begin{center}
\begin{large}
Universit\`a degli Studi di Torino \\
{\bf Scuola~di~Dottorato}
\end{large}
\end{center}
\hrulefill
\begin{center}
\begin{large}
{\bf Dottorato in Fisica}
\end{large}
\end{center}

\vspace{2cm}
\Large{\bf Aspects of Stability, Rigidity and Unitarity in String Vacua}

\vspace{8cm}
\large{\bf Giorgio Leone}

\vspace{1cm}
\large{\bf PhD advisor: Carlo Angelantonj}
    
\end{titlepage}

\frontmatter

\voffset=0.9cm
\textheight= 615pt
\headsep=40pt

\dominitoc%

\microtypesetup{nopatch=item}
\thispagestyle{empty}
\vspace*{\stretch{1}}

\begin{flushright}
\emph{\scriptsize{
"Wovon man nicht sprechen kann, darüber muss man schweigen." \\
"Whereof one cannot speak, thereof one must be silent." \\
{\textbf L. Wittgenstein, Tractus Logico-Philosophicus, (1921). }
}}\end{flushright}

\clearpage
\thispagestyle{empty}
\null\newpage
\clearpage

\setstretch{1.45}

\newpage

\raggedbottom



\chapter*{Acknowledgements}
\thispagestyle{empty}

This thesis could not have been conceived without my advisor, prof. Carlo Angelantonj, to whom I am grateful in so many ways that the present few lines cannot encode to the full extent. First, I am very thankful and honoured for his support in my abilities and ideas even when I felt I did not deserve it. I am grateful for his kindness and availability in everything I have ever needed starting from the first weeks of the pandemic and ending with these last days of my PhD journey. I really enjoyed all the stimulating discussions we had in these years which also allowed me to learn a lot of physics. I want to thank him also for all the time and patience he spent reading and correcting my notes, papers and the present manuscript, and for his suggestions that have surely improved my writing ability. I am also very grateful to him for teaching me how to be a professional scientist and behave towards the scientific community. I will carry all these invaluable experiences with me in the future.

\noindent I would like to thank Prof. Augusto Sagnotti for his comments and interest in my work, which allowed me to continue my research activity. His support has an invaluable meaning to me. I would like to thank Prof. Ralph Blumenhagen for his feedback on the present manuscript and the stimulating discussions during my time in Munich and the conference in Corfou that improved my work.

\noindent I would like to thank my collaborators, Prof. Emilian Dudas, Cezar Condeescu and Prof. Ioannis Florakis, for the stimulating discussions that expanded my knowledge and opened new perspectives on the topics I have been working on.

\noindent A special thank is deserved to my collaborator and friend, Ivano Basile, who taught me a lot of new physics I did not know of and opened my eyes to new worlds to be explored, as well as for his kindness and availability even outside physics. His passion and enthusiasm have been stimulating and inspirational in my research activity.

\noindent During these years I had the opportunity to be involved in interesting conversations, from which my work undoubtedly benefited, with many brilliant colleagues and senior scientists including Prof. Steve Abel, Roberta Angius, Prof. Ignatios Antoniadis, Veronica Collazuol, Niccolò Cribiori, Arun Debray, Matilda Delgado, Prof. Keith Dienes, Markus Dierigl, Bernardo Fraiman, Mariana Grana, Matteo Inglese, Christian Kneissl, Cameron Krulewski, Prof. Carlo Maccaferri, Andriana Makridou, Prof. Marco Meineri, Miguel Montero, Luca Nutricati, Natalia Pacheco-Tallaj, Hector Parra De Freitas, Prof. Herve Partouche, Alessandro Pini, Salvatore Raucci, Flavio Tonioni, Prof. Nicolaos Toumbas, Irene Valenzuela and Matthew Yu.

\noindent I would like to thank also my PhD colleagues and friends with whom I have shared the office and everyday life in these years: Andrea Arduino, Valerio Belocchi, Gloria Bertolotti, Dripto Biswas, Niccolò Brizio, Andrea Bulgarelli, Elia Cellini, Valerio Belocchi, Elia de Sabbata, Jakub Kryss, Luca Orusa, Federica Rossi, Alberto Ruffino, Francesco Sarandrea, Antonio Smecca and Paolo Vallarino. I am honoured to have shared my intellectual space with them.

\noindent I would like to thank my mother, my father and my brother who have always supported me, even during my tough times.

\noindent I am extremely grateful to Claudio Arnetoli, who helped me go through difficult times and grow as a human being and scientist. 

\noindent I would like to thank my old friends Cecilia Belletti, Fabio Buccoliero, Bo De Baene, Stefania Frand-Genisot, Darix Gribaudo, Paolo Racca, Federico Russo and Isabella Sanna who have always been there for me even though our paths took different directions.

\noindent Finally, I would like to thank my partner Micol, for her support in these years. I am very grateful for bearing the burden of my anxiety in the challenges that I had to face, still giving me all the strength and motivation I needed to overcome them and grow as a human being.

\clearpage
\thispagestyle{empty}
\null\newpage
\clearpage

\voffset=0.9cm
\textheight= 615pt
\headsep=40pt

\phantomsection
\pdfbookmark[0]{R\'esum\'e{\textendash}Abstract}{abstract}

 \setstretch{1.45}
\begin{Abstract}

\noindent In this Thesis we investigate properties of stability and unitarity of the string landscape in ten and lower dimensions. The dissertation explores these aspects by intertwining a detailed analysis of string vacua, with and without supersymmetry, with a bottom-up study driven by unitarity. In particular, in Chapter \ref{ChMisSUSY} the possibility of formulating a necessary and sufficient condition for the classical stability of non-supersymmetric string vacua is discussed, emphasising the examples in ten and nine dimensions. In Chapter \ref{ChBSB}, new solutions are presented in six dimensions both for non-supersymmetric BSB vacua and those with minimal $\mathcal{N}=(1,0)$ supersymmetry, arising from a non-trivial cancellation of the R-R tadpoles. In particular, solutions for the $T^4/\mathbb{Z}_6$ supersymmetric and BSB orientifolds are discussed, including a non-trivial non-supersymmetric vacuum that cannot be further deformed, a property absent in its supersymmetric counterpart. In addition, the insertion of probe strings charged with respect to R-R $2$-forms is discussed in both cases, along with the unitarity conditions for the theory living on the world-volume to be adapted to these scenarios. Finally, Chapter \ref{ChUnitarity} discusses the role played by a new kind of global anomaly, arising from the inconsistency of effective field theories under topology change. A systematic analysis of six-dimensional supergravity theories with $\text{SU}(2)$ gauge group and one tensor multiplet and $\text{U}(1)$ gauge group with no tensor multiplets is discussed. Novel constraints are found and their cancellation in string vacua is verified. The latter discussion involves compactifications of supersymmetric heterotic theories whose related anomaly is guaranteed to cancel by a known theorem in the literature. In addition to supersymmetric heterotic theories, we have investigated the $\text{SO}(16)\times \text{SO}(16)$ heterotic theory compactified on the $T^4/\mathbb{Z}_6$ orbifold and the Gepner orientifold with no tensor multiples, showing how such anomalies are cancelled. Even though expected, this result provides a non-trivial check for the consistency of these vacua, which was not guaranteed to hold {\em a priori} by any theorem known in the literature.

\end{Abstract}


\pagenumbering{roman}
\tableofcontents

\phantomsection
\glsaddall 
\newpage 

\mainmatter


\voffset=0.9cm
\textheight= 615pt
\headsep=40pt

\voffset -0.7cm

\chapter{Introduction}

 \setstretch{1.45}

\footskip -20pt

\newpage

\raggedbottom


 \voffset=0.9cm

\section{Introduction}
 
\noindent String theory  \cite{Green:2012oqa, Green:2012pqa, Polchinski:1998rq, Polchinski:1998rr, Becker:2006dvp, Blumenhagen:2013fgp, Kiritsis:2019npv, Johnson:2023onr} has provided a successful framework in which gravity, matter and gauge interactions admit a consistent unified quantum description. Despite its enormous success and despite the impressive progress in the last decades, many questions are still unanswered, especially when trying to make contact with low-energy phenomenology. Indeed, string theory is characterised by universal dynamics in the trans-Planckian regime \cite{Gross:1987ar, Gross:1987kza, Mende:1989wt} constraining the behaviour of scattering amplitudes at the Planck scale, which, however, cannot be directly tested by nowadays experiments. Thus, with the state of the art of our technology, making contact with the current phenomenology requires studying the theory in the low energy regime, aiming to incorporate the well-established standard model in a proper limit. However, the infra-red (IR) physics admits manifold manifestations and the task turns out to be challenging both technically and conceptually. In particular, the need to provide a stringy description of supersymmetry breaking has dramatic consequences on the phenomenology, and on the structure and dynamics of the theory itself. In the last forty years, many ways of breaking supersymmetry have been realised in heterotic string vacua \cite{Rohm:1983aq, Kounnas:1988ye, Ferrara:1988jx, Antoniadis:1988jn, Kounnas:1989dk, Antoniadis:1990ew, Kiritsis:1997ca}, through non-trivial magnetic fields in open strings \cite{Bachas:1995ik, Berkooz:1996km, Angelantonj:2000hi, Blumenhagen:2000wh, Aldazabal:2000cn, Cvetic:2001tj, MarchesanoBuznego:2003axu,  Larosa:2003mz, Blumenhagen:2006ci}, via type 0 \cite{Sagnotti:1995ga, Angelantonj:1998gj} and type II orientifolds with a non-trivial action on the compact directions \cite{Blum:1997gw, Antoniadis:1998ki, Antoniadis:1998ep, Angelantonj:1999gm}, with orientifold planes having positive tension and R-R charges, both in ten \cite{Sugimoto:1999tx} and lower dimensions \cite{Antoniadis:1999xk,  Aldazabal:1999jr, Angelantonj:1999ms, Angelantonj:2000xf, Angelantonj:2024iwi}, and via a combination of all these effects \cite{Angelantonj:2003hr, Angelantonj:2005hs}. However, in most of these models, tachyonic instabilities arise, implying that the background cannot be trusted even classically, since, after tachyon condensation, it is expected to decay into a new, at least metastable, vacuum. This process is relatively under control when the tachyon is associated to the instability of (anti)brane configurations in simple set-ups \cite{Sen:1998sm, Sen:1999nx, Sen:2002in, Sen:2004nf, Schnabl:2005gv}, while only in recent years some progress has been made in understanding the situation when the tachyon belongs to a ten-dimensional heterotic spectrum \cite{Hellerman:2007zz, Kaidi:2020jla, Kaidi:2023tqo}\footnote{A previous study showing similar features has been discussed in \cite{Antoniadis:1991kh} for the type IIB superstring theory at finite temperature.}, and yet a systematic analysis is lacking. Such features are characteristic of tachyonic theories and thus one may wonder whether they underlie an intrinsic property which may distinguish them from non-tachyonic ones.

\noindent In closed-string vacua, the interplay between the ultra-violet (UV) and IR degrees of freedom, guaranteed by the presence of modular invariance, allows one to address this issue by looking at the massive excitations of the string spectrum and thus at its global behaviour. A first attempt in this direction was carried out in \cite{Kutasov:1990sv} and led to the development of the so-called {\em asymptotic supersymmetry}, according to which the IR finiteness of the cosmological constant implies an overall cancellation between on-shell bosonic and fermionic degrees of freedom. A few years later, a refined analysis involving the Rankin-Selberg-Zagier transform \cite{Rank, Selb, Zagier2} revealed that the sum of the degrees of freedom within a given cut-off is proportional to the vacuum energy up to oscillating terms depending on the non-trivial zeros of the Riemann zeta function \cite{Angelantonj:2010ic}. This outcome extends the result of \cite{Kutasov:1990sv}, since it gives an expression for the sum over the massive states for a finite value of the cut-off, but fails to uncover the origin of classical stability. Indeed, although very appealing from a mathematical perspective, it is not clear how to adapt such theorems to the cases in which the integrand grows exponentially at the cusp and admits a non-vanishing Fourier zero mode, the relevant situation for tachyonic vacua\footnote{If the integrand function grows exponentially but admits a vanishing zero mode, there is the possibility of regulating the partition function by subtracting the exponentially divergent piece, according to a conjecture by Zagier. In this case the vacuum admits unphysical tachyons and an explicit realisation of Zagier's conjecture has been shown in \cite{Angelantonj:2010ic}.}. 

\noindent A different route was followed in \cite{Dienes:1994np}, where it was observed that, in a tachyonic-free vacuum, bosons and fermions follow an oscillating pattern as the mass increases\footnote{The same observation has been made in \cite{Cudell:1992bi} where the oscillations between baryons and mesons were traced back to an underlying modular invariance.}, called {\em misaligned supersymmetry}. These oscillations have been described by the large-mass behaviour of the partition function once the mass levels are continued to real values, by defining the so-called {\em sector-averaged sum}. Indeed, such a continuation interpolates among the {\em physical} masses of the tower associated to each character of the conformal field theory (CFT), thus allowing to realise a mutual, if partial, cancellation when bosonic and fermionic characters are added together. In \cite{Dienes:1994np}, the cancellation was proven to hold for the leading term and was conjectured to take place also for the sub-leading ones if and only if tachyons are absent in the spectrum. Moreover, the complete cancellation was conjectured to be reflected in the observed oscillations, which would then provide a necessary and sufficient condition for classical stability. These conclusions have been supported by the analysis in \cite{Cribiori:2020sct}, where a complete cancellation was shown to occur for the $\text{SO}(16) \times \text{SO}(16)$ heterotic string in $D=10$ \cite{Dixon:1986iz, Alvarez-Gaume:1986ghj}, with bosons and fermions following the conjectured oscillatory pattern. Although a significant step forward was taken in \cite{Cribiori:2020sct}, proving Dienes' conjecture calls for a better understanding of the features of tachyonic vacua. Indeed, simple numerical computations (see also \cite{Faraggi:2020fwg, Sara} for previous analyses) actually reveal a {\em slightly} more complicated story. For instance, oscillations have been observed in all the non-supersymmetric heterotic theories in ten dimensions, as well as for the tachyonic region of the moduli space for a Scherk-Schwarz compactification of the heterotic and type IIB superstring in $D=9$. These models provide counter-examples to the presence of oscillations as a necessary and sufficient condition for the absence of tachyons, but still leave the possibility of ascribing the latter to a complete cancellation of the large-mass behaviour of the string spectrum. The issue was finally settled in \cite{Angelantonj:2023egh}, where it was understood that oscillations are reflected in the presence of fermions, which also determines the cancellation of the leading growth of the sector-averaged sum. Furthermore, it was shown that a complete cancellation takes place only when the spectrum has no tachyonic instabilities, thus proving Dienes' conjecture\footnote{It can be shown that the behaviour of the sector-averaged sum is dictated by the deepest tachyon, consistently with the observation in \cite{Dienes:2012dc}.}. In view of these considerations, the oscillations are not reflected in the vanishing of the sector-averaged sum, and thus one needs to differentiate the two concepts, which were originally thought to be linked. However, since the former is trivially a consequence of the presence of fermions, it is more appropriate to define {\em misaligned supersymmetry} as the vanishing of the sector-averaged sum. 

\noindent It is natural to try to extend the analysis to the study of classical stability for orientifold vacua \cite{Sagnotti:1987tw, Pradisi:1988xd, Horava:1989vt, Bianchi:1990yu, Bianchi:1990tb, Bianchi:1991eu, Dudas:2000bn, Angelantonj:2002ct}. However, in such a case, the interplay between the IR and UV properties of the spectrum, previously guaranteed by modular invariance, is lost, and thus the physical meaning of the {\em orientifolded} sector-averaged sums might not encode tachyonic instabilities. Indeed, the modular group maps different descriptions of the same amplitude, while one-loop vacuum amplitudes are mapped to the tree-level propagation of unoriented closed-string states between boundaries or crosscaps (or D-branes or O-planes) by an S or P transformation. This connection, as first noted in \cite{Israel:2007nj, Niarchos:2000kw}, implies that the sector-averaged sums for orientifold amplitudes grow exponentially if and only if a tachyonic character is present in the dual description \cite{Leone:2023qfd}. As a consequence, from such analysis, one can only draw either a necessary condition for the absence of tachyons in the closed-string spectrum or a sufficient criterion for the classical stability of the vacuum. Still, it fails to provide a necessary and sufficient condition, since a non-trivial action of the orientifold projection implies the associated sector-averaged sums are non-vanishing although the tachyon is projected away. It would then be required to realise a mutual cancellation between sector-averaged sum on different Riemann surfaces, which cannot take place. Therefore, misaligned supersymmetry cannot be considered the right tool to uncover the reason behind the classical stability of orientifold vacua, and new ideas and technologies are needed to finally address the issue.   

\noindent Although the vast majority of non-supersymmetric models contain tachyonic instabilities, a few allowed solutions are classically stable, as it happens, in ten dimensions, for the aforementioned heterotic $\text{SO}(16) \times \text{SO}(16)$ model, the type 0'B superstring \cite{Bianchi:1990yu} and the Sugimoto model \cite{Sugimoto:1999tx}. Among the $10D$ tachyon-free vacua, the Sugimoto model \cite{Sugimoto:1999tx} stands out since the tree-level closed-string amplitudes are supersymmetric, as reflected in the presence of the massless ten-dimensional gravitino, while bosons and fermions transform into different representations of the  $\text{USp}(32)$ gauge group, thus breaking supersymmetry. Although the simultaneous presence of the massless gravitino and of non-supersymmetric matter seems at first sight puzzling, a consistent coupling is achieved since supersymmetry is realised non-linearly \cite{Dudas:2000nv} on the open-string sector, with a massless goldstino playing the role of the Volkov-Akulov field \cite{Volkov:1972jx}. Furthermore, in such a setting, the breaking of supersymmetry must be regarded as taking place at the string scale, since there is no order parameter that may recover supersymmetry. Microscopically, such configuration is realised through the presence of orientifold planes with positive tension and R-R charge, O9$_+$, which calls for the introduction of anti-branes, $\overline{\text{D9}}$, to avoid anomalies. As a result, the R-R tadpole of the non-dynamical ten-form vanishes, but the uncancelled NS-NS dilaton tadpole contributes to the scalar potential. Such corrections, however, identify a new vacuum in which the background geometry is modified  \cite{Fischler:1986ci, Fischler:1986tb}, but for which a full-fledged string theory analysis is missing, since it is not known how to quantise the theory on the new background nor how to describe it in the wrong vacuum \cite{Dudas:2004nd, Keller:2007nd, Kitazawa:2008hv, Pius:2014gza}.  This feature is not peculiar to the Sugimoto model but also applies to the other non-supersymmetric vacua that are classically stable. In this scenario, one can thus, at most, develop a low energy effective field theory analysis that can be trusted up to the order of the dilaton tadpole. Although not exhaustive, such approach has been extremely successful in enriching our insights on the geometry characterising these vacua, leading to interesting features such as spontaneous compactification on a space with boundaries or a distinctive cosmological evolution \cite{Dudas:2000ff, DeWolfe:2001nz, Gubser:2001zr, Dudas:2002dg, Dudas:2010gi, Mourad:2016xbk, Basile:2018irz, Antonelli:2019nar, Basile:2020mpt, Basile:2021krk, Mourad:2021qwf, Mourad:2021roa, Raucci:2022bjw, Baykara:2022cwj, Raucci:2022jgw, Mourad:2022loy}.

\noindent A similar situation can be realised in lower dimensions, where however the presence of orientifold planes of different dimensionality provides richer scenarios. For instance, in $D=6$, the standard orientifold projection can be dressed with an involution, $\sigma$, acting on the twisted two-cycles, which trades the standard O5$_-$ planes of the supersymmetric orientifolds \cite{Bianchi:1990yu, Gimon:1996ay, Dabholkar:1996pc} with O5$_+$, with positive tension and charge\footnote{This is not the only combination that can be realised in $D=6$, since the trivial and $\sigma$ involutions are still compatible with the lower dimensional version of the Sugimoto vacuum involving O9$_+$ and O5$_+$ planes and its variation with O9$_+$ and O5$_-$, respectively. However, these cases are straightforward generalisations of the results discussed in this thesis and we will not comment on these possibilities any further.}. Therefore one is forced to introduce $\overline{\text{D5}}$ branes to cancel the R-R tadpoles of the non-dynamical six-form, giving rise to a non-linear realisation of supersymmetry \cite{Pradisi:2001yv}, in a similar way as the $10D$ case. Such configuration realises what in literature is known as Brane Supersymmetry Breaking (BSB), whose simplest setting was first discussed in \cite{Antoniadis:1999xk}. Here, the world-sheet parity operator $\varOmega$ is dressed with the involution $\sigma$ and the $\mathbb{Z}_2$ action arising from the point group of the orbifold compactification $T^4/\mathbb{Z}_2$. As a result, the fixed loci of the orientifold action identify O9$_-$ and sixteen O5$_+$ planes whose contributions to the tadpole of the non-dynamical R-R forms are cancelled by placing $\overline{\text{D5}}$ branes on a single $\mathbb{Z}_2$ fixed point. Such a vacuum is anomaly-free but leaves behind a non-vanishing scalar potential generated by the untwisted NS-NS tadpoles. However, although this is the simplest choice, this is not the only possible one, and one still has the freedom to distribute the $\overline{\text{D5}}$ branes among all the fixed points \cite{Angelantonj:2024iwi}. This induces a non-trivial cancellation of the D9 branes twisted charge against that of $\overline{\text{D5}}$ ones, with a corresponding reduction of the open-string moduli that only allows deformations of the model via complete brane recombination \cite{Angelantonj:2011hs}. Furthermore, uncancelled twisted NS-NS tadpoles are present and determine a richer potential, including contributions from twisted scalars. The latter may lead to a very interesting dynamics since non-vanishing vevs would lead to a spontaneous resolution of the orbifold singularities.

\noindent If for the $T^4/\mathbb{Z}_2$ BSB orientifold placing $\overline{\text{D5}}$ branes on {\em all} the fixed points is optional, and indeed such configuration is connected to the standard solution in \cite{Antoniadis:1999xk} via brane recombination, for the BSB orientifold built on the $T^4/\mathbb{Z}_4$ orbifold it is mandatory. The crucial difference resides in the fact that now O5$_+$ planes carry twisted charges for the non-dynamical R-R six-forms to be cancelled by those of D-branes. Hence, the model in such a case is {\em truly} rigid since D-branes cannot be recombined altogether, without leading to an anomalous model \cite{Angelantonj:2024iwi}. Fractional orientifold planes may seem exotic but are present even in more standard supersymmetric realisations \cite{Gimon:1996ay, Dabholkar:1996pc} involving the $\mathbb{Z}_3$ and $\mathbb{Z}_6$ point groups or the four-dimensional $Z$ orientifold \cite{Angelantonj:1996uy}. The $T^4/\mathbb{Z}_3$ orbifold does not admit a BSB variant, while the BSB version of the  $T^4/\mathbb{Z}_6$  was discussed in \cite{Angelantonj:2024iwi}, and still involves fractional O9$_-$ and O5$_+$ planes. Similarly to the $T^4/\mathbb{Z}_4$ case, the branes configuration guarantees the cancellation of the twisted R-R tadpoles, but, since there is only one $\mathbb{Z}_6$ fixed point, the main properties of this vacuum follow closely those of the $T^4/\mathbb{Z}_2$ model. Nevertheless, such a model admits a solution with rigid branes, with a scalar potential comprising blown-up moduli, potentially leading to a very interesting dynamics. 

\noindent The low-energy effective action for these BSB vacua, with and without additional contributions to the scalar potential, admits a geometric interpretation \cite{Pradisi:2001yv}. Indeed, the gauge couplings can be expressed as a combination of scalars, $J$, which determines the metric of the moduli space in the tensor branch, similarly to what happens for the ${\mathcal N}=(1,0)$ supersymmetric vacua in $D=6$ \cite{Romans:1986er, Sagnotti:1992qw, Ferrara:1997gh, Riccioni:1998th}. In the latter case, however, the $J$ form\footnote{In F-theory vacua the $J$-form plays the role of the K\"ahler form \cite{Kim:2019vuc}, but we will refer to it as such even outside the realm of F-theory.}  is related by supersymmetry to the Wess-Zumino counterterms induced by the generalised Green-Schwarz-Sagnotti (GSS) mechanism \cite{Green:1984sg, Green:1984bx, Sagnotti:1992qw}.  Therefore, with $\mathcal{N}=(1,0)$ supersymmetry, the gauge couplings are linked to the data of the R-R tadpoles, which determine the structure of the reducible anomaly polynomial. However, this connection is only present in supersymmetric settings, since the real {\em stringy} origin of the gauge couplings is tied to the NS-NS tadpoles, and thus one should {\em a priori} use these data to characterise the $J$-form. This is manifest when supersymmetry is broken {\em à la} BSB, so that the na\"ive construction of the $J$ form in terms of the 't-Hooft anomaly coefficients induces ghost-like couplings for scalars and gauge fields in the low energy action \cite{Angelantonj:2020pyr}, spoiling unitarity.  This does not imply that the K\"ahler form cannot be defined, and indeed an explicit solution that preserves unitarity is obtained using the structure of the NS-NS tadpoles, thus solving the puzzle raised in \cite{Angelantonj:2020pyr}. Of course, this observation is consistent with the construction of the K\"ahler form for supersymmetric vacua, where R-R and NS-NS tadpoles are equal, but allows to extend the definition also to cases when supersymmetry is non-linearly realised. 

\noindent Once the $J$-form is found, the unitarity conditions arising from the introduction of probe string defects, demanded by the completeness hypothesis \cite{Polchinski:2003bq, Banks:2010zn}, can be analysed quantitatively adapting to the BSB context the results of \cite{Kim:2019vuc}. These models correspond to {\em bona fide} string theory vacua, and thus are expected to be consistent, so that these unitarity bounds have to be interpreted as a consistency check, in the same spirit as the cancellation of gauge and gravitational anomalies. The original formulation of \cite{Kim:2019vuc} requires the presence of $\mathcal{N}=(0,4)$ supersymmetry on the defect, but the validity of the completeness hypothesis should go beyond the realm of supersymmetry and can be adapted to BSB vacua, since the microscopic theory living on the defect is known. This point of view is also instrumental in extending the conditions of \cite{Kim:2019vuc} to non-supersymmetric set-ups, completing the task initiated in \cite{Angelantonj:2020pyr}. Actually, a generalisation of the latter is also required for supersymmetric models, since in both cases defects admitting an interpretation in terms of instantons of the gauge theory of the corresponding higher dimensional brane realise a Ka$\check{\text{c}}$-Moody algebra on both left and right moving sectors of the CFT in the IR. Therefore, the precise formulation of the unitarity conditions of such {\em instanton} defects would require a detailed knowledge of the CFT realised by the charged non-chiral bosons, which is notoriously difficult to achieve. 

\noindent From a bottom-up perspective, the generalisation of the unitarity constraints of string defects might play a non-trivial role in constraining the landscape of the consistent effective field theories (EFT) that can be coupled to gravity. However, to make such conditions relevant for the swampland program \cite{Vafa:2005ui} (see \cite{Brennan:2017rbf, Palti:2019pca, vanBeest:2021lhn,  Agmon:2022thq} for reviews), one ought to develop an independent low-energy argument through which they can be formulated. In $\mathcal{N}=(1,0)$ supergravity, the values of the central charges $c_{\text{L}}, \, c_{\text{R}}$ of the CFT in the IR can be deduced from the knowledge of the 't-Hooft coefficient of the $\text{SU}(2)$ R-symmetry, since $c_{\text{R}}$ simply follows from the properties of the $\mathcal{N}=(0,4)$ supersymmetry algebra in $D=2$. Therefore, the unitarity bounds translate into a correct identification, in the anomaly polynomial, of the R-symmetry that in \cite{Kim:2019vuc} has been associated with the maximal subgroup of the orthogonal group of the non-compact space transverse to the defect. However, such an identification does not apply to instanton defects, and thus one needs to extend the analysis to spot it correctly\footnote{When supersymmetry is broken, there is no R-symmetry to be realised so that the argument should be rephrased completely.} (determining the central charge realised on the right moving sector of the CFT). Despite the difficulties, unitarity conditions from probe defects supporting the identifications of \cite{Kim:2019vuc} and from anomaly cancellation of the bulk theory are very effective in sharpening the supersymmetric landscape  \cite{Kumar:2010ru, Park:2011wv, Taylor:2018khc, Kim:2019vuc, Lee:2019skh, Kim:2019ths, Katz:2020ewz, Tarazi:2021duw, Morrison:2021wuv, Martucci:2022krl, Baykara:2023plc, Hayashi:2023hqa}, even beyond the $6D$ case \cite{Montero:2020icj, Hamada:2021bbz, Bedroya:2021fbu}. Further constraints can be deduced for theories with at least one tensor multiplet through the {\em null charge} conjecture \cite{Angelantonj:2020pyr}, which requires the existence of a defect whose charge $Q$ satisfies $Q \cdot Q=0$. The latter implies that every consistent realisation of an EFT in $D=6$ should arise from an internal theory admitting a description that is either geometric or dual to a geometric one.  

\noindent Although these unitarity constraints have been extensively discussed, recently it has been understood that the full extent of the consistency conditions arising from unitarity has not yet been harnessed. Indeed, a new perspective on the meaning of anomalies paved the way for the exploration of new consistency conditions, suited for vacua both with and without supersymmetry. Traditionally, anomalies can be traced to the presence of ambiguities in the definition of the partition functions of chiral fields\footnote{For chiral $p$-form fields even the action is ill-defined.}. However, the set of theorems stated and proved by Dai and Freed \cite{Freed:1986hv, Dai:1994kq}, and further extensions \cite{Freed:2016rqq, Yonekura:2018ufj}, allowed to resolve such ambiguities by interpreting chiral fields living on $X$ as a boundary mode of a non-chiral gapped one living in one dimension higher, $Y$, such that $\partial Y=X$. For a generic quantum field theory (QFT), the allowed spaces $Y$ can be properly restricted \cite{Witten:2015aba}, but in a theory of quantum gravity, topology is allowed to change, so that $Y$ can be arbitrarily chosen among the spaces preserving the structure of $X$. In this sense, the anomaly is not related anymore to the failure of gauge invariance but to the dependence of the theory on the arbitrary manifold $Y$. The latter is encoded in a topological quantum field theory, denoted for clear reasons the {\em anomaly theory}, defined on all the {\em closed} spaces $Y_{\text{cl}}$ obtained by glueing two different choices for $Y$. The anomaly theory evaluated on all the possible manifolds $Y_{\text{cl}}$ identifies the so-called  {\em Dai-Freed anomalies} \cite{Garcia-Etxebarria:2018ajm}, which comprise local anomalies \cite{Alvarez-Gaume:1983ict, Alvarez-Gaume:1983ihn, Alvarez-Gaume:1984zlq, Alvarez-Gaume:1984zst, Alvarez-Gaume:2022aak} whenever $Y_{\text{cl}}$ is himself a boundary and global ones \cite{Witten:1982fp, Witten:1985xe, Witten:1985mj, Witten:2019bou} when $Y_{\text{cl}}$ is a mapping torus, but they also include more general anomalies arising from spacetime topology change. Computing the anomaly theory on every closed manifold seems still a daunting task. However, this is simplified by taking advantage of the Dai-Freed theorems. Indeed, it has been shown that, for two manifolds that are equal up to a boundary, the anomaly theory differs only through the local anomaly. Thus a vanishing local anomaly allows to reorganise all closed spaces $Y_{\text{cl}}$ in terms of bordism classes preserving the structure of $X$, for which the anomaly theory is a topological invariant. The set of bordism classes forms a discrete abelian group $\Omega^{X_{\text{str}}}_{D+1}$ under the disjoint union, and thus it is enough to show that the theory of interest is anomaly-free if the anomaly theory vanishes on the generators of $\Omega^{X_{\text{str}}}_{D+1}$.

\noindent In six-dimensional string vacua, local anomalies are cancelled via the GSS mechanism, and thus we shall require the extended $7D$ closed manifolds to preserve the associated Bianchi identity. Demanding that the Bianchi identity be preserved at the level of integer cohomology identifies the {\em twisted string} structure used to organise bordism classes. Specifically in six dimensions, the chiral $2$-form fields entering the GSS mechanism belong to the gravity and tensor multiplets, so that multiple Bianchi identities are to be satisfied at the level of differential forms. Whether all the Bianchi identities have to be uplifted to integer cohomology is the subject of an open debate, since in \cite{Basile:2023zng} it was shown that satisfying only one Bianchi identity at the level of integer cohomology implies anomaly cancellation for known heterotic string theory vacua. In this thesis, following \cite{Basile:2023zng}, we shall make the minimal choice of satisfying at least one of the Bianchi identities, while still requiring the freedom to choose consistently all the ones present at the level of de Rham cohomology. The shape of the bordism groups associated to this choice is not known in the mathematical literature, and therefore a systematic analysis of Dai-Freed anomalies for string theory vacua along the lines of \cite{Basile:2023knk} cannot be performed\footnote{A different route has been taken in \cite{Tachikawa:2021mvw}, where using the technology of the equivariant topological modular forms \cite{hopkins} they have been able to show the absence of such anomalies for all the heterotic supersymmetric string vacua, generalising the $2D$ case of \cite{Tachikawa:2021mby}.}. Nevertheless, it is still possible to perform the analysis on those backgrounds on which the anomaly theory can be exactly computed\footnote{There could still be the possibility to have a bordism group generated by a background on which our computational power is lost.}, aiming on the one hand to verify its cancellation for string vacua and on the other hand to exclude inconsistent low energy effective theories \cite{Basile:2023zng}. Such non-trivial backgrounds are the Lens spaces $L_p^7$, obtained as the quotient $S^7/\mathbb{Z}_p$, satisfying at least one Bianchi identity, for which the anomaly theory of fermions is well known \cite{Debray:2021vob, Debray:2023yrs}. Furthermore, theories admitting chiral $2$-forms contribute through the so-called {\em quadratic refinement}, reproducing locally the GSS term and defined through a characteristic equation linked to the cohomology pairing on the closed $D+1$ manifold. {\em A priori} every solution to the latter is a consistent quadratic refinement, and up to now it is not known how to deduce such information from an independent path. This piece of information is crucial for the anomaly cancellation in $D=6$ supersymmetric theories \cite{Dierigl:2022zll}, and in \cite{Basile:2023zng} we considered all the possible solutions on $L_p^7$ for which only a specific choice leads to a vanishing anomaly theory. In particular, the shape of the quadratic refinement for spin bordism \cite{Hsieh:2020jpj} turns out not to be suited for the consistency of the EFTs, and thus for preserving unitarity. Put differently, since the quadratic refinement characterises the partition function of the chiral $2$-forms, requiring a specific choice of the latter identifies specific classes of chiral forms, excluding all the other vacua characterised by different choices. In \cite{Basile:2023zng}, $\mathcal{N}=(1,0)$ supergravity theories in six dimensions with at most one tensor multiplet were studied for abelian and simply-laced gauge groups, for which theories admitting non-chiral $2$-forms are free of Lens space anomalies, while the same result does not hold for those theories admitting chiral forms unless a specific choice for the quadratic refinement is made. In particular, up to this condition that still requires an explanation, infinite families with arbitrary $\text{U}(1)$ charges and no tensor multiplets turn out to be excluded. Or, phrased differently, Dai-Freed anomalies for a $\text{U}(1)$ gauge group restrict the possible chiral $2$-forms to a very specific choice whose features are still unclear. On the other hand, the anomaly cancellation was verified for explicit string constructions involving the compactification of the supersymmetric heterotic theories on K3 orbifolds and the non-supersymmetric $\text{SO}(16) \times \text{SO}(16)$ heterotic vacuum on the same compactification spaces, as well as the Gepner orientifold with no tensor multiplets. The choice for the quadratic refinement in \cite{Hsieh:2020jpj} is not suited to guarantee the consistency of these string vacua.  

\section*{Outline}

\noindent In this thesis, we will review the results in \cite{Angelantonj:2023egh, Leone:2023qfd, Angelantonj:2024iwi} and \cite{Basile:2023zng}, by discussing the classical stability of non-supersymmetric vacua, the rigidity of brane configuration in $D=6$ orientifolds and unitarity conditions from probe strings and Dai-Freed anomalies, intertwined with a bottom-up analysis for $6D$ supergravities. For each aspect, we shall begin with a general discussion before addressing particular examples in which these features are concretely realised. To this end, the present dissertation is divided into four chapters, organised as follows:

\begin{description}
  	\item[Chapter~\ref{ChMisSUSY}] contains a discussion on the classical stability of non-supersymmetric string vacua in ten and nine dimensions, focusing on the possibility of distinguishing the two cases. In particular, we will show that it is possible to identify a necessary and sufficient condition for the absence of tachyons for closed-string vacua, while a similar criterion cannot be formulated when orientifold projections are performed.  
   
  	\item[Chapter~\ref{ChBSB}] explores new solutions for six-dimensional vacua in cases where supersymmetry is realised both linearly and non-linearly, discussing local anomalies and unitarity constraints from string probes. Particular emphasis will be placed on the $T^4/\mathbb{Z}_6$ orientifolds.
   
  	\item[Chapter~\ref{ChUnitarity}] analyses the role played by Dai-Freed anomalies in six-dimensional string vacua, with and without supersymmetry, and $\mathcal{N}=(1,0)$ supergravity with at most one tensor multiplet. In particular, we shall discuss, as examples, supergravities admitting an $\text{SU}(2)$ or $\text{U}(1)$ gauge group with one and zero tensor multiplets. Gepner orientifolds are then addressed, along with the $\text{SO}(16) \times \text{SO}(16)$ heterotic string compactified on the $T^4/\mathbb{Z}_6$ orbifold.
   
  	\item[Chapter~\ref{conclusion}] summarises our results and provides possible extensions and outlook.
\end{description}

\chapter{Classical stability: Misaligned Supersymmetry} \label{ChMisSUSY}

	\newpage

 \setstretch{1.45}

\section{Roots of Misaligned Supersymmetry: the sector-averaged sum}

Stability is one of the main issues to be faced when trying to characterise the dynamics and the features of a non-supersymmetric string theory vacuum. Indeed, for models admitting space-time supersymmetry, the GSO  projection \cite{Gliozzi:1976jf, Gliozzi:1976qd} eliminates the tachyon from the string spectrum and the vacuum-to-vacuum amplitudes on a generic Riemann surface vanish. As a result, tachyonic instabilities and quantum corrections to the background geometry are absent at each order in perturbation theory, thus implying that these models are perturbatively stable. However, when space-time supersymmetry is absent or broken, nothing prevents tachyons and quantum corrections from being present, and indeed characterising the stability of non-supersymmetric string vacua is still the object of intense activity. Even though understanding the final fate of these vacua is still an open problem, a significant step forward has been taken in recent years in finding a property that protects the classical stability\footnote{Here and in the rest of the thesis classical stability is used as a synonym for the absence of tachyons in the tee-level spectrum.} of the vacuum, known in the literature as {\em misaligned supersymmetry}.

\noindent Misaligned supersymmetry was first introduced by Keith Dienes for closed-string vacua in \cite{Dienes:1994np} and relies on the interplay between the UV and IR regimes implied by modular invariance. The key observation in \cite{Dienes:1994np} was to address the question of classical stability by looking at the large-mass behaviour of the string spectrum, which for closed strings is encoded in the one-loop partition function computed on a Riemann surface with the topology of a torus. We will limit ourselves to points of the moduli space where the vacuum is described by a rational conformal field theory (RCFT), which admits a finite number of conformal primaries $M$, tensored with the CFT given by $D-2$ non-compact free bosons. In general, the characters which identify the holomorphic and anti-holomorphic sectors are different but, for convenience, we shall use the same symbol  $\chi$ to label both. Hence, the contributions from the holomorphic sector will be denoted by $\chi_a$ and those from the anti-holomorphic sectors by $\bar\chi_a$, while non-compact bosons always contribute with suitable powers of the Dedekind eta functions. As a result, we can write the torus partition function as
\begin{equation} \label{eq:toruspartitionfunction}
\begin{aligned}
	\mathcal{Z}&=\int_{\mathcal F} d\mu\, {\mathcal T}
 \\
 &= \int_{\mathcal F} d\mu\, \frac{1}{( \sqrt{\tau_2} \bar\eta\, \eta)^{D-2}} \sum_{a,b} \bar \chi_a \, {\mathcal N}_{ab} \, \chi_b 
	\,, 
 \end{aligned}
\end{equation}
where $\tau=\tau_1 + i \tau_2$ is the complex structure modulus of the world-sheet torus, ${\mathcal F}$ is the fundamental domain of $\text{SL} (2;\mathbb{Z})$ and $d\mu = \tau_2^{-2} \, d\tau_1 \, d\tau_2$ is the modular invariant measure. The so-called GSO matrix ${\mathcal N}_{ab}$ enforces the GSO projection and it is to be invariant under the modular transformations affecting the holomorphic and anti-holomorphic characters. To simplify the notation even further, we define the so-called {\em pseudo-characters}, $\check \chi_a = \eta^{2-D}\, \chi_a$, which, strictly speaking, do not correspond anymore to the characters of the RCFT, since they carry the non-trivial modular weight $1-D/2$. As a consequence, the action of the modular group is dressed with additional phases and powers of $\tau$, identifying the pseudo-characters as {\em vector-valued modular forms}
	\begin{equation}
		\check{\chi}_a (\gamma\cdot \tau ) = (c\tau + d )^{1-D/2}\, M(\gamma )_{ab} \, \check{\chi}_b (\tau )\,, \label{vvmftr}
	\end{equation}
where $\gamma$ is a generic element of $\text{SL} (2;\mathbb{Z})$, $1-D/2$ is the weight of the modular forms, $M(\gamma )$ is a unitary matrix, including a non-trivial multiplier system, representing the action of $\gamma$ on the $\check{\chi}$s, and a sum over the repeated index $b$ is understood.

\noindent The pseudo-characters $\check \chi_a$,  from now on simply referred to as the characters $\chi_a$ with the check omitted, admit power series expansions in the {\em nome} $q=e^{2\pi i \tau}$ of the form
\begin{equation} \label{qexp}
	\chi_a (q) = \sum_{n=0}^\infty d_a (n) q^{H_a + n}\,,
\end{equation}
with 
\begin{equation}
H_a = h_a - c/24 
\end{equation}
expressed in terms of the conformal weight $h_a$ and the central charge $c$ of the full theory, including the $D-2$ non-compact bosons. The variable $\tau$ corresponds to the Teichm\"uller parameter describing the complex structure of the torus.

\noindent The content of the physical spectrum, encoded in \eqref{eq:toruspartitionfunction}, can be directly extracted by imposing the level-matching condition on the $q$-expansion of the characters, which in the large-mass limit simply reduces to  
\begin{equation} \label{eq:dn}
    d(n) \equiv \sum_{a,b} {\mathcal N}_{ab}\,  \bar d_a (n+ H_b -\bar H_a) \, d_b (n) \, .
\end{equation}
Therefore the behaviour of \eqref{eq:dn} depends on the expressions for the degeneracies $d_a(n)$ of the characters $\chi_a$ which is dictated by the Rademacher {\em exact} formula  \cite{Rademacher:1937a,Rademacher:1937b,Rademacher:1938} for $D >2$\footnote{The formula is conjectured to hold also for the zero weight $D=2$ case, in which a refined estimate for the error determining the convergence of the expression is believed to exist \cite{Dijkgraaf:2000fq, Manschot:2007ha}.}
\begin{equation}
	d_a (n) = 2\pi \, \sum_{\ell=1}^\infty \sum_{b\, | \, H_b <0}\, Q^{(\ell , n)}_{ab} \, f_{b} (\ell , n) \, ,\label{finalRes}
\end{equation}
adapted for the vector-valued modular forms \cite{Dijkgraaf:2000fq, Manschot:2007ha}, through the known {\em Circle Method} \cite{Hardy, Kani:1989im}. In the expression reported above, $|H_b| \le 1$\footnote{One can easily modify the previous expression whenever $|H_b| \le h$, with $h$ a positive integer. In this case, the sum over $b$ includes all Fourier modes $d_b (m)$ which have $m+H_b <0$ for which we have to replace $|H_b|\to |m+H_b|$ into \eqref{finalRes}.} and we have defined the {\em generalised Kloosterman sum} 
\begin{equation} \label{eq:genkloost}
	Q ^{(\ell , n)}_{ab} = i^{1-D/2} \, \sum_{p=0\atop (p,\ell )=1}^{\ell -1} e^{\frac{2\pi i}{\ell} \left( H_b p' - p (n+H_a )\right)}\, (M^{-1}_{\ell ,p})_{ab} \, ,
\end{equation}
where the generic modular transformation is parametrised as 
\begin{equation}
	\gamma_{p,\ell} = \begin{pmatrix} - p' & \frac{1 + pp'}{\ell} \\ -\ell & p\end{pmatrix} \,,
\end{equation}
with $p'$ fixed by the condition $\gamma_{p,\ell}\in \text{SL} (2;\mathbb{Z})$. Furthermore, the behaviour of the degeneracy \eqref{finalRes} is dictated by 
\begin{equation} \label{bessel}
	f_{b} (\ell , n) = \frac{2\pi d_b (0)}{\ell}\, \left(\frac{|H_b|}{n+H_a}\right)^{D/4} \, I_{D/2} \left( \frac{4\pi}{\ell} \sqrt{ |H_b | (n+ H_a )}\right)\,,
\end{equation}
which includes the modified Bessel function of the first kind $I_\alpha (x)$, admitting the asymptotic expansion 
\begin{equation} \label{eq:besselasy}
	I_\alpha (x) \sim \frac{e^x}{\sqrt{2\pi x}} \, \left( 1- \frac{4\alpha^2-1}{8x} + \ldots \right)\,.
\end{equation}
\noindent Therefore to deduce the large-mass behaviour of the spectrum, it is enough to plug \eqref{finalRes} into \eqref{eq:dn} and take the limit $n \gg |H_a|$. However, one subtlety has to be kept in mind: the connection between the asymptotic growth of states and the absence of tachyons cannot be formulated in terms of the physical $d (n)$. A comparison between them requires the continuation of the variable $n$ to the reals, thus "filling" the leftover mass levels. As a result, this implies the introduction of the continuous enveloping functions $\Phi_{a} (n)$, which reproduce the $d_{a} (n)$ for the special values of $n$ associated with the actual masses \cite{Dienes:1994np}. Using these functions, we can define the {\em sector-averaged sum} as introduced in \cite{Dienes:1994np}
\begin{equation} \label{eq:sas}
	\langle d (n) \rangle ( \mathcal{T}) \equiv \sum_{a,b} {\mathcal N}_{ab}\, \bar\Phi_a (n+ H_b -\bar H_a) \Phi_b (n )\,, 
\end{equation}
where the enveloping functions behave asymptotically
\begin{equation} \label{degasy}
\begin{aligned}
	\Phi_a (n) \sim \sum_{b \, |\, H_b <0} \frac{d_b (0)}{\sqrt{2}}\, \frac{|H_b|^{(d-1)/4}}{n^{(d+1)/4}}\,
	& \left[ Q^{(1,n)}_{ab} e^{4\pi \sqrt{|H_b| n}} + \frac{Q^{(2,n)}_{ab}}{\sqrt{2}}\, e^{2\pi \sqrt{|H_b| n}}   \right. 
 \\
 &\left. + \frac{Q^{(3,n)}_{ab}}{\sqrt{3}}\,  e^{\frac{4\pi }{3}\sqrt{|H_b| n}} +\ldots \right]\,,
 \end{aligned}
\end{equation}
with the first two terms dictated by the $S$ and $P$ modular transformations 
\begin{equation} \label{Qmod}
	\begin{split}
		&Q_{ab}^{(1,n)} = i^{1-d/2}\, S_{ab} \, ,
		\\
		&Q_{ab}^{(2,n)} = i^{1-d/2}\, (-1)^n P_{ab} \, ,
		\\
		& \qquad \qquad \vdots
	\end{split}
\end{equation}
\noindent With the definition above, it has been possible to show in \cite{Dienes:1994np} that without tachyons the leading term of the expression vanishes and it has been conjectured that actually a complete cancellation among the enveloping functions would entail the stability of the vacuum. To prove the conjecture, hence, it is required to extend the analysis to the sub-leading orders. However, the study of the sub-dominant contributions to $\langle d (n) \rangle ( \mathcal{T})$ is a bit more involved and we shall employ a refinement of the expression \eqref{eq:sas} along the lines of \cite{Cribiori:2020sct, Cribiori:2021gbf}. To this end, we notice that the $Q$s are periodic functions of $n$, with period $\ell$, 
\begin{equation} \label{eq:kloostper}
	Q^{(\ell,n+\ell m)}_{ab} = Q^{(\ell,n)}_{ab}\,,
\end{equation}
for any integer $m$.  Moreover, since the number of physical states \eqref{eq:sas} depends on the product of the $Q$ functions from the holomorphic and anti-holomorphic characters,  it is convenient to decompose the degrees of freedom into classes organised by the common periodicity $v_{\ell,\bar \ell} = \ell\, \bar \ell /\text{gcd} (\ell,\bar \ell)$ of the two $Q$s and associate to each class its own enveloping function $\Phi_a (n , \{ w\} )$. As a result, we can write the {\em refined} sector-averaged sum
\begin{equation} \label{eq:sasrefined}
	\begin{split}
		\langle d (n ) \rangle  ( \mathcal{T}) &= \sum_{a,b}\sum_{\{w  \}} {\mathcal N}_{ab} \, \bar \Phi_a (n, \{ w  + H_b - \bar H_a \} )\, \Phi_b (n, \{ w  \} )
		\\
		&\equiv \sum_{a,b} \sum_{{c,d \atop \, H_c, \bar H_d <0}}  \sum_{\ell,\bar \ell =1}^{\infty} \sum_{w=0}^{v_{\ell,\bar \ell}-1} {\mathcal N}_{ab}\, Q^{(\ell,w)}_{bc}\,  \bar Q^{(\bar \ell ,  w+H_b - \bar H_a)}_{ad} \, f_c (\ell,n )\,  \bar f_d (\bar \ell , n) \,.
	\end{split}
\end{equation}
It can be noticed that the refinement does not affect the leading growth so that the considerations made in \cite{Dienes:1994np} still apply.

\noindent The definition of the sector-averaged sums can be extended to the vacuum-to-vacuum amplitudes relevant for orientifold vacua. These are given by the remaining (un)oriented Riemann surfaces with boundaries and cross-caps and vanishing Euler characteristic \cite{farkas2012riemann}: the Klein bottle, the annulus and the M\"obius strip. We can write them in a compact way  as  
\begin{equation} \label{ampld}
	\mathcal{Z}_I=  \ \sum_{a=0}^{M-1} \text{Z}_I^a \ \chi_a = \sum_{a=0}^{M-1} \text{Z}_I^a \sum_{n=0}^{\infty} d_a(n) q^{n+H_a} \,,
\end{equation}  
with $I=1,2,3$ labels the Klein bottle, annulus and Moebius strip amplitudes, respectively, while the $\text{Z}_I^a$ are suitable integers. In \eqref{ampld}, $\chi_a$ are the {\em real} characters depending on the modulus $\tau$ of the double covering torus whose real part is fixed. For $\mathcal{K}$ and $\mathcal{A}$ the Teichm\"uller parameter is purely imaginary ($\tau= 2 i \tau_2$ and $\tau = \frac{i\tau_2}{2}$, respectively) so that the characters defined in \eqref{qexp} are real, whereas for the M\"obius strip surface the modulus of the covering torus is  $\frac{1}{2} + i \frac{\tau_2}{2}$, and one needs to introduce the extra phase $e^{-i \pi H_a}$ to make them real \cite{Bianchi:1990yu}. This extra phase will not play any role in the following discussion and thus, with an abuse of notation, we shall omit it. These amplitudes describe the one-loop propagation of closed and open strings, which in $\mathcal{K}$ and $\mathcal{M}$ flip their orientation. It is conventional to refer to this description as the {\em loop-channel} which involves a {\em vertical proper-time} \cite{Bianchi:1990yu,Angelantonj:2002ct}. Similarly, by exchanging the length $\sigma^1$ of the string with the proper time $\sigma^0$ on the world-sheet through the $S$ or $P$ transformation\footnote{Actually, for the M\"obius strip this inversion of proper time is realised via the $P=T^{1/2} S T^2 S T^{1/2}$ transformation \cite{Bianchi:1990yu}.}, we can describe the free propagation of closed strings between boundaries ({\em i.e.} D-branes) and cross-caps ({\em i.e.} O-planes) \cite{Bianchi:1990yu,Angelantonj:2002ct} via the {\em tree-level} amplitudes 
\begin{equation} \label{ampltr}
	\Tilde{\mathcal{Z}}_I=  \sum_{a=0}^{M-1} \Tilde{\text{Z}}_I^a \ \chi_a= \sum_{a=0}^{M-1} \Tilde{\text{Z}}_I^a \sum_{n=0}^{\infty} d_a(n) q^{n+H_a}  \,,
\end{equation}   
where $q=e^{-2 \pi \ell} $ is written in terms of the new horizontal proper time $\ell$. For the M\"obius amplitudes it carries an extra minus sign, which originates from the non-vanishing real part of $\tau$. As a result, the coefficients  $\text{Z}_I^a$ and $\Tilde{\text{Z}}_I^a$ entering the eqs. \eqref{ampld} and \eqref{ampltr} are not independent but are connected to each other by the $i^{1-D/2}S$ or $i^{1-D/2}P$ modular transformations. 

\noindent From the expressions \eqref{ampld} and \eqref{ampltr}, it is straightforward to extract the large mass behaviour in the two channels encoded in the {\em sector-averaged sums} associated to the corresponding amplitudes
\begin{equation} \label{sasor}
	\left \langle d(n) \right \rangle  \left ( \mathcal{Z}_I \right ) =  \sum_{a=0}^{M-1} \ \text{Z}_I^a \ \Phi_a(n) \ , \qquad 	\left \langle d(n) \right \rangle \left ( \Tilde{\mathcal{Z}}_I \right ) =  \sum_{a=0}^{M-1} \ \Tilde{\text{Z}}_I^a \ \Phi_a(n ) \ .
\end{equation}   
As in the closed-string case, 
\begin{equation}
	\Phi_a (n) = \sum_{\ell =1}^\infty \sum_{w=0}^{\ell -1} \, \Phi_a (n,w )\ ,
\end{equation}
are the enveloping functions associated with the degeneracies of the real characters, once the continuation of $n$ to the reals and the refinement on the spectrum are employed. Employing the same refinement used for closed-string vacua, we can define the refined enveloping functions $\Phi_a (n, w)$ for the Klein bottle and annulus amplitudes as 
\begin{equation}
	\Phi(n, w )= \sum_{b \, | \,  H_b < 0}  \sum_{\ell=1}^{\infty} \sum_{w=0}^{\ell-1} Q^{(\ell,w)}_{a b} f_b(\ell, n) \, .
\end{equation} 
In the M\"obius strip amplitude, they involve the new Kloosterman sums
\begin{equation} \label{hattedkloost}
	\hat{Q}_{ab}^{( \ell,w)}= (-1)^w Q^{(\ell,w)}_{a b}\,,
\end{equation}   
where the alternating sign is clearly associated to the fixed real part of the Teichm\"uller parameter. The presence of $(-1)^w$ affects their periodicity in $w$, which is now $\hat \ell= \text{lcm}(2,\ell)$. This little change has remarkable consequences for the property of the sector-averaged sum associated with the M\"obius strip amplitudes both in the direct and transverse channel, and reflects itself in the need of using the $P$ transformations to connect the two channels, in the same way as the $S$ transformation does for the Klein bottle and the annulus amplitudes.  

\noindent Now that we have defined the sector-averaged sums for all relevant amplitudes, we can explore their properties and their links to the presence of tachyons in the spectrum. This will allow us, on one hand, to clarify the role of oscillations and also to prove Dienes' conjecture for oriented closed-string vacua. On the other, we shall only establish sufficient conditions for the classical stability of the vacuum, as well as a necessary criterion for the absence of closed-string tachyons, for orientifolds.

\section{Misaligned Supersymmetry for Closed String Vacua} \label{MSoriented}

Let us now dig into the study of the properties of the refined sector-averaged sum for closed-string vacua. The leading order of \eqref{eq:sasrefined} can be straightforwardly deduced from the expression for the Rademacher expansion \eqref{finalRes} and the asymptotic behaviour of the modified Bessel function \eqref{eq:besselasy}. As can be extracted from \eqref{eq:besselasy}, the leading contribution is dictated by the deepest tachyon, with the smallest conformal weight, included in $\chi_0$, associated with the ubiquitous NS vacuum which plays the role of the identity in the RCFT. As a result, the leading growth of each character is universal and reads
\begin{equation}
	d_a (n) \sim e^{4\pi \sqrt{c n/24}} + \ldots \,, \label{leadgrw}
\end{equation}
in agreement with the Cardy formula \cite{Cardy:1981fd}. Plugging \eqref{leadgrw} into the definition of the sector-averaged sum \eqref{eq:sasrefined} allows one to identify the leading contribution to the growth of states degeneracies     
\begin{equation}
	\langle d (n) \rangle ( \mathcal{T}) \sim  \, e^{4\pi \sqrt{n} \left( \sqrt{c_\text{L}/24}+\sqrt{ c_\text{R} /24}\right)}\, \sum_{a,b} {\mathcal N}_{ab} = e^{C_\text{tot} \, \sqrt{n}} \, \, \sum_{a,b} {\mathcal N}_{ab} \, ,
\end{equation}
determined by the $\text{SL} (2;\mathbb{C})$ invariant vacuum, where we have defined the {\em total} central charge\footnote{This reflects the terminology used in literature since the work of \cite{Dienes:1994np}. Nevertheless, it is an abuse of notation since the total central charge should correspond to $c_\text{L}+c_{\text{R}}$.} $C_{\text{tot}}= \sqrt{c_\text{L}/24}+\sqrt{ c_\text{R} /24}$.
However, whether or not the asymptotic growth is controlled by $C_\text{tot}$ depends entirely on the particular properties of ${\mathcal N}$, since the GSO matrix must yield a modular invariant partition function. Consequently, we can relate the entry for the NS-NS vacuum to the sum over all the characters appearing in the partition function,
\begin{equation}
	{\mathcal N}_{00} = \frac{1}{M} \, \sum_{a,b} {\mathcal N}_{ab}
\end{equation}
since $S_{a0}=i^{D/2-1}/\sqrt{M}$, for a RCFT with all the $M$ characters resolved. Moreover, the entries ${\mathcal N}_{ab}$ can only be $\pm 1$ or zero, which yields the following two scenarios:
\begin{enumerate}
	\item If the spectrum {\em does} contain the deepest tachyon, {\em i.e.} ${\mathcal N}_{00}=1$, one has 
	\begin{equation}
		\sum_{a,b} {\mathcal N}_{ab} \neq 0 \qquad \text{and} \qquad \langle d (n) \rangle ( \mathcal{T}) \sim e^{C_\text{tot} \, \sqrt{n}}\,.
	\end{equation}
	\item If the spectrum {\em does not} contain the deepest tachyon, {\em i.e.} ${\mathcal N}_{00}=0$,  then
	\begin{equation}
		\sum_{a,b} {\mathcal N}_{ab} = 0 \qquad \text{and} \qquad \langle d (n) \rangle ( \mathcal{T}) \sim e^{C_\text{eff} \, \sqrt{n}}\label{FBNzero}
	\end{equation}
	with an {\em effective central charge} $C_\text{eff} < C_\text{tot}$. 
\end{enumerate}
The condition ${\mathcal N}_{00}=0$ clearly requires that fermions be present in the spectrum and, in particular, the number of bosonic sectors must equal the number of fermionic ones. On the contrary, in the absence of extended symmetries, namely when each character appears only once in  \eqref{eq:toruspartitionfunction}, the condition ${\mathcal N}_{00}=1$ implies that the spectrum only contains bosonic excitations. These observations were already contained in \cite{Dienes:1994np} and, in fact, represent its main result. However, it is important to stress that $C_\text{eff} < C_\text{tot}$ does {\em not} imply classical stability since, although the leading tachyon must indeed be absent, the condition ${\mathcal N}_{00} =0$ does not automatically exclude the possibility that other tachyons be present, as happens in most non-supersymmetric theories. This also implies that the mere presence of oscillations, which have been observed in {\em all} non-supersymmetric theories with space-time fermions, has to be ascribed to the presence of the latter states and thus cannot be considered as a necessary and sufficient condition for classical stability.   

\noindent On the other hand, the full extent of this simple analysis can lead to a further consideration. Indeed, as things stand now, the presence of fermions is not enough of course to guarantee the absence of {\em all} tachyons but it is only a necessary condition. Thus, on a perturbative level, requiring the vacuum to be at least classically stable unavoidably introduces fermions, justifying their presence in a phenomenologically meaningful universe.      

\noindent We can now turn to the analysis of the sub-leading contributions. This is where the refinement becomes effective and useful, since taking advantage of the expression in \eqref{eq:sasrefined}, we can arrange the mass-levels in sub-classes  $w=k_\ell + \ell r$, determined by the periodicity of the generalised Kloosterman sum \eqref{eq:genkloost} reported in \eqref{eq:kloostper}
\begin{equation}
\begin{aligned}
    \langle d (n) \rangle ( \mathcal{T})= \sum_{a,b} \sum_{{c,d \atop \, H_c, \bar H_d <0}} \sum_{\ell,\bar \ell =1}^{\infty}\sum_{k_\ell=0}^{\ell-1} \sum_{r=0}^{\frac{\bar \ell}{\text{gcd} (\ell,\bar \ell)}-1} &{\mathcal N}_{ab} \, Q^{(\ell,k_\ell )}_{bc}\,  \bar Q^{(\bar \ell ,  k_\ell + r \ell + H_b - \bar H_a)}_{ad} \, 
    \\
    & \times f_c (\ell,n )\,  \bar f_d (\bar \ell , n) \, ,\label{maineq}
\end{aligned}
\end{equation}
where we have used
\begin{equation}
	\sum_{w=0}^{v_{\ell,\bar \ell} -1} = \sum_{k_\ell =0}^{\ell-1} \sum_{r=0}^{\frac{\bar \ell}{\text{gcd} (\ell,\bar \ell)}-1} \, .
\end{equation}
This is the expression from which we can extract our main results. We have to distinguish the two cases $\ell = \bar \ell$ and $\ell\neq \bar \ell$. In the latter case, if $\ell$ does not divide $\bar\ell$, there is no contribution to the sector-averaged sum, since
\begin{equation}
	\sum_{r=0}^{\frac{\bar \ell}{\text{gcd} (\ell,\bar \ell)}-1} \bar Q^{(\bar \ell ,  k_\ell + r \ell + H_b - \bar H_a)}_{ad} =0\,. \label{eqllbdiff}
\end{equation}
This property follows from 
\begin{equation}
	\begin{split}
		\sum_{r=0}^{\frac{\bar \ell}{\text{gcd} (\ell,\bar \ell)}-1} \bar Q^{(\bar \ell ,  k + r \ell )}_{ad} &= \sum_{r=0}^{\frac{\bar \ell}{\text{gcd} (\ell,\bar \ell)}-1} \sum_{\bar p=0\atop (\bar p,\bar \ell )=1}^{\bar \ell -1} e^{-\frac{2 \pi i}{\bar \ell} (\bar p ( k + r \ell)+  \bar p ' \bar H_d )}\, \left( M^{-1}_{\bar \ell , \bar p}\right)_{ad}
		\\
		&=\sum_{\bar p=0\atop  (\bar p, \bar \ell)=1}^{\bar \ell -1}  e^{-\frac{2 \pi i}{\bar \ell} (\bar p k +  \bar p ' \bar H_d )}\, \left( M^{-1}_{\bar \ell , \bar p}\right)_{ad}\, \frac{1-e^{-2\pi i \bar p \frac{\bar \ell}{\text{gcd} (\ell,\bar \ell)}}}{1-e^{- 2\pi i \frac{\bar p \ell}{\bar \ell}}} 
		\\
		&=0 \,,
	\end{split}
\end{equation}
since $\bar \ell$ is an integer multiple of $\text{gcd} (\ell, \bar \ell)$. Similarly, if $\bar\ell = m \ell$, for some integer $m$, eq. \eqref{eqllbdiff} still holds since $\bar p$ and $\bar\ell$ must be co-prime, and thus
\begin{equation}
	\sum_{r=0}^{m-1} e^{-2 \pi i \bar p r/m} = 0\,.
\end{equation}
The case $\ell=\bar \ell$ is a bit more involved and, as we shall see, discriminates between tachyonic and non-tachyonic string vacua. The $Q$ functions have now the same periodicity, which implies that in eq. \eqref{maineq} the sum over $r$ is trivial, since nothing depends on $r$, and
\begin{equation}
	\begin{split}
		\langle d (n) \rangle ( \mathcal{T} ) &= \sum_{a,b} \sum_{{c,d \atop \, H_c, \bar H_d <0}} \sum_{\ell =1}^{\infty} \sum_{ p=0\atop (p, \ell )=1}^{ \ell -1} \sum_{\bar p=0\atop (\bar p,  \ell )=1}^{ \ell -1} {\mathcal N}_{ab} \, \left( M^{-1}_{\ell ,  p}\right)_{bc} \, \left( M^{-1}_{ \ell , \bar p}\right)_{ad}^*
		\, \\
		&\qquad \times e^{-\frac{2\pi i}{\ell}  [ (p-\bar p )  H_b - (p' H_c - \bar p ' \bar H_d )]}\,
		f_c (\ell,n )\,  \bar f_d ( \ell , n)  \sum_{k_\ell=0}^{\ell-1} e^{- \frac{2 \pi i}{\ell} (p-\bar p ) k_\ell }  
		\,.
	\end{split}
\end{equation}
The sum over $k_\ell$ imposes the condition $p=\bar p$ which identifies the holomorphic and anti-holomor\-phic modular transformations. For large $n$, using the modular invariance of the GSO matrix ${\mathcal N}_{ab}$,  and taking into account that in string theory tachyons have $-1\le \bar H_a , H_b <0$, one arrives at the final result
\begin{equation}
	\langle d(n) \rangle ( \mathcal{T} ) = \sum_{\ell=1}^{\infty} 
	\sum_{{a, b\atop \bar H_a = H_b <0}}  \,\ell\, \varphi (\ell ) \, {\mathcal N}_{ab}\,  f_b (\ell,n )\,  \bar f_a ( \ell , n)  \,,
\end{equation}
where $\varphi (\ell)$ is a function counting the number of co-prime integers with $\ell$, the {\em so-called} Euler totient function.   {\em The exponential growth of the sector-averaged sum is thus directly linked to the presence of tachyons in the physical string spectrum.} In fact,  if physical tachyons, with $H_b = \bar H_a <0$, are present in the string spectrum, then
\begin{equation}\label{largemasssas}
	\langle d(n) \rangle ( \mathcal{T} ) \sim \sum_{a,b \atop H_b = \bar H_a <0} {\mathcal N}_{ab}\,  d_b (0) \, \bar d _a (0)\, \frac{|H_b|^{(d-1)/2}}{2\, n^{(d+1)/2}}\, 
	\sum_{\ell=1}^{\infty} \, \varphi (\ell) \, e^{\frac{8\pi}{\ell} \sqrt{|H_b|\, n}} \,.
\end{equation}
The exponential growth is then dictated by the conformal weight $H_b$ of the {\em deepest} physical tachyon,  $C_\text{eff} = 8 \pi \sqrt{|H_b|}$. Instead, in the absence of on-shell tachyons the sector-averaged sum $\langle d (n) \rangle ( \mathcal{T} ) $ vanishes, and thus
\begin{equation}
	C_\text{eff} =0\,.
\end{equation}

\noindent All in all, thanks to modular invariance and employing the clever refinement used in \cite{Cribiori:2020sct}, we have proved that the asymptotic growth rate of the sector-averaged sum is dictated by the mass of the {\em lightest} states, whether tachyonic or massless,
\begin{equation}
	C_\text{eff} = 4 \pi \sqrt{|\alpha' m^2_\text{lightest} |} \le C_\text{tot}\,,
\end{equation}
and thus the {\em necessary and sufficient} condition for classical stability is the vanishing of both the sector-averaged sum and the effective central charge, as conjectured by Dienes in \cite{Dienes:1994np}. Notice that the deepest tachyon also determines the {\em effective} Hagedorn temperature $T_\text{eff}$ of strings at finite temperature \cite{Dienes:2012dc}, and thus links it to $C_\text{eff}$. 

\noindent The analysis is fully general and applies to any vacuum of oriented closed strings. It extends the discussion of \cite{Cribiori:2020sct}, which heavily relies on the representation of the characters in terms of eta quotients of special type, which is a rather restrictive requirement, not met by most non-supersymmetric string vacua. Obviously, when both representations are possible the two results coincide.

\section{Misaligned Supersymmetry for Orientifold Vacua} \label{MSunoriented}

For orientifold vacua, the conclusion reached in the previous Section cannot hold since modular invariance is no longer a property of the vacuum-to-vacuum amplitudes. This means that the large-mass behaviour encoded in the sector-averaged sum does not reflect anymore the presence of physical tachyons in the tree-level spectrum. Nevertheless, one may na\"ively expect that since the closed-string UV/IR mixing translates into the connection between dual descriptions in orientifold amplitudes, the sector-averaged sum may encode the presence of tachyonic characters in the dual channel. This expectation is going to be confirmed in the following analysis, in which the properties of the generalised Kloosterman sums \eqref{eq:genkloost}, valid for the Klein bottle and annulus amplitudes, and the \eqref{hattedkloost}, valid for the M\"obius strip amplitude, play a key role.    

\noindent Indeed, the properties of the Kloosterman sum drastically simplify the expressions of the enveloping functions. In fact, as shown previously, 
\begin{equation}\label{Kloostann}
	\begin{split}
		\sum_{w=0}^{ \ell-1}  Q^{( \ell , w )}_{ab} &= 	\sum_{w=0}^{ \ell-1} \sum_{ p=0\atop ( p, \ell )=1}^{ \ell -1} e^{-\frac{2 \pi i}{ \ell} ( p w +  p '  H_b )}\, \left( M^{-1}_{ \ell , p}\right)_{ab} \ ,
		\\
		&= \sum_{ p=0\atop ( p, \ell )=1}^{ \ell -1}  e^{-\frac{2 \pi i}{ \ell}   p '  H_b }\, \left( M^{-1}_{ \ell , p}\right)_{ab}	\sum_{w=0}^{ \ell-1} e^{-\frac{2 \pi i}{ \ell}  p w} \  ,
		\\
		&= \sum_{ p=0\atop ( p, \ell )=1}^{ \ell -1}  e^{-\frac{2 \pi i}{ \ell}   p '  H_b }\, \left( M^{-1}_{ \ell , p}\right)_{ab} \ell \delta_{p,0} \, ,
	\end{split}
\end{equation}
which vanishes identically unless $\ell=1$, since only in this case $p=0$ is coprime with $\ell$. Therefore, the only contribution to the growth of the sector-averaged sum comes from $\ell=1$, 
\begin{equation}\label{leadPhiKA}
	\Phi_a (n) = \sum_{b \ | \ H_b <0} \, Q^{ (1, 0)}_{ab} \, f_b (1,n ) =   i^{1-D/2}\, \sum_{b \ | \ H_b <0} S_{ab} \, f_b (1,n)\,,
\end{equation}
and, in the last equality, we have used the explicit formula \eqref{Qmod} for $Q^{ (1, 0)}_{ab}$. Notice that this expression only involves the $S$ modular transformation that relates the $1$-loop amplitudes, $\mathcal{K}$ and $\mathcal{A}$, to their dual counterparts,  $\Tilde{\mathcal{K}}$ and $\Tilde{\mathcal{A}}$, describing the tree-level propagation of closed strings between pairs of O-planes and D-branes.

\noindent The M\"obius strip amplitudes depend on the modified Kloosterman sum, which also vanishes unless $\ell=2$. In fact,
\begin{equation}\label{Kloostmob}
	\begin{split}
		\sum_{w=0}^{ \hat \ell-1}  \hat Q^{( \ell , w )}_{ab} &= \sum_{w=0}^{\hat\ell -1} (-1)^w \, \sum_{p=0 \atop (p,\ell)=1}^{\ell-1}\, e^{- \frac{2\pi i}{\ell} (pw + p' H_b )}\,  \left( M^{-1}_{ \ell , p}\right)_{ab}
		\\
		&=\sum_{ p=0\atop ( p, \ell )=1}^{ \ell -1}  e^{-\frac{2 \pi i}{ \ell}   p '  H_b }\, \left( M^{-1}_{ \ell , p}\right)_{ab}	\sum_{r=0}^{\frac{\hat \ell}{\text{gcd}(\hat \ell,\ell)}-1} \sum_{k=0}^{ \ell-1} e^{-\frac{2 \pi i}{ \ell} ( p- \frac{\ell}{2}) (k+ \ell r)} \ 
		\\
		&= \sum_{ p=0\atop ( p, \ell )=1}^{ \ell -1}  e^{-\frac{2 \pi i}{ \ell}   p '  H_b }\, \left( M^{-1}_{ \ell , p}\right)_{ab} \sum_{r=0}^{\frac{\hat \ell}{\text{gcd}(\hat \ell,\ell)}-1}  e^{-2 \pi i  ( p- \frac{\ell}{2})  r} \ell \, \delta_{p,\frac{\ell}{2}} \ ,
	\end{split}
\end{equation}
where in the second line we have written $w=k+\ell r$ and converted the sum over $w$ into the sums over $k$ and $r$. Clearly, the only non-vanishing term is for $\ell=2$ and $p=1$, so that
\begin{equation} \label{leadPhiM}
	\sum_{w=0}^{1}\hat{Q}_{ab}^{(2,w)} = 2 i^{1-D/2} \  \left (T^{\frac12} S T^2 S T^{\frac12} \right )_{ab} = 2 i^{1-D/2}\, P_{ab} \,,
\end{equation}
and we recognise the modular transformation $P$ which connects the dual amplitudes $\mathcal{M}$ and $\Tilde{ \mathcal{M}}$, describing the orientifold projection of the open-string spectrum and the propagation of closed-string states between a boundary and a cross-cap, respectively.

\noindent With these considerations, we can now state the main result of this Section, which extends misaligned supersymmetry of oriented closed strings to the case of orientifold constructions. In fact, eq. \eqref{leadPhiKA}  implies
\begin{equation} 
	\left \langle d(n) \right \rangle \left (\mathcal{Z}_I \right )  = i^{1-D/2}  \ \sum_{a=0}^{M-1} \sum_{b \, |\, H_b <0} Z_{I}^{a} \, S_{ab} \, f_b (1 , n ) =   \sum_{b \,|\, H_b <0} \ \Tilde{\text{Z}}_I^b \   f_b(1, n) \ ,
\end{equation}
for the Klein bottle and annulus amplitudes, $I=1,2$, while eq. \eqref{leadPhiM} implies that
\begin{equation} 
	\left \langle d(n) \right \rangle \left (\mathcal{M} \right )  =2 i^{1-D/2}  \ \sum_{a=0}^{M-1} \sum_{b \, |\, H_b <0} Z^a_3 \, P_{ab} \, f_b (2 , n ) = 2  \sum_{b \,|\, H_b <0} \ \Tilde{\text{Z}}_3^b \   f_b(2, n) \ ,
\end{equation}
for the M\"obius strip amplitude. In both cases, the sector-averaged sum associated to the loop-channel amplitudes grows exponentially if and only if a closed-string tachyons propagates in the tree-level channel. Using the explicit expression \eqref{bessel}, one can write
\begin{equation}\label{dsas}
	\left \langle d(n) \right \rangle \left (\mathcal{Z}_I \right ) =  \sqrt{\tfrac{\ell_I }{2}}  \sum_{b\, |\, H_b <0} \ \Tilde{\text{Z}}_I^b  \, d_b(0)  \, \frac{\left | H_b \right |^{(d-1)/4}}{n^{(d+1)/4}} \, e^{ \frac{4\pi}{\ell_I} \, \sqrt{\left | H_b \right | \, n}} \ ,
\end{equation}
with $\ell_I =1$ ($\ell_I =2$) for $I= 1, 2$ ($I=3$). Similarly, the sector-averaged sum associated to the tree-level amplitudes reads
\begin{equation} \label{trsas}
	\begin{split}
		\left \langle d(n) \right \rangle  ( \Tilde{\mathcal{Z}}_I  )  &= \ell_I  \sum_{a\, |\, H_a <0} \ \text{Z}_I^a \ f_a(\ell_I,n)
		\\
		&=  \sqrt{\tfrac{\ell_I}{2}}  \sum_{a\, |\, H_a <0} \ \text{Z}_I^a  \, d_a(0)  \, \frac{\left | H_a \right |^{(d-1)/4}}{n^{(d+1)/4}} \, e^{ \frac{4\pi}{\ell_I} \, \sqrt{\left | H_a \right | \, n}} \ ,
	\end{split}
\end{equation}
and grows exponentially if and only if a tachyonic character is present in the associated loop-channel amplitudes. If no tachyons are present in the tree-level (loop) channel, the sector-averaged sum $\left \langle d(n) \right \rangle \left (\mathcal{Z}_I \right ) $ ($\left \langle d(n) \right \rangle  ( \Tilde{\mathcal{Z}}_I  ) $) vanishes identically. This means that $\langle d(n) \rangle (\mathcal{Z}_I)$ carries information about the coupling of closed-string tachyons to D-branes or orientifold planes, while $\langle d(n) \rangle  ( \Tilde{\mathcal{Z}}_I ) $ encodes the presence of tachyonic characters in the one-loop partition functions. Therefore the latter must be considered for the study of the classical stability of the vacuum. This result clarifies the physical meaning of the sector-averaged sum for orientifold vacua and thus allows to extend the notion of misaligned supersymmetry to this context, matching our expectations. 

\noindent Notice that such considerations only provide either a necessary condition for the absence of closed-string tachyons or a sufficient criterion for the stability of the vacuum. Indeed, closed-string tachyons could be present in the spectrum even though their coupling to D-branes or O-planes vanishes, which implies that $\left \langle d(n) \right \rangle \left (\mathcal{Z}_I \right ) =0$ cannot guarantee the classical stability of the vacuum. Furthermore, albeit the vanishing of the sector-averaged sum of the parent closed oriented string and of the $\left \langle d(n) \right \rangle  ( \Tilde{\mathcal{Z}}_I  ) $s automatically guarantees the classical stability of the orientifold vacua, it is not true in general that a $\left \langle d(n) \right \rangle  ( \Tilde{\mathcal{Z}}_I  ) \neq 0$ implies that physical tachyons are present in the spectrum. In fact, tachyons could be projected away by $\Omega$, and when this happens each individual sector-averaged sum can experience a non-vanishing exponential growth. Although one would na\"ively expect that upon summing the contributions from all amplitudes ${\mathcal T}$ and $\Tilde{\mathcal Z}_I$ this exponential growth should disappear, this cannot occur. In fact, the sector-averaged sum \eqref{largemasssas} associated to the torus amplitude contains infinite contributions with $\ell \geq 1$, which clearly cannot be cancelled by $\langle d(n) \rangle ( \Tilde{\mathcal{K}})  $, which has only the leading term, $\ell=1$.

\noindent A similar story goes on for the open-string sector. Because of the properties \eqref{Kloostann} and \eqref{Kloostmob} of the generalised Kloosterman sums, only one term contributes to the sector-averaged sum, and has $\ell=1$ for the transverse annulus but $\ell =2$ for the transverse M\"obius strip amplitude, which yields different growth rates so that they can never cancel. 

\noindent This suggests that the sector-averaged sums associated with different Riemann surfaces cannot be directly compared. This is not surprising since they compute different properties of the vacuum. On the one hand, $\langle d(n) \rangle ({\mathcal T})$ controls the asymptotic growth of physical degrees of freedom and therefore it is tied to the spectrum of oriented closed strings. On the other hand, the quantities $\langle d(n) \rangle (\Tilde{\mathcal Z}_I)$ carry no global information on the spectrum, but rather to the one-point couplings of closed-string states to D-branes or O-planes. 

\noindent  Therefore, we can only conclude that it is enough for the classical stability of the vacuum to satisfy $\langle d(n) \rangle (\mathcal{T})=0$ and $\langle d(n) \rangle (\Tilde{\mathcal Z}_I)=0$ . Actually, because the mere role of the Klein bottle and M\"obius strip amplitudes is to enforce the orientifold projection on ${\mathcal T}$ and ${\mathcal A}$, the simultaneous vanishing of the sector-averaged sums associated to the torus and to the tree-level channel annulus amplitudes is a sufficient condition for the classical stability of the orientifold vacuum. On the contrary, no information can be drawn from the vanishing of $\langle d(n) \rangle ( \Tilde{\mathcal{K}}) $ and $\langle d(n) \rangle ( \Tilde{\mathcal{M}})$, since tachyons could be associated to oriented strings. 

\noindent All in all, these considerations show that the sector-averaged sums have nothing to say for vacua in which classical stability is a consequence of a non-trivial action of the orientifold projection. Some concrete examples will be now provided through which the general discussions of the previous Sections can be shown at work.

\section{An example in ten dimensions}

A simple arena where to test our results is given by ten-dimensional string vacua with no space-time supersymmetry. Indeed, in ten dimensions the number of consistent constructions realised in string theory is finite 
\cite{Dixon:1986iz, Kawai:1986vd, Antoniadis:1986rn, Horava:2007hg, Kaidi:2019pzj, Kaidi:2019tyf, BoyleSmith:2023xkd} and thus a systematic analysis can be carried out. In all these cases, the spectrum is organised according to the conjugacy classes of the ten-dimensional little group $\text{SO}(8)$ combined with themselves or with characters describing the representations of the gauge group for the heterotic string. Furthermore, such characters are dressed with the Dedekind $\eta$ functions from the non-compact bosons  \cite{Kani:1989im} giving rise to irrational CFT. Still, following \cite{Kani:1989im}, we can overcome this problem by defining the pseudo-characters 
\begin{equation} \label{pseudochar}
	(O_{2n} , V_{2n} , S_{2n} , C_{2n} ) \to \left(\frac{O_{2n} }{\eta^8}, \frac{V_{2n}}{\eta^8} , \frac{S_{2n}}{\eta^8} , \frac{C_{2n}}{\eta^8}\right)\,,
\end{equation}
and including suitable phases in the modular transformations 
\begin{equation}
	T \, :\quad   (O_{2n} , V_{2n} , S_{2n} , C_{2n} ) \to e^{- i \pi (n+8)/12}\,  (O_{2n} , - V_{2n} , e^{i \pi n/4} \, S_{2n} , e^{i \pi n/4} \, C_{2n} ) \,, \label{TSO2n}
\end{equation}
and 
\begin{equation}
	S\,:\quad \begin{pmatrix} O_{2n} \\ V_{2n} \\ S_{2n} \\ C_{2n}\end{pmatrix} \to \tau^{-4} \, \frac{1}{2}\begin{pmatrix} 1 & 1 & 1 & 1 \\ 1 & 1 & -1 & -1 \\ 1 & -1 & i^{-n} & - i^{-n}\\ 1 & -1 & - i^{-n} &  i^{-n} \end{pmatrix} \, \begin{pmatrix} O_{2n} \\ V_{2n} \\ S_{2n} \\ C_{2n}\end{pmatrix}\,. \label{SSO2n}
\end{equation}

\noindent All these theories can be studied at once by noticing that their partition functions can be compactly written as
\begin{equation}
	{\mathcal T}_A = \sum_{a,b=0}^3 \bar \chi^A_a\, {\mathcal N}_{ab} \, \chi_b \,,
\end{equation}
where $\chi = (O_8 , V_8 , S_8 , C_8)$ denotes the left-moving characters, which are common to all ten-dimen\-sional non-supersymmetric theories, and $\bar \chi_a^A$ denotes the right-moving characters, which depend on the specific model $A$. Thus, once the characters are identified, the partition functions are determined by the choice of the GSO matrices that read
\begin{equation}
	{\mathcal N}_\text{het} = \begin{pmatrix} 0 & 1 & 0 & 0 \\ 1 & 0 & 0 & 0 \\ 0 & 0 & -1 & 0 \\ 0 & 0 & 0 &-1 \end{pmatrix}\,,
\end{equation}
for the heterotic theories, while 
\begin{equation} \label{type0GSO}
	{\mathcal N}_{\text{0A}} = \begin{pmatrix} 1 & 0 & 0 & 0 \\ 0 & 1 & 0 & 0 \\ 0 & 0 & 0 & 1 \\ 0 & 0 & 1 & 0 \end{pmatrix}\,,
	\qquad 
	{\mathcal N}_{\text{0B}} = \begin{pmatrix} 1 & 0 & 0 & 0 \\ 0 & 1 & 0 & 0 \\ 0 & 0 & 1 & 0 \\ 0 & 0 & 0 & 1 \end{pmatrix}\,,
\end{equation}
for the type 0A and type 0B theories and 
\begin{equation}
	{\mathcal N}_{\text{IIA}} = \begin{pmatrix} 0 & 0 & 0 & 0 \\ 0 & 1 & 0 & -1 \\ 0 & -1 & 0 & 1 \\ 0 & 0 & 0 & 0 \end{pmatrix}\,,
	\qquad 
	{\mathcal N}_{\text{IIB}} = \begin{pmatrix} 0 & 0 & 0 & 0 \\ 0 & 1 & -1 & 0 \\ 0 & -1 & 1 & 0 \\ 0 & 0 & 0 & 0 \end{pmatrix}\,,
\end{equation}
for the type IIA and type IIB theories.  

\noindent The type IIB, 0A and 0B superstring vacua are left-right symmetric on the world-sheet and thus can be modded out by the world-sheet parity $\Omega $ \cite{Sagnotti:1987tw, Pradisi:1988xd, Horava:1989vt, Bianchi:1990yu, Bianchi:1990tb, Bianchi:1991eu}, possibly combined with extra symmetries. 
Following the rules of the orientifold construction (see, for instance, \cite{Angelantonj:2002ct}), the Klein bottle projection for the type IIB and type 0A superstrings is unique \cite{Bianchi:1990yu,Sagnotti:1995ga}, while four independent choices are allowed for the type 0B \cite{Bianchi:1990yu,Sagnotti:1995ga, Sagnotti:1996qj}.

\noindent The analysis for the ten-dimensional heterotic models has been carried out in great detail in \cite{Angelantonj:2023egh} reproducing the results of \cite{Cribiori:2020sct} for the $\text{SO}(16) \times \text{SO}(16)$ model. The vacua with $\mathcal{N}=(1,1)$ world-sheet supersymmetry and their related orientifolds have been discussed in \cite{Leone:2023qfd} reproducing the results of \cite{Cribiori:2021txm} for the type 0'B superstring. The analysis for the type 0A superstring in \cite{Leone:2023qfd} is somewhat quick and schematic, and thus we take advantage of this opportunity to provide a more detailed discussion.

\noindent The type 0A superstring  \cite{Bianchi:1990yu,Sagnotti:1995ga,Sagnotti:1996qj} combines the left and right moving $\text{SO}(8)$ characters via the GSO matrix \eqref{type0GSO}. In this case, the spectrum is purely bosonic, and thus the tachyon related to the identity of the algebra is present. In addition, the massless states comprise the dilaton, the metric, the Kalb-Ramond field, two RR $1$-forms and two RR $3$-form fields.

\noindent Compared to the type 0B case, the only changes occur for the bosonic content in the RR sector, and thus play no role in the computation of the sector-averaged sum associated with the torus amplitude with the one reported in \cite{Leone:2023qfd} for the type 0B superstring,
\begin{equation} \label{0Aavsumtor}
	\langle d(n) \rangle \left ( \mathcal{T} \right ) = \sum_{\ell=1}^\infty \varphi(\ell) \frac{e^{\frac{8 \pi}{\ell}\sqrt{n/2}}}{(2n)^{11/2}} \ ,
\end{equation} 
where the exponential growth is dictated by the NS vacuum associated to the tachyon with mass $H_0=-1/2$.

\noindent As already mentioned, the type 0A superstring is left-right symmetric, but admits only one type of orientifold projection \cite{Bianchi:1990yu}, resulting in the Klein bottle amplitude 
\begin{equation}
	\mathcal{K}= O_8 + V_8 \, .
\end{equation}     
Under the action of the $S$ modular transformation, the tree-level propagation of unoriented closed-string states is encoded in the transverse channel 
\begin{equation}
	\tilde{\mathcal K}= 2^5 \left \{ O_8 + V_8 \right \} \, .
\end{equation}
The structure of the amplitude is not affected when moving from the one-loop and tree-level channel and {\em vice versa}. Furthermore, in both cases, there is a tachyonic character, which shows that the closed-string tachyon is kept by the orientifold projection\footnote{It eliminates the Kalb-Ramond field and one copy of the RR $1$-form and $3$-form.} and couples to orientifold planes. 
Although there are only NS NS tadpoles, and thus the introduction of the open sector is not mandatory, we shall be a bit more general and include D-branes into the game to exploit the properties of the sector-averaged sums associated to each amplitude. Since in the transverse channel only the $O_8$ and $V_8$ characters may appear, the only available choice for the transverse annulus amplitude is given by
\begin{equation}
	\tilde{\mathcal A}= 2^{-5} \left \{ (n_B-n_F)^2 O_8 +  (n_B+n_F)^2 V_8\right \} \, ,
\end{equation} 
where we have introduced the stack of branes $n_B$ and $n_F$, leading in the direct channel to 
\begin{equation}
	\mathcal A= (n_B^2+n_F^2) (O_8 + V_8) \, - \, 2 \, n_B \, n_F (S_8 + C_8) \, .
\end{equation}  
Aside for an overall sign $\varepsilon$, we can compute the M\"obius strip amplitude in the transverse channel as 
\begin{equation}
	\tilde{\mathcal M}= 2 \varepsilon \left \{ (n_B-n_F) \hat O_8 +  (n_B+n_F) \hat V_8\right \} \, ,
\end{equation} 
that after a $P$ transformation gives 
\begin{equation}
	\mathcal M= - \varepsilon \left \{ (n_B-n_F) \hat O_8 -  (n_B+n_F) \hat V_8\right \} \, .
\end{equation} 
Therefore we can read the massless open-string spectrum according to the choice of $\varepsilon$. If $\varepsilon=1$, the open sector provides a gauge boson in the adjoint of $\text{USp}(n_B) \times \text{USp}(n_F)$ gauge group, with a tachyon transforming into the $(\smalltableau{  \null \\ \null \\ },1)+(1,\smalltableau{  \null \& \null \\ })$ representations and left and right fermion transforming in the bi-fundamental representation. If $\varepsilon=-1$ the gauge group becomes a product of orthogonal ones $\text{SO}(n_B) \times \text{SO}(n_F)$ with a tachyon in  $(\smalltableau{  \null \& \null \\ },1)+(1,\smalltableau{  \null \\ \null \\ })$ and still with fermions transforming in the bi-fundamental representations.  

\noindent The sector-averaged sums associated to the transverse amplitudes come straightforwardly 
\begin{equation}
	\begin{array}{ll}
		\left \langle d(n) \right \rangle\left (\mathcal{K} \right )= \frac{e^{4 \pi \sqrt{n/2}}}{  (2 n)^{11/4}} \, ,
		\\ 
		\\
		\left \langle d(n) \right \rangle\left (\mathcal{A} \right )= \frac{e^{4 \pi \sqrt{n/2}}}{(2n)^{11/4}} \left ( n_B -n_F \right )^2 \, ,
		\\
		\\
		\left \langle d(n) \right \rangle\left (\mathcal{M} \right )= - \varepsilon \frac{e^{2 \pi \sqrt{n/2}}}{  2^{9/4} n^{11/4}} \left ( n_B -n_F \right ) \, ,
	\end{array}
\end{equation}
reflecting the coupling of the closed-string tachyon to boundaries and cross-caps. Since all the terms are non-vanishing, this is enough to conclude that the closed-string tachyon survives the orientifold projection. Indeed, this can be seen from the one-loop Klein bottle amplitude and is reflected into the computation of the sector-averaged sum associated to the transverse channel
\begin{equation}
	\left \langle d(n) \right \rangle  ( \Tilde{\mathcal{K}} ) =\frac{e^{4 \pi \sqrt{n/2}}}{ (2 n)^{11/4}} \ .
\end{equation}
The open-string tachyon is also unoriented, implying non-vanishing contributions to the sector-averaged sums of both transverse annulus and M\"obius strip amplitudes  
\begin{equation} \label{eq:opentransversesas}
	\begin{array}{ll}
		\left \langle d(n) \right \rangle (\Tilde{\mathcal{A}}  )= \frac{e^{4 \pi \sqrt{n/2}}}{(2 n)^{11/4}} \left ( n_B^2 + n_F^2 \right ) \, ,
		\\
		\\
		\left \langle d(n) \right \rangle (\Tilde{\mathcal{M}}  )= \varepsilon \frac{e^{2 \pi \sqrt{n/2}}}{ 2^{9/4} n^{11/4}} \left ( n_B -n_F \right ) \, .
	\end{array}
\end{equation}
Choosing $\varepsilon=-1$ and the Chan-Paton factors to be $n_B=1$ and $n_F=0$, the tachyon is projected away from the open spectrum. Note, however, that there is no way to provide a mutual cancellation between the associated sector-averaged sums, since the powers and the coefficients characterising the expressions in \eqref{eq:opentransversesas} are different.	

\section{An example in \texorpdfstring{$D=9$}{D=9}} \label{10dexamples}

Aside from the ten-dimensional examples described so far, there are other ways to build non-supersymmetric string vacua. A notable option is breaking supersymmetry also via continuous deformations, employing the Scherk-Schwarz mechanism \cite{Scherk:1979zr,Ferrara:1987es}, or its T-dual version: M-theory breaking \cite{Antoniadis:1998ki,Antoniadis:1998ep}. Focusing on the simplest scenario with the internal manifold described by a circle $S^1(R)$, one can realise both constructions as freely acting orbifolds \cite{Kounnas:1989dk}, where the supersymmetry breaking generator $g$ is combined with a suitable shift $ \delta$ along the compact direction. Explicitly, this means that $g = (-1)^F$, with $F$ the space-time fermion number, while $\delta$ acts as $\delta: y \to y+\pi R$ for the Scherk-Schwarz case and $\delta: \ y_{L,R} \to y_{L,R} \pm \frac{\alpha'}{2 R} \pi$ for the M-theory breaking. The latter action has a natural interpretation in terms of the T-dual variable $\tilde y = y_L-y_R$ implying that the two oriented closed-string vacua are equivalent. However, new physics emerges in the orientifold case, since T-duality changes the dimensionality of the orientifold planes and D-branes, as well as the $\Omega$ projection, providing for the M-theory breaking construction an additional projection which eliminates the tachyon \cite{Dudas:2000sn, Dudas:2002dg,  Dudas:2003wp}.  Therefore, we shall focus on the latter\footnote{For a parallel discussion of misaligned supersymmetry for the Scherk-Schwarz and M-theory breaking compactification see \cite{Angelantonj:2023egh, Leone:2023qfd}.} by first discussing the role of misaligned supersymmetry for the oriented spectrum and then for the two orientifold projections. 

\noindent Let us start from the torus partition function obtained from a freely acting orbifold of the type IIB superstring generated by $(-1)^F\delta $
\begin{equation}
	\begin{split}
		{\mathcal T}_\text{M} &= \tfrac{1}{2} |V_8 - S_8|^2 \, \sum_{m,n} \Lambda_{m,n} (R) + \tfrac{1}{2} |V_8 + S_8 |^2 \, \sum_{m,n} (-1)^n \, \Lambda_{m,n} (R)
		\\
		& + \tfrac{1}{2} |O_8 - C_8|^2 \, \sum_{m,n} \Lambda_{m+\frac{1}{2},n} (R) + \tfrac{1}{2} |O_8 + C_8|^2 \, \sum_{m,n} (-1)^n\, \Lambda_{m+\frac{1}{2},n} (R)  \,,
	\end{split}
\end{equation}
where the action of $\delta$ on the winding modes and its modular completion are shown. The Kaluza-Klein momenta and windings associated to the compact direction contribute with the standard Narain lattice
\begin{equation}
	\Lambda_{m,n} = \frac{q^{\frac{\alpha'}{4} \left( \frac{m}{R} + \frac{nR}{\alpha '}\right)^2}}{\eta} \, \frac{\bar q^{\frac{\alpha'}{4} \left( \frac{m}{R} - \frac{nR}{\alpha '}\right)^2}}{\bar \eta}\,.\label{NarainL}
\end{equation}
In the $R\to 0$ limit, the orbifold action is trivialised and one recovers the supersymmetric  IIB theory, while for generic values of the radius $R$ supersymmetry is spontaneously broken and the gravitini acquire a mass $m\simeq R$. The excitations of the NS-NS vacuum in $| O_8|^2$ survive the GSO projection in the twisted sector, and the lightest state has mass
\begin{equation}
	m^2_{|O_8|^2} =-\frac{1}{2 \alpha'} +\frac{1 }{4} \left( \frac{1}{2 R}\right)^2\,. \label{masstachyon}
\end{equation}
This scalar is then massive for small values of $R$, but becomes tachyonic above the critical radius $R_c = \sqrt{\alpha'} / 2 \sqrt{2}$. As the radius increases, more and more states become tachyonic, and it is clear that these models represent an ideal ground to study the realisation of {\em misaligned supersymmetry} in string theory\footnote{See for instance \cite{Abel:2015oxa} for a first study of {\em misaligned supersymmetry} in the context of the T-dual Scherk-Schwarz reductions.}. 

\begin{figure}[]
	\centering
	\includegraphics[width=7cm]{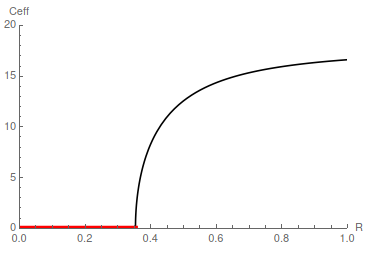}
	\caption{The  figure displays the dependence of the effective central charge on $R$ with $\alpha'=1$ and shows that above the critical radius the M-theory breaking reduction is not a deformation of the original theory.}\label{figCeff}
\end{figure}

\noindent To illustrate the analysis of Section \ref{MSoriented} on the degeneracies of states, we need to select rational values for $ R^2/\alpha ' = s/t \in\mathbb{Q}$ since, in this case, the Narain lattice reduces to a RCFT 
\begin{equation}
	\sum_{m,n} \Lambda_{m,n} (R) \to \sum_{a=0}^{2st-1} \lambda_a \, \bar\lambda_{a l}\,, \label{NarainRCFT}
\end{equation}
with the $2st$ characters defined as
\begin{equation}
	\lambda_a (q) = \sum_m \frac{q^{st \left( m + \frac{a}{2st}\right)^2}}{\eta (q)}\,. \label{defl}
\end{equation}
Note that $\lambda_0$ and $\lambda_{st}$ are real, while $\lambda_a$ and $\lambda_{2st-a}$, $a=1,\ldots , st-1$,  form conjugate pairs. 
The $\lambda$s have conformal weight $h_a = a^2/4st$, and thus $H_a=a^2/4st-1/24$, with conjugate pairs carrying the same weight.  Notice that $H_a <0$ for $a < \sqrt{2s}$ and $p=0$, so that the associated characters can describe tachyonic states. In eq. \eqref{NarainRCFT} the anti-holomorphic characters have index $a l$, where $l = rt + s v$, with the integers $r$ and $v$ satisfying the relation 
\begin{equation}\label{eq:primecond}
    rt - vs =1 \, ,
\end{equation}
and the label $a l$ is defined modulo $2st$. 
As shown in \cite{Angelantonj:2023egh}, these characters are eigenstates of the shift operator $\delta$ only for even $t$, but must be broken into sub-characters for odd $t$. For simplicity, here we shall restrict the discussion  to the even-$t$ case, where
\begin{equation}
	\delta\ :\quad \lambda_a \to (-1)^{a/2s} \, \lambda_a\,,
\end{equation}
and we  also take $s=1$, so that the condition \eqref{eq:primecond} can be easily solved by $r=0$ and $v=-1$, for any $t$\footnote{Other choices for $s$ and $t$ yield equivalent results.}. In this setting, the action of the modular group on these characters is encoded in the $T$ and $S$ matrices
\begin{equation}
	T_{a b} = e^{i \pi \left( \frac{a^2}{2t}-\frac{1}{12}\right)}\, \delta_{a b}\,,\qquad
	S_{a b} = \frac{e^{2 \pi i \frac{a b}{2t}}}{\sqrt{2t}}\,.
\end{equation} 
The new characters that enter the partition function are thus \cite{Angelantonj:2023egh}
\begin{equation} \label{eq:characters}
	\{\chi_\alpha\}_{\alpha=0}^{8t-1} = (O_8 , V_8 ,  S_8 ,  C_8 ) \otimes \{\lambda_a \}_{a=0}^{2t-1}\,.
\end{equation}
where the $\chi_\alpha$ have shifted conformal weights  $H_{2tp+a} = \frac{a^2}{4t} - \frac{1}{2} \delta_{p,0}$, with $p=0,\ldots ,3$. In terms of the characters \eqref{eq:characters}, the torus amplitude associated to the $(-1)^F \,  \delta$ orbifold of the type IIB superstring reads
\begin{equation} \label{torM}
	\begin{split}
		{\mathcal T}_{\text{M}} &= \sum_{a=0}^1 \left ( \left | \chi_{2t+at} \right |^2 + \left | \chi_{4t+at} \right |^2 \right )
		\\
		&+ \sum_{b=1}^{\frac{t}{2}-1}\left(  \chi_{2t+2b} \bar{\chi}_{2t-2b}+ \chi_{4t+2b}\bar{\chi}_{4t-2b} \right.
		\\
		&\qquad\qquad\qquad \left.
		+ \chi_{2t+t+2b}\, \bar\chi_{2t-t-2b} + \chi_{4t+t+2b}\, \bar\chi _{4t-t-2b} \right)
		\\
		&-\sum_{b=0}^{\frac{t}{2}-1}\left(  \chi_{2t+2b+1} \bar{\chi}_{4t-2b-1}+ \chi_{4t+2b+1}\bar{\chi}_{2t-2b-1} \right.
		\\
		&\qquad\qquad\qquad \left.
		+ \chi_{2t+t+2b+1}\, \bar\chi_{4t-t-2b-1} + \chi_{4t+t+2b+1}\, \bar\chi _{2t-t-2b-1} \right)
		\\
		&+  \sum_{a=0}^{1} \left ( \left | \chi_{\frac{2a+1}{2} t} \right |^2 + \left | \chi_{6t+\frac{2a+1}{2}t}\right |^2 \right )
		\\
		&+ \sum_{b=1}^{\frac{t}{2}-1}\left(  \chi_{\frac{t}{2}-2b} \bar{\chi}_{\frac{t}{2}+2b}+ \chi_{6t+\frac{t}{2}-2b} \bar{\chi}_{6t+\frac{t}{2}+2b}\right.
		\\
		&\qquad\qquad\qquad \left.
		+ \chi_{\frac{3t}{2}-2b}\, \bar\chi_{-\frac{t}{2}+2b} + \chi_{6t+\frac{3t}{2}-2b}\, \bar\chi_{6t-\frac{t}{2}+2b} \right)
		\\
		&- \sum_{b=0}^{\frac{t}{2}-1}\left(  \chi_{\frac{t}{2}-2b-1} \bar{\chi}_{\frac{t}{2}+2b+1}+ \chi_{6t+\frac{t}{2}-2b-1} \bar{\chi}_{6t+\frac{t}{2}+2b+1}\right.
		\\
		&\qquad\qquad\qquad \left.
		+ \chi_{\frac{3t}{2}-2b-1}\, \bar\chi_{-\frac{t}{2}+2b+1} + \chi_{6t+\frac{3t}{2}-2b-1}\, \bar\chi_{6t-\frac{t}{2}+2b+1} \right) \, .
	\end{split}
\end{equation}
\noindent Given the expression for the torus partition function, we can proceed by computing the sector-averaged sum, which reads, for $t < 8$,   
\begin{equation} \label{sastorMth}
	\langle d(n) \rangle ( \mathcal{T})= \left (\frac{8-t}{16} \right )^4 \frac{1}{ n^{5}}\, 
	\sum_{\ell=1}^{\infty} \, \varphi (\ell) \, e^{\frac{2\pi}{\ell} \sqrt{(8-t)\, n}} \ ,
\end{equation} 
while it vanishes for $t \geq 8$, as expected. The Scherk-Schwarz reduction at this level can be obtained via a T-duality operation, in which the radius of compactification is sent to $R \to \alpha'/R$. This means that the sector-averaged sum for this vacuum is related to \eqref{sastorMth} by exchanging $s \leftrightarrow t$. We can therefore define the effective central charge in terms of the radius $R$
\begin{equation}
	C_\text{eff} = \begin{cases}
		2 \pi \sqrt{8-\alpha'/R^2 } & \text{for}\quad R^2 > \alpha'\frac{1}{8}\,,
		\\
		0 & \text{for}\quad  R^2 \le \alpha'\frac{1}{8} \, ,
	\end{cases} \label{CeffRirratMth}
\end{equation} 
where the limits $R \to 0$ and $R \to \infty$ recover the results for the type IIB and type 0B superstrings respectively, as shown in figure \ref{figCeff} once the radius is continued to real values. Recalling that the sector-averaged sum is tied to the number of degrees of freedom, the behaviour of \eqref{CeffRirratMth} suggests that the M-theory breaking compactification does not correspond to a continuous deformation which, as such, should imply that the number of degrees of freedom does not change with the radius $R$. It would be interesting to make this observation more quantitative, by deforming the partition function whose free energy can be shown to depend quadratically on $C_{\text{eff}}$  \cite{Angelantonj:2023egh}. This would characterise the phase transition as a first order and it would be interesting to look for a holographic dual in AdS$_3$ for which, in the large central charge limit, the Hagedorn-like phase transitions of the CFT are mapped to Hawking-Page transitions of the gravity theory, dominated by the entropy of BTZ black holes \cite{Keller:2011xi}. However, such a study would require a clearer picture on the meaning of the averaging procedure in this context, and lies beyond the scope of this thesis. 

\noindent The amplitude described in \eqref{torM} is invariant under world-sheet parity, and can be orientifolded by adding the Klein bottle amplitude that, for the standard choice of $\Omega$, reads
\begin{equation} \label{kleinM}
	\begin{split}
		{\mathcal K}_{\text{M}} &=  \sum_{a=0}^{1}\left(  \chi_{a t+ 2t} - \chi_{at+4t} \right)
		+ \sum_{a=0}^{1} \left( \chi_{\frac{2a+1}{2} t}\, - \chi_{\frac{2a+1}{2}t+ 6 t}\right ) \,.
	\end{split}
\end{equation}
Notice that aside from the first two terms associated to the $V_8$ and $S_8$ characters dressed with the KK and winding excitations, two extra contributions appear associated to $O_8$ and $C_8$.  In the transverse channel
\begin{equation} \label{trklein}
	\Tilde{\mathcal{K}}_{\text{M}}=  2^5 \frac{2}{\sqrt{t}} \sum_{b=0}^{t/2-1} \left \{ \chi_{2t+4b}- \chi_{4t+4b+2} \right \} \ ,
\end{equation}
so that this vacuum involves pairs of O9$_-$ and $\overline{\text{O9}}_-$ planes which do not carry a net RR charge, as can be seen from the fact that the characters $\chi_{4t+4b+2}= S_8 \, \lambda_{4b+2}$ are massive, $m^2\propto \frac{(4b+2)^2}{4t} = \frac{1}{4} R^2\, (4 b+2)^2 $. In the formal limit $R\to 0$ they become massless, the $\overline{\text{O9}}_-$ planes decouple and a net RR tadpole emerges from \eqref{trklein}. Taking this limiting case into account,  we shall also impose the RR tadpoles associated to the $\chi_{4t+4b+2}$ characters.

\noindent Following \cite{Antoniadis:1998ki}, we add $(n_1,n_2)$ stacks of branes and $(n_3,n_4)$ stacks of antibranes so that
\begin{equation}\label{trannM}
	\begin{split}
		\Tilde{\mathcal{A}}_{\text{M}}= 2^{-5} \frac{2}{\sqrt{t}} \sum_{b=0}^{t/2-1} & \left \{  \left ( n_1 + n_2 + n_3 + n_4 \right )^2  \chi_{2t + 4 b}  + \left ( n_1 - n_2 + n_3 - n_4 \right )^2 \chi_{2t+ 2 + 4b }  \right.  
		\\
		& \left.  -  \left ( n_1 + n_2 - n_3 - n_4 \right )^2 \chi_{4t +4b}- \left ( n_1 - n_2 - n_3 + n_4 \right )^2  \chi_{4t+4b+2}\right \} \ ,  
	\end{split}
\end{equation}
and
\begin{equation} 
	\begin{split}
		\Tilde{\mathcal{M}}_{\text{M}}= - 2 \frac{2}{\sqrt{t}} \sum_{b=0}^{t/2-1} & \left \{  \left ( n_1 + n_2 + n_3 + n_4 \right )  \hat{\chi}_{2t + 4 b}   -   \left ( n_1 - n_2 - n_3 + n_4 \right ) \hat{\chi}_{4t+4b+2}\right \}.   
	\end{split}
\end{equation}  

\noindent The tadpole conditions 
\begin{equation}
	n_1 + n_2 + n_3 + n_4 = 32\ , \qquad n_1 + n_2 - n_3 - n_4 = 0\ ,
\end{equation}
associated to the massless characters $\chi_{2t}$ and $\chi_{4t}$, together with the ``massive'' tadpole $n_1 - n_2 - n_3 + n_4 = 32$ for $\chi_{4t+4b+2}$, admit the unique solution
\begin{equation}
	n_1 = n_4 =16\ , \qquad n_2 = n_3 =0\,.
\end{equation}
The open spectrum can be then deduced by employing the $S$ and $P$ modular transformations, yielding 
\begin{equation} \label{annM}
	\begin{split}
		\mathcal{A}_{\text{M}}= \sum_{a=0}^{1} \bigg \{   & \left ( n_1^2+n_2^2 + n_3^2+n_4^2\right )  \left (   \chi_{2t + \frac{2 a }{2} t}  -\chi_{4t + \frac{2 a }{2} t}\right )     
		\\
		& + 2 \left ( n_1 n_2 + n_3 n_4 \right )  \left (   \chi_{2t + \frac{2 a +1}{2} t}  -\chi_{4t + \frac{2 a +1 }{2} t}  \right )     
		\\
		& + 2 \left ( n_1 n_3 + n_2 n_4 \right )  \left (   \chi_{ \frac{2 a }{2} t}  -\chi_{6t + \frac{2 a  }{2} t}  \right )     
		\\
		& +  2 \left ( n_2 n_3 + n_1 n_4 \right ) \ \left (  \chi_{\frac{2 a+1 }{2} t}  - \chi_{6t+\frac{2 a+1 }{2} t} \right ) \bigg \} \ ,
	\end{split}
\end{equation}
and 
\begin{equation}
	\begin{split}
		\mathcal{M}_{\text{M}}= -  \sum_{a=0}^{1}& \left \{ \left ( n_1 + n_2 + n_3 + n_4\right )  \hat{\chi}_{2t+\frac{2 a }{2} t}  -  \left ( n_1 - n_2 - n_3 + n_4\right ) (-1)^a \hat{\chi}_{4t+\frac{2 a }{2} t}  \right \} \ .
	\end{split}
\end{equation}

\noindent At low energy, this M-theory breaking orientifold describes a graviton, the dilaton and a RR two-form from the closed-string sector together with an $\text{SO}(16) \times \text{SO}(16)$ gauge group coupled to a left-handed fermion in the adjoint representation. For large values of the compactification radius, {\em i.e.} for $t<8$ a real closed-string tachyon and an open-string one in the $(\mathbf{16},\mathbf{16})$ representation emerge. 

\noindent The amplitudes can be shown to be compatible with those of \cite{Antoniadis:1998ki} once the value $R^2 = \alpha '/t$ is chosen, and the momentum and winding sums are written in terms of the $\lambda$ characters. 

\noindent We can now study the classical stability for this M-theory breaking vacuum by computing the associated sector-averaged sums. An inspection of the direct channel amplitudes gives $\left \langle d(n) \right \rangle \left ( \mathcal{Z}_I \right )=0$, reflecting the absence of closed-string tachyons that freely propagate between D-branes or O-planes. This, however, does not guarantee the classical stability of the whole construction, since a tachyon can be present in ${\mathcal T}_{\text{M}}$ for special values of $t$. A completely different story regards the sector-averaged sum associated with the transverse channel. 
In fact, from $\Tilde{\mathcal K}$ one finds
\begin{equation}
	\left \langle d(n) \right \rangle  ( \Tilde{\mathcal{K}}_{\text{M}}  )= 2 \frac{1}{\sqrt{2} n^{\frac{5}{2}}} \left ( \frac{8-t}{16} \right )^2 e^{\pi \sqrt{(8-t)\, n}}\ ,
\end{equation}
for $t<8$, while it vanishes if $t \geq 8$. This behaviour is consistent with the fact that the closed-string spectrum is tachyon-free for $t \geq 8$, while a tachyon is present when $t<8$.

\noindent The sector-averaged sum from the transverse annulus amplitude has a similar behaviour
\begin{equation} \label{sastrannM}
	\begin{split}
		\left \langle d(n) \right \rangle ( \Tilde{\mathcal{A}}_{\text{M}})=  2 \frac{1}{\sqrt{2} n^{\frac{5}{2}}}  \ 2 n_1 n_4 \  \left ( \frac{8-t}{16}  \right )^2 e^{ \pi \sqrt{(8-t)\, n} }  ,
	\end{split}
\end{equation}
reflecting the presence of a tachyon in the bi-fundamental representation $(\boldsymbol{16}, \boldsymbol{16})$ when $t<8$. When $t \geq 8$, $\left \langle d(n) \right \rangle ( \Tilde{\mathcal{A}}_{\text{M}})=0$ since all would-be tachyons are actually massive. Finally, the sector-averaged sum of the transverse Moebius strip amplitude vanishes identically, consistently with the fact that the open-string tachyon, when present, is oriented and valued in the bi-fundamental representations of $\text{SO}(16) \times \text{SO}(16)$. Therefore, in this case, the vanishing of the sector-averaged sum $\langle d (n) \rangle (\tilde{\mathcal A}_{\text{M}})$ is a necessary and sufficient condition for the classical stability of the open-string spectrum, while for the stability of the closed-string sector we must require that both $\langle d (n)\rangle ({\mathcal T}_{\text{M}})$ and  $\langle d (n)\rangle (\tilde{\mathcal K}_{\text{M}})$ vanish. 

\noindent As discussed at the beginning of this Section, the Klein bottle of eq. \eqref{kleinM} is not the only orientifold projection compatible with \eqref{torM}. Indeed, in \cite{Dudas:2002dg, Dudas:2000sn, Dudas:2003wp} it was shown that other options for $\Omega$ are possible, and in particular one can choose to project away the closed-string tachyon by using 
\begin{equation} \label{kleinM'}
	{\mathcal K}'_{\text{M}} =  \sum_{a=0}^{1}\left(  \chi_{a t+ 2t} - \chi_{at+4t} \right)
	- \sum_{a=0}^{1} \left( \chi_{\frac{2a+1}{2} t}\, - \chi_{\frac{2a+1}{2}t+ 6 t}\right ) \, .
\end{equation}
This change of sign affects the geometry of the orientifold planes, since now
\begin{equation} \label{trkleinM'}
	\Tilde{\mathcal{K}}'_{\text{M}}=  2^5 \frac{2}{\sqrt{t}} \sum_{b=0}^{t-1} \left \{ \chi_{2t+4b+2}- \chi_{4t+4b} \right \} \ ,
\end{equation}
which reveals the presence of O9$_-$ and $\overline{\text{O9}}_+$ planes which have a net RR charge but vanishing tension. The annulus amplitude is still given by eqs. \eqref{annM} and \eqref{trannM}, while the tree-level channel M\"obius strip amplitude reads
\begin{equation} \label{trmoebiusM'} 
	\begin{split}
		\Tilde{\mathcal{M}}'_{\text{M}}= - 2 \frac{2}{\sqrt{t}} \sum_{b=0}^{t-1} & \left \{  \left ( n_1 - n_2 + n_3 - n_4 \right )  \hat{\chi}_{2t + 4 b+2}   -   \left ( n_1 + n_2 - n_3 - n_4 \right ) \hat{\chi}_{4t+4b}\right \}.   
	\end{split}
\end{equation}  
The RR tadpole conditions
\begin{equation}
	n_1 + n_2 - n_3 - n_4 =32 \ , \qquad n_1 - n_2 - n_3 + n_4 =0 
\end{equation}
are clearly incompatible with the NS-NS tadpole $n_1 + n_2 + n_3 + n_4 =0$ since the O-planes are tensionless. Upon a $P$ modular transformation, we get  
\begin{equation} \label{moebiusM'}
	\begin{split}
		\mathcal{M}'_{\text{M}}= -  \sum_{a=0}^{1}& \left \{ \left ( n_1 - n_2 + n_3 - n_4 \right ) (-1)^a \hat{\chi}_{2t+\frac{2 a }{2} t} - \left ( n_1 + n_2 - n_3 - n_4 \right ) \hat{\chi}_{4t+\frac{2 a }{2} t}  \right \} \ .
	\end{split}
\end{equation}    

\noindent The light spectrum from the closed strings comprises a graviton, the dilaton and a RR two-form, and no extra light states emerge at special values of the compactification radius, since the would-be tachyon is removed by the new orientifold projection. From \eqref{annM} and \eqref{moebiusM'} we read the gauge group $\text{SO}(n_1) \times \text{USp}(n_2) \times \text{SO}(n_3) \times \text{USp}(n_4)$. Left-handed fermions and tachyons transform in the representations $(\smalltableau{  \null \\ \null \\ },1;1,1)+(1,\smalltableau{  \null \\ \null \\ };1,1)+(1,1;\smalltableau{  \null \& \null \\ },1)+(1,1;1,\smalltableau{  \null \& \null \\ })$ and $(\smalltableau{ \null \\},1;\smalltableau{\null \\},1) + (1,\smalltableau{ \null \\};1,\smalltableau{\null \\}) $, respectively. When $t<8$ additional tachyons in the representations $(\smalltableau{ \null \\},1;1,\smalltableau{\null \\}) + (1,\smalltableau{ \null \\};\smalltableau{\null \\},1) $ appear. All open-string ta\-chyons can be eliminated by taking the minimal solution $n_1=n_2=16$ and $n_3=n_4=0$ of the RR tadpoles, but we shall focus on the more general set-up allowing the simultaneous presence of branes and anti-branes to fully analyse the behaviour of the sector-averaged sums. 
It is possible to show that, choosing the value $R^2 = \alpha '/t$ in the model of \cite{Dudas:2002dg} one obtains the amplitudes presented in this section.

\noindent As in the case of the standard M-theory breaking, $\langle d(n) \rangle (\mathcal{Z}'_I) \equiv 0$ for every loop amplitude, since no closed-string tachyon propagates between D-branes or O-planes. The sector-averaged sums associated with the tree-level channels are, instead, slightly modified and read 
\begin{equation} \label{sastrkleinM'}
	\left \langle d(n) \right \rangle ( \Tilde{\mathcal{K} }'_{\text{M}} )= - 2 \frac{1}{\sqrt{2} n^{\frac{5}{2}}} \left ( \frac{8-t}{16} \right )^2 e^{\pi \sqrt{(8-t)\, n} } \ ,
\end{equation}
and 
\begin{equation} \label{sastrannM'}
\begin{aligned}
	\left \langle d(n) \right \rangle ( \Tilde{\mathcal{A}}'_{\text{M}} )=   \frac{1}{\sqrt{2} n^{\frac{5}{2}}}  & \left \{ 2\left (  n_1 n_4 + n_2 n_3 \right )   2  \left ( \frac{8-t}{16} \right )^2 e^{\pi \sqrt{ (8-t) \, n}} \right.
 \\
 & \left. + 2\left (  n_1 n_3 + n_2 n_4 \right ) \frac{1}{4}\,  e^{4 \pi \sqrt{n/2}} \right \},
 \end{aligned}
\end{equation}
while $\left \langle d(n) \right \rangle ( \Tilde{\mathcal{M}}'_{\text{M}})=  0 $ since, also in this case, open-string tachyons are oriented. Even though the only difference with the standard M-theory breaking analysis is in the minus sign in eq. \eqref{sastrkleinM'}, the physical interpretation is {\em deeply} different. It tells us that the tachyon that appears in the spectrum when $t < 8$ is actually projected out. However, eq. \eqref{sastrkleinM'} fails to cancel the sector-averaged sum  \eqref{sastorMth} of the torus because of the appearance of the sub-leading contributions, $\ell >1$.

\noindent To conclude,  eq. \eqref{sastrannM'} reflects the presence of tachyons in the spectrum, as long as anti-branes are present. Also in this case, since all tachyons transform in bi-fundamental representations, a necessary and sufficient condition for the stability of the open-string sector is the vanishing of  $\left \langle d(n) \right \rangle ( \Tilde{\mathcal{A}}'_{\text{M}} )$.

\chapter{Rigidity: New solutions for BSB and SUSY K3 orientifolds}\label{ChBSB}

\newpage

\section{Orientifold vacua in six dimensions} \label{Sec:OrientifoldandKSV}

Orientifold vacua in six dimensions have received much attention, since their dynamics is highly constrained by anomaly cancellation, and thus are an ideal playground for model building and for studying supersymmetry breaking. In particular, we will focus on orientifolds obtained from the orbifold limits of K3 manifolds, $T^4/\mathbb{Z}_N$, with $N=2,3,4,6$ \cite{Dixon:1985jw, Dixon:1986iz, Dixon:1986jc, Aspinwall:1996mn}, on which the nonlinear sigma model is exactly solvable, and thus we can use an exact world-sheet description to investigate these vacua. In particular, in $D=6$ the standard world-sheet parity operator can be dressed with the orbifold action with fixed loci given by orientifold planes with different dimensionalities, namely O9$_-$ and O5$_-$. One must thus introduce D9 and D5 branes to avoid local anomalies \cite{Bianchi:1990yu, Gimon:1996ay, Gimon:1996rq, Dabholkar:1996pc}. This configuration preserves $\mathcal{N}=(1,0)$ supersymmetry, and the spectrum is organised in suitable super multiplets. However, this choice of the orientifold action is also compatible with the presence of O9$_+$ and O5$_+$ planes requiring $\overline{\text{D9}}$ and $\overline{\text{D5}}$ branes, which thus provide a compactification of the Sugimoto model, where, as in the ten-dimensional case, supersymmetry is broken in the full open-string sector. Actually, this is not the only way to break supersymmetry, since it is possible to dress the orientifold projection with an involution that flips the twisted $2$-cycles of K3 which implies the simultaneous presence of orientifold planes with positive and negative tension and charge\footnote{The case of a non-vanishing background $B$-field \cite{Bianchi:1991eu, Angelantonj:1999jh, Kakushadze:1998bw, Angelantonj:1999xf} will not be presented in this thesis.}. The possible configurations in this scenario are given by O9$_+$ and O5$_-$ or, alternatively, O9$_-$ and O5$_+$ planes, requiring the introduction of $\overline{\text{D9}}$ and D5 or D9 and $\overline{\text{D5}}$ branes, respectively. These scenarios are completely equivalent and are known in the literature as {\em Brane Supersymmetry Breaking} (BSB) \cite{Antoniadis:1999xk} and will be the subject (with its supersymmetric counterpart) of this Chapter\footnote{We will focus on the configuration with O9$_-$ and O5$_+$ planes. The other aforementioned choices are straightforward.}. The original construction \cite{Antoniadis:1999xk} was based on the $T^4/\mathbb{Z}_2$ orbifold, while generalisations to the $\mathbb{Z}_4$ and $\mathbb{Z}_6$ cases\footnote{The BSB version of the $T^4/\mathbb{Z}_3$ cannot be realised since there are no O5 planes identified by the orientifold action.} have been recently discussed in \cite{Angelantonj:2024iwi}. These models are particularly interesting, since they do not have tachyonic instabilities and admit a tree-level supersymmetric closed-string sector coupled to non-supersymmetric open strings. 
The consistency of the vacuum is guaranteed since supersymmetry is, actually, non-linearly realised \cite{Dudas:2000nv, Pradisi:2001yv}, with the massless goldstino in the singlet of the gauge group playing the role of the Volkov-Akulov field \cite{Volkov:1972jx}. Its presence, together with the non-vanishing dilaton tadpole, is hence crucial in writing down a consistent interaction term for the massless gravitino, which is coupled to a conserved current. In this setting supersymmetry is broken in the open-string sector, but such a breaking has to be thought of as occurring at the string scale, since no order parameter is present which may restore the symmetry.  

\noindent In \cite{Antoniadis:1999xk}, the simplest realisation of such class of vacua was presented, where, for instance, the cancellation of R-R tadpoles was achieved by placing $\overline{\text{D5}}$ branes on a single $\mathbb{Z}_2$ fixed point. However, this is not the only option at our disposal and one can still choose to place $\overline{\text{D5}}$ branes on all the fixed points. In this case, the charges of D9 branes can be cancelled against those of $\overline{\text{D5}}$ branes, following an observation made in \cite{Bianchi:1990yu, Aldazabal:1999nu} in a different setting. Such a mechanism may lead to non-trivial solutions, since the D9 and $\overline{\text{D5}}$ branes involved are fractional as one can see from the particle interpretation of the vacuum-to-vacuum orientifold amplitudes.    
On the other hand, this way of cancelling twisted charges may leave behind additional uncancelled NS-NS tadpoles, thus providing new contributions to the scalar potential. Therefore in addition to the usual dilaton contribution from the untwisted tadpoles, new terms arise involving blown up moduli, which could be responsible for an eventual spontaneous resolution of orbifold singularities, in case the dynamics forces them to get non-trivial vevs. Furthermore, when BSB orientifolds are built from higher order point group orbifolds, O-planes become fractional, as reflected in the absence of open-string moduli that would allow partial recombination of branes leaving the charge of orientifold planes uncancelled. As explained in \cite{Angelantonj:2024iwi}, this is manifest for the BSB orientifold built upon the $T^4/\mathbb{Z}_4$ orbifold, for which the presence of fractional O5$_+$ planes located on the four $\mathbb{Z}_4$ fixed points enforces the introduction of four stacks of $\overline{\text{D5}}$ branes on the top of them and local charge cancellation forbids $\overline{\text{D5}}$ branes to be moved away. A similar situation applies for the $T^4/\mathbb{Z}_6$ orbifold in which both O9$_-$ and O5$_+$ planes are fractional but since the $\mathbb{Z}_6$ action identifies only one fixed point, such feature is less manifest. Nevertheless, a solution without open-string moduli is also available in this case and will be discussed in the present Chapter.

\noindent Although this way of cancelling D-brane twisted charges seems to be specific to BSB vacua, a similar situation applies to supersymmetric vacua as well\footnote{For instance, the observation pointed out in \cite{Aldazabal:1999nu} was made for $D=4$ supersymmetric $\mathbb{Z}_6$ orientifolds.}. Indeed, also for the supersymmetric $T^4/\mathbb{Z}_4$ and $T^4/\mathbb{Z}_6$ orientifolds D9 and D5 branes are fractional, and one can choose to cancel the R-R charges of D9 branes against those of the D5 ones, once the latter are placed over all the fixed points. As a result, we obtain a finite number of non-trivial solutions to the R-R tadpole conditions that determine different {\em bona fide} string vacua. However, when $\mathcal{N}=(1,0)$ space-time supersymmetry is present, orientifold planes are fractional only for the $\mathbb{Z}_6$\footnote{Also in the $T^4/\mathbb{Z}_3$ orientifold O9$_-$ planes are fractional, but we will not discuss the features of this vacuum.} point group, for which the previous considerations apply. In the rest of this Chapter, we will present a parallel discussion of the BSB and supersymmetric vacua built from the $T^4/\mathbb{Z}_6$ orbifold and we will enlighten the features just described. 

\noindent For simplicity, we consider a factorised $T^4 = T^2 \times T^2$, with complex coordinates $(z_1,z_2)$. The $\mathbb{Z}_6$ group acts as a $60$-degree rotation on each $T^2$, which is then a symmetry for a torus with fixed complex structure $U= e^{\pi i /3}$. Compatibility with supersymmetry requires an equal or opposite rotation on the two $T^2$s. Our choice is
\begin{equation}
g:\quad (z_1 , z_2) \to  (e^{2\pi i/6} z_1 , e^{-2\pi i /6} z_2)\,,
\end{equation}
where $g$ is the generator of $\mathbb{Z}_6$. 

\noindent In orbifold compactifications, the structure of fixed points plays a crucial role, since it gives the multiplicities of the twisted states and determines the structure of the associated Hilbert spaces. The action of $g$ has a single fixed point with coordinates $\zeta=(0,0)$, which is also fixed under the action of $g^5$. The elements $g^2$ and $g^4$ form a $\mathbb{Z}_3$ subgroup. Aside from the origin, eight of the nine fixed points arrange in four $\mathbb{Z}_6$ doublets, and the associated Hilbert spaces only support $\mathbb{Z}_3$ projectors. Similarly, $g^3$ generates a $\mathbb{Z}_2$ subgroup and the fifteen fixed points different from the origin, arrange into five $\mathbb{Z}_6$ triplets, with only a $\mathbb{Z}_2$ projection.  

\noindent This structure of fixed points is reflected in the various terms of the torus partition function, which is well known and is explicitly written in \cite{Angelantonj:2024iwi}. Keeping only the massless states, we can write the partition function in terms of the $\text{SO}(4)$-characters associated to the little group of the six-dimensional Minkowski space, yielding
\begin{equation}
\begin{split}
{\mathcal T}_\text{IIB} =& \left | V_4 - 2 \, S_4 \right |^2 + 2 \left | 2 \, O_4 -  C_4 \right |^2  
\\
&+  \, \left | 2 \, O_4 -  C_4 \right |^2 + \, \left | 2 \, O_4 -  C_4 \right |^2  
\\
&+ (1+4)\left[ \, \left | 2 \, O_4 -  C_4 \right |^2 + \, \left | 2 \, O_4 -  C_4 \right |^2   \right]
\\ 
&+(1+5) \, \left | 2 \, O_4 -  C_4 \right |^2  \,.
\end{split} \label{eq:torZ6}
\end{equation}
The first line corresponds to the contribution from the untwisted sector, the second one encodes the contribution from the unique $\mathbb{Z}_6$ fixed point from the $g$ and $g^5$-twisted sectors, while in the third the extra multiplicity $4$ takes into account the four $\mathbb{Z}_3$ doublets form the $g^2$ and $g^4$-twisted sectors. Similarly for the last line associated to the $g^3$-twisted sector, where the $5$ encodes the five $\mathbb{Z}_2$ triplets. This partition function yields the (unique) massless spectrum with $\mathcal{N}=(2,0)$ supersymmetry, with one gravity multiplet and twenty-one tensor multiplets. 

\noindent Starting from this oriented closed-string sector, one can implement the two different orientifold projections giving rise to the supersymmetric and BSB vacua.

\noindent Although in eq. \eqref{eq:torZ6} all the massless characters appear left-right symmetric, when the full massive excitations are considered this parity symmetry only survives for the first contribution in the untwisted sector and the last one associated with the $g^3$-twisted sector. Indeed, the geometric action of the orbifold combines the $g^k$ and the $g^{N-k}$ twisted characters, resulting into a charge conjugate modular invariant partition function. This means that the orientifold projection yields the Klein bottle amplitude with massless characters
\begin{equation}
{\mathcal K}^{(\sigma)} = \left ( V_4- 2 \, S_4 \right ) + (-1)^\sigma (1+5) \, \left ( 2 \, O_4 - C_4 \right )\,,
\end{equation}
where we have introduced the involution $\sigma$, which acts trivially on the $g^3$-twisted cycles in the supersymmetric case, $\sigma=0$, while it flips the parity of these cycles in the BSB scenario $\sigma=1$. In the supersymmetric case, this means that the orientifold projection selects a $\mathcal{N}=(1,0)$ hyper multiplet on each $\mathbb{Z}_2$ fixed point, while for the BSB case the projection selects a tensor mulitplet. As a result, we have one tensor multiplet and two hyper multiplets arising from the untwisted sector, one tensor multiplet and one hyper multiplet from the $g$ and $g^5$ twisted sectors, five tensor multiplets and five hyper multiplets from the $g^2$ and $g^4$ twisted sectors, while the $g^3$ twisted sector yields six hyper multiplets in the supersymmetric case or six tensor multiplets for the BSB construction, as a consequence of the non-trivial involution $\sigma$. To summarise, the choice $\sigma=0$ provides fourteen hyper multiplets and seven tensor multiplets, while $\sigma=1$ leads to thirteen tensor multiplets and eight hyper multiplets.

\noindent The nature of orientifold planes is encoded in the transverse channel Klein bottle amplitude, which reads
\begin{equation}
\begin{aligned} \label{eq:trkleinZ6}
\tilde{\mathcal K}^{(\sigma)} =& \frac{2^{-5}}{6}\, \left(-2^5 \sqrt{v} - (-1)^\sigma \frac{2^5}{\sqrt{v}} \right)^2\, ( V_4- 2 \, S_4 ) 
\\
& + \frac{2^{-5}}{6}\, \left( - 2^5\, \sqrt{v} + (-1)^\sigma \frac{2^5}{\sqrt{v}} \right)^2\,  ( 2 \, O_4-  C_4 + 2 \, O_4-  C_4  ) 
\\
&+ \frac{2^{-4}}{9} \left[ 8\cdot  q_{2,t}^2 + (q_{2,t} -(-1)^\sigma 3 q_{2,t} )^2 \right] 
 ( 2 \, O_4-  C_4)
 \\
 & + \frac{2^{-4}}{9} \left[ 8\cdot q_{4,t}^2 + (q_{4,t} -(-1)^\sigma 3 q_{4,t} )^2 \right] 
 ( 2 \, O_4-  C_4) \, .
\end{aligned}
\end{equation}
In the first line, we have presented the contribution to the untwisted tadpoles. It  explicitly shows that the choice of the involution determines which orientifold planes are considered: $\sigma=0$ reflects the presence of O-planes carrying negative tension and charge O9$_-$ and O5$_-$, while $\sigma=1$ requires orientifold planes with opposite tension and charge, namely O9$_-$ and O5$_+$. The last two lines encode the contribution from the $g^2$ and $g^4$-twisted sectors, and reflect the geometry of the configuration of orientifold planes. Indeed, the O9$_-$ planes fill all space-time and are thus charged under the twisted R-R forms located on all the $\mathbb{Z}_3$ fixed points. The O5$_+$ planes are instead placed at the $\mathbb{Z}_2$ fixed points, and thus only the one located at the origin (which corresponds to the unique $\mathbb{Z}_6$ fixed point) carries a non-trivial charge. The $g^{2,4}$-twisted charges of O9$_-$ planes are given by $q_{2,t} = -(-1)^{\sigma} q_{4,t}= (-1)^\sigma 8$, while the O5$_\mp$ planes are coupled to the six-form through  $ \mp 3 q_{\alpha,t}$, following the usual formula \cite{Angelantonj:1999ms}
\begin{equation} \label{normalisationcharges}
q =  q^{\text{9}} \sqrt{\frac{\# \text{ fixed points}}{\#\text{ occupied fixed points}}} \,.
\end{equation}
To cancel irreducible anomalies \cite{Aldazabal:1999nu, Bianchi:2000de}, one must introduce D9 and D5 branes with the appropriate charges required to cancel the R-R tadpoles. In general, the Chan-Paton factors do feel the action of the orbifold group, so that each element of the point group $g^\alpha$ identifies the corresponding brane images, and we label with $N_\alpha$ ( $D_\alpha$) the action of $g^\alpha$ on D9 (D5 or $\overline{\text{D5}}$) branes. Furthermore, since the $\mathbb{Z}_6$ orbifold has only one fixed point, we introduce a single stack of D5 or $\overline{\text{D5}}$ branes on it. Consequently, the massless contribution to the transverse channel of the annulus amplitude reads 
\begin{equation}
\begin{split}\label{eq:trannulusZ6}
\tilde{\mathcal A}^{(\sigma)} = & \frac{2^{-5}}{6}  \Bigg \{ V_4 \left( N_0 \sqrt{v} + \frac{D_0}{\sqrt{v}}\right)^2 \, - 2 \, S_4 \left( N_0 \sqrt{v} + (-1)^\sigma \frac{D_0}{\sqrt{v}}\right)^2 \, 
\\
&+ 4\, O_4  \left( N_0 \sqrt{v} - \frac{D_0}{\sqrt{v}}\right)^2 \, - 2 \, C_4 \left( N_0 \sqrt{v} -(-1)^\sigma \frac{D_0}{\sqrt{v}}\right)^2 \, 
\\
&+ 2^2 \, \sum_{\alpha=1,5} \left[ 2 \, O_4 (N_\alpha - D_\alpha)^2  - C_4 (N_\alpha - (-1)^\sigma D_\alpha )^2 \right]
\\
&+ \frac{4}{3} \sum_{\alpha =2,4} \left[ (2 \, O_4 - C_4  ) \, 8\, N_\alpha^2 \,  + 2 \, O_4 (N_\alpha - 3 \, D_\alpha)^2 - C_4 (N_\alpha -(-1)^\sigma 3\, D_\alpha)^2  \right] 
\\
& + \,\left[ (2 \, O_4 - C_4)  15\,  N_3^2 \,  + 2 \, O_4 (N_3 - 4\, D_3 )^2 \,  - C_4 (N_3 -(-1)^\sigma 4\, D_3 )^2 \, \right] \Bigg \} \,, 
\end{split}
\end{equation}
while the massless terms in the transverse channel of M\"obius strip amplitude can be computed as geometric means of $\tilde{\mathcal K}$ and $\tilde{\mathcal A}$, as required by compatibility
\begin{equation}
\begin{split} \label{eq:trmoebiusZ6}
\tilde{\mathcal M}^{(\sigma)} = - \frac{1}{3} & \Bigg \{ \hat{V}_4 \left( \sqrt{v} +(-1)^\sigma \frac{1}{\sqrt{v}} \right) \left( N_0 \sqrt{v} + \frac{D_0}{\sqrt{v}}\right) 
\\
&+ 4 \, \hat{O}_4\left( \sqrt{v} -(-1)^\sigma \frac{1}{\sqrt{v}} \right)\left( N_0 \sqrt{v} - \frac{D_0}{\sqrt{v}}\right) 
\\
&- 2 \, \hat{S}_4 \left( \sqrt{v} +(-1)^\sigma \frac{1}{\sqrt{v}} \right) \left( N_0 \sqrt{v} +(-1)^\sigma \frac{D_0}{\sqrt{v}}\right) 
\\
& - 2 \, \hat{C}_4 \left( \sqrt{v} -(-1)^\sigma \frac{1}{\sqrt{v}} \right)\left( N_0 \sqrt{v} -(-1)^\sigma \frac{D_0}{\sqrt{v}}\right) 
\\
& - \frac{2^{-3}}{3} \sum_{\alpha=2,4} \left[ 8\,  q_{\alpha,t}\, N_\alpha \, (2 \, \hat{O}_4 - \hat{C}_4  ) \right.
\\
& \qquad \qquad \quad + 2 \hat{O}_4 \, (N_\alpha - 3\, D_\alpha) (1-(-1)^\sigma 3 ) q_{\alpha,t}  
\\
&\left. \qquad \qquad \quad - \hat{C}_4 \, (N_\alpha - (-1)^\sigma 3\, D_\alpha) (1-(-1)^\sigma 3) q_{\alpha,t}\right] \,.
\end{split}
\end{equation}
Given all the amplitudes \eqref{eq:trkleinZ6}, \eqref{eq:trannulusZ6} and \eqref{eq:trmoebiusZ6}, one can straightforwardly deduce the R-R tadpole conditions which then read
\begin{equation}
N_0 = 32\,, \qquad D_0 = 32\,,
\end{equation}
from the untwisted sector, 
\begin{equation}
N_{1,5} -(-1)^\sigma D_{1,5} = 0\,,
\end{equation}
from the $g$ and $g^5$ twisted sectors,
\begin{equation}
N_{2} =-(-1)^\sigma N_4 =- (-1)^\sigma  8\,, \qquad  N_{2,4} -(-1)^\sigma 3 D_{2,4} = \pm(3-(-1)^\sigma ) \, 8\,,  
\end{equation}
from the $g^2$ and $g^4$ twisted sectors, and, finally,
\begin{equation}
N_3 -(-1)^\sigma 4 D_3 =0\,, \qquad N_3 =0\,,
\end{equation}
from the $g^3$ twisted sector.

\noindent Note that for $\sigma=0$ bosonic and fermionic degrees of freedom in the eqs. \eqref{eq:trannulusZ6} and \eqref{eq:trmoebiusZ6} have the same tadpoles, consistently with the presence of supersymmetry in this vacuum, while they are different for the choice $\sigma=1$, when supersymmetry is broken. This means that, for the supersymmetric case, satisfying R-R tadpoles necessarily implies the cancellation of the NS-NS ones as well. However, in the BSB case, one is left with an uncancelled tension that contributes to the scalar potential. In general, we will choose to satisfy the twisted R-R tadpole conditions via a mutual cancellation between the twisted charges of D9 and $\overline{\text{D5}}$ branes, which schematically implies 
\begin{equation}
V (\phi , \xi ) = (D_0 + 32)\, e^{-\phi} + e^{-\phi} \left[ (N_1-D_1) \, (\xi_1 + \xi_5 ) + (N_2-3 \, D_2)  \, (\xi_2 + \xi_4 ) + O (\xi^2 )\right]\,,
\label{Z6DilPot}
\end{equation}
where the field $\phi$ corresponds to the ten-dimensional dilaton while $\xi_\alpha$ correspond to massless scalars localised on the $\mathbb{Z}_6$ and $\mathbb{Z}_3$ fixed points. Such scalars are massless and they could play a non-trivial role in the dynamics since a non-vanishing vev would imply a spontaneous blown-up of the orbifold singularity. 

\noindent In eq. \eqref{Z6DilPot}, the first and last terms are always present since they are fixed by the solution to the corresponding R-R tadpoles, $D_0+32=64$ and $N_2-3 \, D_2=-16$, while the second term could be present or not according to our choice to cancel the $g$($g^5$)-twisted R-R tadpoles.

\subsection{The \texorpdfstring{$T^4/\mathbb{Z}_6$}{T4/Z6} BSB orientifold}
\label{SSec:BSBZ6}

In this case $\sigma=1$ and the involution acts non-trivially, calling for the presence of $\overline{\text{D5}}$ branes. Furthermore, the choice of $\sigma$ also affects the structure of the Chan-Paton labels. In particular, in our setting the Chan-Paton factors are real and one can decompose $N_\alpha$ and $D_\alpha$ into labels that are eigenvalues of the orbifold action
\begin{equation}
N_\beta = \sum_{\gamma=0}^{5} e^{2 \pi i \beta \gamma/6}\, n_\gamma\,,
\qquad
D_{(k)\,,\, \beta} = \sum_{\gamma=0}^{5} e^{2 \pi i \beta \gamma/6}\, d_{(k)\,,\,\gamma}\,.
\label{CPparamb}
\end{equation}
The solution to the tadpole conditions is not unique and reads
\begin{equation}
\begin{split}
& n_0 = 8+2a\,, \quad n_1 = n_5 = 4+a\,, \quad n_2 = n_4 = 4-a\,, \quad n_3 =8-2a\,,
\\
& d_0 = 8-2a\,, \quad d_1 = d_5 = 4-a\,, \quad d_2 = d_4 = 4+a\,, \quad d_3 =8+2a\,,
\end{split}
\end{equation}
with $a=0,\ldots, 4$. This parameter controls the way the twisted charges are cancelled between D9 and $\overline{\text{D5}}$ branes.
Performing $S$ and $P$ modular transformations on \eqref{eq:trannulusZ6} and $\eqref{eq:trmoebiusZ6}$, we can extract the direct channel amplitudes
\begin{equation}
\begin{aligned}
{\mathcal A}_{99} =& (n_0^2 + 2 n_1 n_5 + 2 n_2 n_4 + n_3^2 ) ( V_4 - 2 \, S_4 ) 
\\
&+ 2 (n_0 n_5 + n_1 n_4 + n_2 n_3 ) (2 \, O_4 - C_4 ) 
\\
&+ 2  (n_0 n_1 + n_5 n_2 + n_4 n_3 ) (2 \, O_4 - C_4 )\,,
\end{aligned}
\end{equation}
and
\begin{equation}
{\mathcal M}_9 = - (n_0 + n_3 ) ( \hat{V}_4 - 2 \, \hat{S}_4 )\,,
\end{equation}
for open strings stretched between D9 branes,
\begin{equation}
\begin{aligned}
{\mathcal A}_{\bar 5 \bar 5} =& (d_0^2 + 2 d_1 d_5 + 2 d_2 d_4 + d_3^2 ) ( V_4 - 2 \, S_4 )  
\\
&+ 2 (d_0 d_5 + d_1 d_4 + d_2 d_3 ) (2 \, O_4 - C_4 )  
\\
&+ 2  (d_0 d_1 + d_5 d_2 + d_4 d_3 ) (2 \, O_4 - C_4 ) \,,
\end{aligned}
\end{equation}
and
\begin{equation}
{\mathcal M}_{\bar 5} = (d_0 + d_3 ) ( \hat{V}_4 + 2 \, \hat{S}_4)\,,
\end{equation}
for open strings stretched between $\overline{\text{D}5}$ branes, and
\begin{equation}
\begin{split}
{\mathcal A}_{9\bar 5} =& 2 (n_0 d_0 + n_1 d_5 + n_2 d_4 + n_3 d_3 + n_4 d_2 + n_5 d_1 )(- S_4)
\\
&+ 2 (n_0 d_5 + n_1 d_4 + n_2 d_3 + n_3 d_2 + n_4 d_1 + n_5 d_0) O_4
\\
&+2 (n_0 d_1 + n_5 d_2 + n_4 d_3 + n_3 d_4 + n_2 d_5 + n_1 d_0) O_4 \,,
\end{split}
\end{equation}
for open strings stretched between D9s and $\overline{\text{D}5}$s. Supersymmetry is thus broken on the $\overline{\text{D}5}$ branes, since the gauge bosons and would-be gauginos come in different representations in $({\mathcal A}_{\bar 5\bar 5} + {\mathcal M}_{\bar 5})/2$, while in ${\mathcal A}_{9\bar 5}$ a different GSO projection determines the presence of a left-handed symplectic Majorana-Weyl fermion and two reals scalars in different representations of the gauge group. The Chan-Paton gauge group is then
\begin{equation}
\begin{aligned}
G_\text{CP} =& \text{SO} (8+2a ) \times \text{U} (4+a) \times \text{U} (4-a) \times \text{SO} (8-2a ) \Big|_{\text{D}9} 
\\
&\times 
\text{USp} (8-2a ) \times \text{U} (4-a) \times \text{U} (4+a) \times \text{USp} (8+2a ) \Big|_{\overline{\text{D}5}}\,.
\end{aligned}
\label{CPZ6}
\end{equation}
The charged matter can be readily extracted from the previous partition functions, and comprises vectors in the adjoint representation of $G_\text{CP}$ together with a left-handed Majorana-Weyl fermions in the representations
\begin{equation}
\smalltableau{  \null \\ \null \\ }_{9_1} + \left( \smalltableau{  \null \\}\times \overline{\smalltableau{  \null \\}} \right)_{9_2} +  \left(\smalltableau{  \null \\}\times \overline{\smalltableau{  \null \\}} \right)_{9_3} +
\smalltableau{  \null \\ \null \\ }_{9_4}  +\smalltableau{  \null \\ \null \\ }_{\bar 5_1} + \left(\smalltableau{  \null \\}\times \overline{\smalltableau{  \null \\}} \right)_{\bar 5_2} +  \left( \smalltableau{  \null \\}\times \overline{\smalltableau{  \null \\}} \right)_{\bar 5_3} + \smalltableau{  \null \\ \null \\ }_{\bar 5_4} \,,
\end{equation}
four scalars and a right-handed Majorana-Weyl fermion ({\emph{i.e.} a full hyper multiplet) in the representations
\begin{equation}
(\smalltableau{  \null \\ }_{9_1} , \overline{ \smalltableau{  \null \\ }}_{9_2}) +
(\smalltableau{  \null \\ }_{9_2} , \overline{ \smalltableau{  \null \\ }}_{9_3}) +
(\smalltableau{  \null \\ }_{9_3} ,  \smalltableau{  \null \\ }_{9_4})  +
(\smalltableau{  \null \\ }_{\bar 5_1} , \overline{ \smalltableau{  \null \\ }}_{\bar 5_2})+
(\smalltableau{  \null \\ }_{\bar 5_2} , \overline{ \smalltableau{  \null \\ }}_{\bar 5_3})+
(\smalltableau{  \null \\ }_{\bar 5_3} ,  \smalltableau{  \null \\ }_{\bar 5_4})
\,,
\end{equation}
a left-handed symplectic Majorana-Weyl fermion in the representations 
\begin{equation}
(\smalltableau{  \null \\ }_{9_1} ,  \smalltableau{  \null \\ }_{\bar 5_1}) +
(\smalltableau{  \null \\ }_{9_2} , \overline{ \smalltableau{  \null \\ }}_{\bar 5_2}) +
(\smalltableau{  \null \\ }_{9_3} , \overline{ \smalltableau{  \null \\ }}_{\bar 5_3}) +
(\smalltableau{  \null \\ }_{9_4} , \smalltableau{  \null \\ }_{\bar 5_4}) +
(\overline{\smalltableau{  \null \\ }}_{9_2} ,  \smalltableau{  \null \\ }_{\bar 5_2}) +
(\overline{\smalltableau{  \null \\ }}_{9_3} ,  \smalltableau{  \null \\ }_{\bar 5_3})\,,
\end{equation}
and two scalars in the representations
\begin{equation}
(\smalltableau{  \null \\ }_{9_1} ,  \overline{\smalltableau{  \null \\ }}_{\bar 5_2}) +
(\smalltableau{  \null \\ }_{9_2} , \overline{ \smalltableau{  \null \\ }}_{\bar 5_3}) +
(\smalltableau{  \null \\ }_{9_3} ,  \smalltableau{  \null \\ }_{\bar 5_4}) +
(\smalltableau{  \null \\ }_{9_4} , \smalltableau{  \null \\ }_{\bar 5_3}) +
(\overline{\smalltableau{  \null \\ }}_{9_2} ,  \smalltableau{  \null \\ }_{\bar 5_1}) +
(\overline{\smalltableau{  \null \\ }}_{9_3} ,  \smalltableau{  \null \\ }_{\bar 5_2}) 
 \,.
\end{equation}
To lighten the notation, the representation $\smalltableau{  \null \\ }_{p_i}$ refers to the fundamental of the $i$-th factor in the gauge group on the $\text{D}p$ (anti-)brane, as ordered in \eqref{CPZ6}, and similarly for the antisymmetric representations. 

\noindent For $a=4$, this vacuum configuration is rigid since branes cannot be recombined further and moved into the bulk. Indeed, although scalars in bi-fundamental representations are always present, moving branes into the bulk would lead to an anomalous model, since tadpoles cannot be satisfied anymore. A more detailed analysis would require the computation of the $F$ and $D$ terms, which however lies beyond the scope of this Thesis. However, other solutions allow a deformation to take place. For instance, one can consider the example where D9 and $\overline{\text{D}5}$ branes are recombined to 
\begin{equation}
G_\text{CP} = \text{SO} (4) \Big|_{\text{D}9\ \text{bulk}} \times \text{USp} (4) \Big|_{\overline{\text{D}5}\ \text{bulk}} \times \text{SO} (4)^2 \Big|_{\text{D}9}\times \text{USp} (4)^2 \Big|_{\overline{\text{D}5}}\,,
\end{equation}
with left-handed Majorana-Weyl fermions in the antisymmetric representation of all group factors, four scalars in 
\begin{equation}
    (\boldsymbol{10}, \boldsymbol{1}; \boldsymbol{1}, \boldsymbol{1}; \boldsymbol{1},\boldsymbol{1}) + (\boldsymbol{1}, \boldsymbol{6}; \boldsymbol{1}, \boldsymbol{1}; \boldsymbol{1},\boldsymbol{1}) \, ,
\end{equation} 
a right-handed Majorana-Weyl fermion in 
\begin{equation}
    (\boldsymbol{10}, \boldsymbol{1}; \boldsymbol{1}, \boldsymbol{1}; \boldsymbol{1},\boldsymbol{1}) + (\boldsymbol{1}, \boldsymbol{10}; \boldsymbol{1}, \boldsymbol{1}; \boldsymbol{1},\boldsymbol{1}) \, ,
\end{equation}
a left-handed symplectic Majorana-Weyl fermion in 
\begin{equation}
    (\boldsymbol{1}, \boldsymbol{1}; \boldsymbol{4}, \boldsymbol{1}; \boldsymbol{4},\boldsymbol{1}) + (\boldsymbol{1}, \boldsymbol{1}; \boldsymbol{1}, \boldsymbol{4}; \boldsymbol{1},\boldsymbol{4}) \, ,
\end{equation}
two scalars and a left-handed symplectic Majorana-Weyl fermion in 
\begin{equation}
\begin{aligned}
    &
    (\boldsymbol{4}, \boldsymbol{1}; \boldsymbol{1}, \boldsymbol{1}; \boldsymbol{4},\boldsymbol{1})
+ (\boldsymbol{4}, \boldsymbol{1}; \boldsymbol{1}, \boldsymbol{1}; \boldsymbol{1},\boldsymbol{4})+ (\boldsymbol{1}, \boldsymbol{4}; \boldsymbol{4}, \boldsymbol{1}; \boldsymbol{1},\boldsymbol{1})
+ (\boldsymbol{1}, \boldsymbol{4}; \boldsymbol{1}, \boldsymbol{4}; \boldsymbol{1},\boldsymbol{1}) \, ,
\end{aligned}
\end{equation} 
and in six copies of 
\begin{equation}
\begin{aligned}
    & (\boldsymbol{4}, \boldsymbol{4}; \boldsymbol{1}, \boldsymbol{1}; \boldsymbol{1},\boldsymbol{1}) \, .
\end{aligned}
\end{equation} 
The branes on the fixed points cannot be further deformed, unless the scalars associated with strings stretching between the bulk branes and those on the fixed points are turned on. This would imply a recombination of the various branes, although one cannot bring everything in the bulk, since this would leave the $g^2$ and $g^4$ twisted R-R charge of the O-planes un-matched. As before, the analysis of the allowed higgsing requires a detailed knowledge of the $F$ and $D$ terms and of the allowed magnetisations of this model, which are beyond the scope of this study.

\subsection{The \texorpdfstring{$T^4/\mathbb{Z}_6$}{T4/Z6} supersymmetric orientifold} \label{SSec:SUSYZ6}

The choice of $\sigma=0$ implies instead a "trivial" action on the twisted two-cycles which induces O9$_-$ and O5$_-$ planes requiring D9 and D5 branes to cancel anomalies. The action of the orbifold group on the Chan-Paton labels is summarised in
\begin{equation}
N_\beta = e^{ \pi i \beta /6} \sum_{\gamma=0}^{5} e^{2 \pi i \beta \gamma/6}\, n_\gamma\,,
\qquad
D_{\, \beta} = e^{ \pi i \beta /6} \sum_{\gamma=0}^{5} e^{2 \pi i \beta \gamma/6}\, d_{\,\gamma}\,,
\label{CPparam}
\end{equation}
The extra phase implies that the Chan-Paton labels are all complex, leading to unitary gauge groups. The general solution of the R-R tadpoles is 
\begin{equation}
\begin{aligned}
    &n_0=n_5= 4+a \, , \quad n_1=n_4= 8 \, , \quad n_2=n_3= 4-a \, ,
    \\
     &d_0=d_5= 4+a \, , \quad d_1=d_4= 8 \, , \quad d_2=d_3= 4-a\, ,
\end{aligned}
\end{equation}
with gauge group 
\begin{equation} \label{eq:CPSUSYZ6}
    G_{\text{CP}}= \text{U}(4+a) \times \text{U}(8) \times \text{U}(4-a) \Big|_{\text{D9}} \times \text{U}(4+a) \times \text{U}(8) \times \text{U}(4-a) \Big|_{\text{D5}} \, .
\end{equation}
The parameter $a=0,\ldots, 4$ depends on how the twisted R-R charges are cancelled between D9 and D5 branes.
The contribution of the massless states to the open-string amplitudes reads 
\begin{equation}
\begin{aligned} \label{eq:annulusSUSYZ6D9D9}
    \mathcal{A}_{99}= &(2 n_0 \,  n_5 + 2 n_1 \, n_4 + 2 n_2 \, n_3) ( V_4 - 2 \, S_4 ) 
    \\
    &+  ( 2 n_0 \, n_4+ 2 n_1 \, n_3 + n_2^2 + n_5^2) (2 \, O_4 - C_4 )
    \\
    &+ (n_0^2 +  2 n_1 \, n_5 + 2 n_2 \, n_4 + n_3^2 ) (2 \, O_4 - C_4 ) 
    \end{aligned}
\end{equation}
and
\begin{equation} \label{eq:moebiusSUSYZ6D9}
    \mathcal{M}_{9}=  - (  n_2 + n_5) (2 \, \hat{O}_4 - \hat{C}_4 )- (n_0 + n_3) (2 \, \hat{O}_4 - \hat{C}_4 )
\end{equation}
for open strings stretched between D9 branes, 
\begin{equation} \label{eq:annulusSUSYZ6D5D5}
\begin{aligned}
    \mathcal{A}_{5 5}= &(2 d_0 \,  d_5 + 2 d_1 \, d_4 + 2 d_2 \, d_3) ( V_4 - 2 \, S_4 )  
    \\
    &+ ( 2 d_0 \, d_4+ 2 d_1 \, d_3 + d_2^2 + d_5^2)  (2 \, O_4 - C_4 )
    \\
    &+ (d_0^2 +  2 d_1 \, d_5 + 2 d_2 \, d_4 + d_3^2 ) (2 \, O_4 - C_4 ) 
\end{aligned}
\end{equation}
and
\begin{equation}\label{eq:moebiussSUSYZ6D5}
    \mathcal{M}_{5}= - (  d_2 + d_5) (2 \, \hat{O}_4 - \hat{C}_4 ) - (d_0 + d_3) (2 \, \hat{O}_4 - \hat{C}_4 )  
\end{equation}
for open strings stretched between D5 branes, and 
\begin{equation}\label{eq:annulusSUSYZ6D9D5}
\begin{aligned}
    \mathcal{A}_{95}= & (n_0 \, d_{5}  + n_1 \, d_{4} + n_2 \, d_{3} + n_3 \, d_{2} + n_4 \, d_1 + n_5 \, d_0) (2 \, O_4 - C_4 )
\end{aligned}
\end{equation}
for open strings between D9 and D5 branes. The light spectrum is supersymmetric and comprises a vector multiplet in the adjoint representation of the gauge group \eqref{eq:CPSUSYZ6}, one hyper multiplet in the representations
\begin{equation}
\begin{aligned}
    &( \smalltableau{ \null \\ \null \\}, 1, 1;1, 1, 1)  + ( \overline{\smalltableau{ \null \\}}, \smalltableau{ \null \\},1; 1, 1, 1)  + (1,  \overline{\smalltableau{ \null \\}},\smalltableau{ \null \\}; 1, 1, 1) + ( 1, 1, \overline{\smalltableau{ \null \\ \null \\}}; 1, 1, 1) \, ,
        \\
 &( 1, 1, 1;\smalltableau{ \null \\ \null \\}, 1, 1)  + ( 1, 1, 1; \overline{\smalltableau{ \null \\}}, \smalltableau{ \null \\},1)  + (1, 1, 1; 1,  \overline{\smalltableau{ \null \\}},\smalltableau{ \null \\}) + ( 1, 1, 1; 1, 1, \overline{\smalltableau{ \null \\ \null \\}}) \, ,
 \\
&(\smalltableau{ \null \\},1,1; \overline{\smalltableau{ \null \\}},1 , 1) + (1,\smalltableau{ \null \\},1;1, \overline{\smalltableau{ \null \\}}, 1) + (1,1,\smalltableau{ \null \\};1,1, \overline{\smalltableau{ \null \\}})  \, ,   
\end{aligned}
\end{equation}
from open strings stretched between D9/D9, D5/D5 and D9/D5 branes, respectively.
In this supersymmetric vacuum, the tadpole conditions are weaker than in its BSB counterpart, since a recombination of branes is always possible, even with the minimal choice $a=4$. Indeed, in such a case, one can move branes in the bulk yielding the gauge group
\begin{equation}\label{eq:CPSUSY6bulk}
G_\text{CP} = \text{USp} (4) \Big|_{\text{D}9\ \text{bulk}} \times \text{USp} (4) \Big|_{\text{D}5\ \text{bulk}} \times \text{U} (4) \Big|_{\text{D}9}\times \text{U} (4) \Big|_{\text{D}5}\,,
\end{equation}
whose light spectrum can be deduced by setting $n_0=n_2=0$ in the amplitudes of eqs.\eqref{eq:annulusSUSYZ6D9D9}, \eqref{eq:moebiusSUSYZ6D9}, \eqref{eq:annulusSUSYZ6D5D5},   \eqref{eq:moebiussSUSYZ6D5}, \eqref{eq:annulusSUSYZ6D9D5} and by adding the contribution from the bulk branes. All in all, the spectrum comprises a vector multiplet in the adjoint of \eqref{eq:CPSUSY6bulk}, an hyper multiplet in the representations
\begin{equation}
\begin{aligned}
&(\boldsymbol{6}, \boldsymbol{1}; \boldsymbol{1}; \boldsymbol{1}) + (\boldsymbol{1}, \boldsymbol{6};  \boldsymbol{1}; \boldsymbol{1})  +  (\boldsymbol{1}, \boldsymbol{1}; \boldsymbol{4} ; \boldsymbol{\bar{4}}) + (\boldsymbol{4}, \boldsymbol{1}; \boldsymbol{1}; \boldsymbol{4}) + (\boldsymbol{1}, \boldsymbol{4}; \boldsymbol{4} ; \boldsymbol{1})  \, ,
\end{aligned}
\end{equation}
and in three copies of 
\begin{equation}
    (\boldsymbol{4}, \boldsymbol{4}; \boldsymbol{1} ; \boldsymbol{1}) \, .
\end{equation}

\section{Anomaly polynomials and the K\"ahler \texorpdfstring{$J$}{J} form}
\label{Sec:Anomaly}

Tadpoles in orientifold constructions are deeply tied to the structure of the low-energy effective action. In particular, the R-R tadpoles determine the charge through which D-branes and O-planes are coupled to R-R fixed forms, including the topological Chern-Simons terms entering the Green-Schwarz-Sagnotti mechanism \cite{Green:1984bx,Green:1984sg, Sagnotti:1992qw}. All these terms are captured in the expansion of the Wess-Zumino coupling for D-branes  \cite{Douglas:1995bn, Green:1996dd} and O-planes \cite{Morales:1998ux, Stefanski:1998he}
\begin{equation} \label{eq:DpactionWZ}
    S_{\text{Dp}}  = - \int_{M_p} \text{tr} \, C \wedge e^{i F_{p}} \wedge \sqrt{ \hat{A}(R)} \, ,
\end{equation}
and 
\begin{equation} \label{eq:OpactionWZ}
    S_{\text{Op}_\pm} = - \pm 2^{p-4} \, \int_{M_p} C \wedge \sqrt{ \hat{L} \left (\frac{R}{4} \right )} \, ,
\end{equation}
where $C= \sum_p C_{(p)}$ collects the various $p$-form fields, $R$ and $F_{p}$ describe the curvature of the tangent and D$p$-brane gauge bundle, $\hat{A}(R)$ corresponds to the A-roof genus and $\hat{L} \left (R \right )$ the Hirzebruch polynomial. From eqs. \eqref{eq:DpactionWZ} and \eqref{eq:OpactionWZ}, one can directly read the R-R tadpoles from the coefficients multiplying the non-dynamical $6$-forms and the topological coupling dictated by the Chern-Simons term   
\begin{equation}
S_\text{GS} = \int \Omega_{\alpha \beta} \, C_2^\alpha \wedge X_4^\beta \, ,
\label{Green-Schwarz}
\end{equation} 
where  $\Omega_{\alpha\beta}$ is an $\text{SO} (1,n_\text{T})$ invariant metric with mostly minus signature, and $n_\text{T}$ is the number of tensor multiplets. The term $X_4^\alpha$ is a polynomial in the curvatures of the tangent and gauge bundles  
\begin{equation}
     X_4^\alpha= \frac12 \Big ( a^\alpha \, \text{tr} \, R^2 + \sum_i \frac{b_i^\alpha}{\lambda_i}\, \text{tr} \, F_i^2 \Big ) \, ,
    \label{factor-2}
\end{equation}
where the 't Hooft coefficients $a^\alpha$ and $b_i^\alpha$ depend on the charges of D$p$ branes, $q^{\text{Dp}}$, and O$p$ planes, $q^{\text{Op}}$, 
\begin{equation} \label{eq:DpOpcharges}
b_i^\alpha = \frac{\ell_\alpha}{2 \sqrt{N}} \, \lambda_i \, q^{D_i}_\alpha\,, \qquad 
a^\alpha = \frac{\ell_\alpha}{2 \sqrt{N}} \, \sum_{p} q^{\text{O}p}_\alpha \,,
\end{equation}
as follows from eqs. \eqref{eq:DpactionWZ} and \eqref{eq:OpactionWZ}. 
In the  expression above, $\lambda$ is the group theoretic factor tied to the highest root and $\ell_\alpha$ is a geometric factor given by
\begin{equation}
\begin{aligned}
\ell &= \left(1,1; \frac{1}{\sqrt{2} \, \sin (\pi/2)} \boldsymbol{1}_{16 \sigma} \right)\,, & \text{for}\quad N=2\,,
\\
\ell &= \left(1,1; \frac{1}{\sin (2 \pi /6 )} \, \boldsymbol{1}_{9} \right)\,, & \text{for}\quad N=3\,,
\\
\ell &= \left(1,1; \frac{1}{\sin (\pi /4 )} \, \boldsymbol{1}_{4} ; \frac{1}{\sqrt{2} \, \sin (2\pi/4)} \boldsymbol{1}_{10 \sigma} \right)\,, & \text{for}\quad N=4\,,
\\
\ell &= \left(1,1; \frac{1}{\sin ( \pi /6 )} ; \frac{1}{\sin (2\pi/6)} \, \boldsymbol{1}_{5} ; \frac{1}{\sqrt{2} \, \sin (3\pi/6)} \boldsymbol{1}_{6 \sigma} \right)\,, & \text{for}\quad N=6\,,
\end{aligned}
\end{equation}
with the coefficients $\sin (a\pi/N)$ clearly related to the number of fixed points in the various twisted sectors, and $\boldsymbol{1}_d$ a 
$d$-dimensional vector whose entries are all equal to one. For the BSB $N=6$ orientifold the charges are given by
\begin{equation}
\begin{split}
q^{\text{O}9_-} &= ( -4,-4 ; 0; - \boldsymbol{1}_5 ; \boldsymbol{0}_{6} )\,, \\
q^{\text{D}9_1} &= (1,1; 1; \boldsymbol{1}_5 ; \boldsymbol{1}_{6} )\,, \\
q^{\text{D}9_2} &= (2,2; 1 ; - \boldsymbol{1}_5 ; -2\, \boldsymbol{1}_{6} )\,, \\ 
q^{\text{D}9_3} &= (2,2; -1 ;  - \boldsymbol{1}_5 ;  2\, \boldsymbol{1}_{6} )\,, \\
q^{\text{D}9_4} &= (1,1; -1;  \boldsymbol{1}_5 ; - \boldsymbol{1}_{6} ) \,,
\end{split}
\qquad
\begin{split}
q^{\text{O}5_+^i} &= ( \tfrac{1}{4},-\tfrac{1}{4} ; 0 ; -3\, \boldsymbol{\delta}_5^1 ; \boldsymbol{0}_{6}  ) \,,  \\
q^{\overline{\text{D}5}_{1}} &= (-1,1;1;  3\, \boldsymbol{\delta}_5^1 ; 4\, \boldsymbol{\delta}^1_{6} )\,, \\ 
q^{\overline{\text{D}5}_{2}} &= (-2,2; 1; - 3\, \boldsymbol{\delta}_5^1 ; -8\, \boldsymbol{\delta}^1_{6} ) \,, \\ 
q^{\overline{\text{D}5}_{3}} &= (-2,2;-1;  -3\, \boldsymbol{\delta}_5^1 ; 8\, \boldsymbol{\delta}^1_{6} )\,, \\ 
q^{\overline{\text{D}5}_{4}} &= (-1,1; -1; 3 \, \boldsymbol{\delta}_5^1 ; - 4\, \boldsymbol{\delta}^1_{6} ) \,,
\end{split} \label{eq:DCVBSBZ6}
\end{equation}
while for the supersymmetric $N=6$ orientifold the charges are
\begin{equation}
\begin{split}
q^{\text{O}9_-} &= \left( -4,-4 ; 0;  \boldsymbol{1}_5 \right)\,, \\
q^{\text{D}9_1} &= (2,2; \sqrt{3}; \boldsymbol{1}_5  )\,, \\
q^{\text{D}9_2} &= (2,2; 0 ; - 2\boldsymbol{1}_5  )\,, \\ 
q^{\text{D}9_3} &= (2,2; - \sqrt{3} ;   \boldsymbol{1}_5 )\,, 
\end{split}
\qquad
\begin{split}
 q^{\text{O}5_-^i} &= \left( -\tfrac{1}{4},-\tfrac{1}{4} ; 0 ; -3\, \boldsymbol{\delta}_5^1  \right) \,, \\
q^{\text{D}5_{1}} &= (2,-2;- \sqrt{3};  3\, \boldsymbol{\delta}_5^1 )\,, \\ 
q^{\text{D}5_{2}} &= (2,-2; 0; - 6\, \boldsymbol{\delta}_5^1  ) \,, \\ 
q^{\text{D}5_{3}} &= (2,-2; \sqrt{3};  3\, \boldsymbol{\delta}_5^1  )\,. \\ 
\end{split} \label{eq:DCVSUSYZ6}
\end{equation}
Here $\boldsymbol{0}_d$ denotes a $d$-dimensional null vector and $\boldsymbol{\delta}_d^i$ is a $d$-dimensional vector whose only non-vanishing component is $\{\boldsymbol{\delta}^i_d\}_j = \delta^i_j$.

\noindent In eq. \eqref{Green-Schwarz}, we have chosen to write the topological coupling using a general basis, although from the expansion of the terms \eqref{eq:DpactionWZ} and \eqref{eq:OpactionWZ} it appears in diagonal form. This choice has the advantage of reflecting the geometry of D-branes and O-planes \cite{Sagnotti:1992qw,Angelantonj:2020pyr} but obscures some features reflecting other properties of these vacua. Indeed, it is always possible to find a basis where the self-duality of the lattice is manifest, which follows \cite{Seiberg:2011dr} from the identification of the dyonic strings as gauge instantons of charges $b_i$ for the gauge group $G_i$ \cite{Duff:1996cf} and a gravitational instanton with charge $a$. Although this requirement plays a non-trivial role in constraining the landscape of low-energy effective theories and will be discussed in great detail in the next Chapter, in what follows we shall rely on the diagonal basis for the anomaly polynomial.

\noindent The cancellation of R-R tadpoles \cite{Aldazabal:1999nu, Bianchi:2000de} implies the vanishing of all irreducible anomalies and allows to factorise the anomaly polynomial as   
\begin{equation}
I_8 = \tfrac12 \Omega_{\alpha \beta} X_4^\alpha \wedge X_4^\beta \, ,
\label{factor-1}
\end{equation} 
which is then cancelled by the contribution from \eqref{Green-Schwarz}, via the celebrated Green-Schwarz-Sagnotti mechanism \cite{Green:1984bx,Green:1984sg, Sagnotti:1992qw}. In fact, requiring a non-trivial transformation of the $2$-forms
\begin{equation}
    \delta C_2^\alpha= {\omega_{2}^{(1)}}^\alpha \, ,
\end{equation}
provides the counter-terms that cancel the reducible anomaly, where ${\omega_{2}^{(1)}}^\alpha$ is tied to the Chern-Simons form $\omega_{3}^\alpha$ and $X_4^\alpha$ via the descent equations
\begin{equation}
    \begin{aligned}\label{eq:Bianchi}
       & \delta \omega_3^\alpha= d \, {\omega_{2}^{(1)}}^\alpha 
       \\
    &d  H^\alpha = X_4^\alpha \, ,
    \end{aligned}
\end{equation}
with $H^\alpha= d \, C_2^\alpha - \omega_3^\alpha$.
In our case, the anomaly polynomial for the BSB $\mathbb{Z}_6$ orientifold is given by
\begin{equation}
\begin{split}\label{eq:bsb6pol8}
    I_8 &=\tfrac{1}{192}  \left (  \text{tr} F_{9,1}^2 + 2\text{tr} F_{9,2}^2 + 2 \text{tr} F_{9,3}^2 + \text{tr} F_{9,4}^2 \right. 
    \\
    & \left. \qquad \qquad \qquad - \text{tr} F_{\bar 5,1}^2  - 2\text{tr} F_{\bar 5,2}^2 - 2 \text{tr} F_{\bar 5,3}^2 - \text{tr} F_{\bar 5,4}^2 \right )^2 
    \\
    &- \tfrac{1}{192}  \left ( -8 \text{tr} R^2 + \text{tr} F_{9,1}^2 + 2 \text{tr} F_{9,2}^2 + 2 \text{tr} F_{9,3}^2 +  \text{tr} F_{9,4}^2  \right.
    \\
    & \left.  \qquad \qquad  \qquad +  \text{tr} F_{\bar 5,1}^2 + 2 \text{tr} F_{\bar 5,2}^2 + 2 \text{tr} F_{\bar 5,3}^2 +  \text{tr} F_{\bar 5,4}^2 \right )^2  
     \\
    &  - \tfrac{1}{48}  \left (   \text{tr} F_{9,1}^2 + \text{tr} F_{9,2}^2 -  \text{tr} F_{9,3}^2 -  \text{tr} F_{9,4}^2 
    \right.
    \\
    & \left.  \qquad \qquad \qquad + \text{tr} F_{\bar 5,1}^2 + \text{tr} F_{\bar 5,2}^2 - \text{tr} F_{\bar 5,3}^2-  \text{tr} F_{\bar 5,4}^2    \right )^2
    \\
    & - \tfrac{1}{144} \left [ \left ( - 4 \text{tr} R^2 + \text{tr} F_{9,1}^2 - \text{tr} F_{9,2}^2 -  \text{tr} F_{9,3}^2 + \text{tr} F_{9,4}^2  
    \right. \right.
    \\
    & \left.  \qquad \qquad \qquad + 3 \left ( \text{tr} F_{\bar 5,1}^2 - \text{tr} F_{\bar 5,2}^2 -  \text{tr} F_{\bar 5,3}^2 +  \text{tr} F_{\bar 5,4}^2 \right ) \right )^2   
    \\
    & - \tfrac{1}{144} \left (  -\text{tr} R^2 + \text{tr} F_{9,1}^2 - \text{tr} F_{9,2}^2 - \text{tr} F_{9,3}^2 + \text{tr} F_{9,4}^2 \right )^2 
    \\
    & -\tfrac{1}{284} \left ( \text{tr} F_{9,1}^2 - 2 \text{tr} F_{9,2}^2 + 2\text{tr} F_{9,3}^2 - \text{tr} F_{9,4}^2 
    \right. 
    \\
    & -\tfrac{1}{96}  \left ( \text{tr} F_{\bar 5,1}^2 - 2 \text{tr} F_{\bar 5,2}^2 + 2\text{tr} F_{\bar 5,3}^2 -  \text{tr} F_{\bar 5,4}^2 \right )^2  
    \\
    &   -\tfrac{15}{284}  \left ( \text{tr} F_{9,1}^2 - 2 \text{tr} F_{9,2}^2 + 2\text{tr} F_{9,3}^2 - \text{tr} F_{9,4}^2 \right )^2 \, , 
 \end{split}
\end{equation}
while for the supersymmetric one it reads
\begin{equation}\label{eq:susy6pol8}
	\begin{aligned}
		I_8& =\tfrac{1}{48}  \left ( -4 \text{tr} R^2 + \text{tr} F_{9,1}^2 + \text{tr} F_{9,2}^2 + \text{tr} F_{9,3}^2 + \text{tr} F_{5,1}^2 + \text{tr} F_{5,2}^2 + \text{tr} F_{5,3}^2 \right )^2 \\
		&- \tfrac{1}{48} \left ( \text{tr} F_{9,1}^2 + \text{tr} F_{9,2}^2 + \text{tr} F_{9,3}^2 - \text{tr} F_{5,1}^2 - \text{tr} F_{5,2}^2 - \text{tr} F_{5,3}^2 \right )^2   
  \\
		&   - \tfrac{1}{16} \left (   \text{tr} F_{9,1}^2 - \text{tr} F_{9,3}^2 - \text{tr} F_{5,1}^2 + \text{tr} F_{5,3}^2  \right )^2 
  \\
		& - \tfrac{1}{144}  \left ( - 2 \text{tr} R^2 + \text{tr} F_{9,1}^2 - 2 \, \text{tr} F_{9,2}^2 +  \text{tr} F_{9,3}^2 - 3 \left ( \text{tr} F_{5,1}^2 -2 \, \text{tr} F_{5,2}^2 +  \text{tr} F_{5,3}^2 \right ) \right )^2  
  \\
		& - \tfrac{4}{72} \left ( \text{tr} R^2 + \text{tr} F_{9,1}^2 - 2 \,  \text{tr} F_{9,2}^2 + \text{tr} F_{9,3}^2  \right )^2  \, .
	\end{aligned}
\end{equation}
In both expressions, we notice that the presence of fractional orientifold planes is reflected in the appearance of the first Pontryagin class (roughly $\text{tr} R^2$) in the last two lines of eq \eqref{eq:susy6pol8} for the supersymmetric case and in the seventh and ninth lines of eq. \eqref{eq:bsb6pol8}, which are associated with the coupling to the twisted R-R forms.

\noindent All in all, there is a deep connection between the R-R tadpoles, encoding the charges of O-planes and D-branes, and the  Wess-Zumino coupling in the low-energy effective action, which is encoded in the expression \eqref{eq:DpOpcharges} for the 't Hooft coefficients. For vacua with $\mathcal{N}=(1,0)$ supersymmetry, these in turn also determine the kinetic term for the gauge fields \cite{Sagnotti:1992qw}
\begin{equation}
\tfrac{1}{2} \, J_\alpha b_i^\alpha \, \text{tr} \, F_{i} \wedge\star  F_i  \,. \label{GKT} 
\end{equation}
The $J_\alpha$\footnote{Following \cite{Kim:2019vuc}, the symbol $J_\alpha$ is used to identify the vector $v^r$ of \cite{Romans:1986er}. $J$ plays the role of the K\"ahler form determining the geometry of the scalar manifold.} function depends on the $n_\text{T}$ scalars $\tau$ in the tensor multiplets, which parametrise the coset $\text{SO} (1 , n_\text{T})/\text{SO} (n_\text{T} )$. Positivity of the kinetic term for the scalar fields and the gauge vectors then requires \cite{Sagnotti:1992qw}
\begin{equation}
J\cdot J >0\qquad \text{and}\qquad J\cdot b_i >0 \,, \label{JKSV}
\end{equation}
where the inner product is taken with $\Omega_{\alpha \beta}$. 

\noindent Although the supersymmetry invariance of the low energy effective action requires that the gauge kinetic function be given in terms of the anomaly vector $b_i$, its origin in string theory is different. In orientifold vacua, the gauge fields originate from the open-string sector, and it should thus be possible to extract the coupling \eqref{GKT} from the tadpoles. Indeed, turning on a background magnetic field $H_i$ on the D-branes, and expanding the $J$ function as $J_\alpha \sim \tau_\alpha +  O(\tau^2)$,  eq. \eqref{GKT} becomes schematically
\begin{equation}
\tau_\alpha b_i^\alpha \, H_i^2. \label{GKTtadpole}
\end{equation}
This equation identifies the coefficients $b_i^\alpha$ with the tadpoles of the scalars in the tensor multiplets, \emph{i.e.} the one-point functions of the $\tau$ fields with the various D-branes. Now, while the anti-self-dual tensors $C_2^-$ originate from the R-R sector, their scalar super partners are of NS-NS type, and therefore the $b_i^\alpha$ in \eqref{GKT} and \eqref{GKTtadpole} are actually related to the NS-NS tadpoles, and not to the R-R ones. Clearly, supersymmetry identifies them, and this is the reason why the same anomaly vectors enter both in eqs. \eqref{factor-2} and \eqref{GKT}. 

\noindent This observation suggests that some care is needed when dealing with BSB vacua. In fact, in this class of orientifolds, R-R and NS-NS tadpoles are, in general, no longer related by supersymmetry, since for the $\overline{\text{D}5}$ branes the tension and charges are not equal. The kinetic terms for the gauge fields should then involve new coefficients $\tilde b_i^\alpha$,
\begin{equation}
\tfrac{1}{2}\, J \cdot \tilde b_i \, \text{tr}\, F_i \wedge \star F_i\,,
\end{equation}
which can be extracted from the NS-NS tadpoles and are related to the "tension" of branes $t^{D_i}$ via
\begin{equation}
\tilde{b}_i^\alpha = \frac{\ell_\alpha}{2 \sqrt{N}} \, \lambda_i \, t^{D_i}_\alpha\,.
\end{equation} 
Now $t^{D_i}= q^{D_i}$ for D-branes, but $t^{D_i}=-q^{D_i}$ for anti-branes, and thus they differ from the anomaly vectors by a sign flip in the entries associated with the gauge factors living on the antibranes. This is fully compatible with supersymmetry, since on the $\overline{\text{D}5}$ branes (or, in general, in the non-supersymmetric sector) supersymmetry is realised non-linearly, which implies that each term in the corresponding Lagrangian is invariant by itself \cite{Pradisi:2001yv}. 

\noindent Replacing the coefficients $b_i$ with $\tilde{b}_i$ in the eq. \eqref{GKT} allows to define a $J$ function also for the $T^4/\mathbb{Z}_N$ BSB orientifolds \cite{Angelantonj:2024iwi}, since the conditions 
\begin{equation}
J\cdot J > 0 \qquad \text{and}\qquad J\cdot \tilde b_i >0
\end{equation}
are now compatible. This solves the puzzle on the non-existence of the $J$ form in \cite{Angelantonj:2020pyr}, where the original condition \eqref{GKT} led to the presence of ghosts.  If, following \cite{Kim:2019vuc}, we also impose the positivity of the Gauss-Bonnet term, $J\cdot a <0$, the components of the $J$ vector must satisfy $J_0 > - J_1$ and $J_1 <0$, where, for simplicity, we have set to zero the remaining entries
\begin{equation}
J=  (J_0 , - |J_1 |; \boldsymbol{0}_4 ; \boldsymbol{0}_{10} )\,, \qquad J_0 > | J_1|\,.
\label{eq:Jform}
\end{equation}

\section{Defects and (new) unitarity constraints}
\label{Sec:defects}

In six dimensions the term described in \eqref{Green-Schwarz} arises form the coupling of the R-R $2$-forms $C_2^\alpha$ to D9 and D5 or $\overline{\text{D5}}$ branes and allows to cancel the local anomaly of the theory via an anomalous transformation of $C_2^\alpha$. However, the latter fields admit also a $\text{SO}(1,n_T)$ invariant coupling
\begin{equation}
S_{2D} = - Q^\alpha \, \Omega_{\alpha\beta} \int C_2^\beta\,,
\end{equation}
involving the presence of one-dimensional defect for this six-dimensional vacuum, which in our case consist of D1 branes localised on the internal manifolds or $\text{D}5'$ branes wrapping the entire $T^4/\mathbb{Z}_N$ space\footnote{Clearly, one can also have $\overline{\text{D}1}$, $\overline{\text{D}5}'$ and magnetised $\text{D}5'$ branes. In these cases, the discussion would follow a similar pattern.}. However, the non-trivial behaviour of $C_2^\alpha$ under local gauge and Lorentz transformations, as required by the Green-Schwarz-Sagnotti mechanism, induces an anomaly inflow on the world-volume of the defect which, in a consistent vacuum, must be cancelled by the anomalous contribution of its two-dimensional degrees of freedom \cite{Green:1996dd}
\begin{equation}
\begin{aligned}
I_4  &= Q\cdot X_4 +\tfrac{1}{2} Q\cdot Q \, \chi (N) 
\\
&= \tfrac{1}{2} \left( Q\cdot a \,\text{tr}\, R^2 +Q\cdot Q\, \chi (N) +\sum_i \frac{Q\cdot b_i}{\lambda_i} \, \text{tr} \, F_i^2 \right)= - I_\text{Inflow} \, .
\end{aligned}
\end{equation}
The pontryagin class in the above equation encodes the curvature of the tangent bundle of the whole six-dimensional space-time which can be decomposed in terms of the curvature of the tangent bundle of the $2D$ world-sheet and its $\text{SO}(4)$-normal bundle $N$ via
\begin{equation}
\text{tr}\, R^2 = -\tfrac{1}{2} p_1 (T_2 ) + c_2 (u) + c_2 (v) \,, \qquad \text{and}\quad \chi (N) = c_2 (v) - c_2 (u)\,,
\end{equation}
where we have used the decomposition $\text{SO} (1,5) = \text{SO} (1,1) \times \text{SU} (2)_u \times \text{SU} (2)_v$ and the definition of the Euler class  $\chi(N)$ of the normal bundle. Thus, the anomaly polynomial can be written as
\begin{equation}
I_4 = - \tfrac{1}{12} 3 Q\cdot a \, p_1 (T_2)  + \tfrac{1}{2} (Q\cdot Q + Q \cdot a )\, c_2 (v)  + \tfrac{1}{2} (Q\cdot Q - Q \cdot a)\, c_2 (u)  +\tfrac{1}{2} \sum_i \frac{Q\cdot b_i}{\lambda_i} \, \text{tr} F_i^2\,.
\end{equation}
When the theory on the defect flows to the IR, we can interpret the coefficient of $p_1 (T_2)$ as the difference of the central charges from the left and right movers, $c_\text{R} - c_\text{L} =  6 Q\cdot a $. In supersymmetric vacua, the correct identification of the $\text{SU}(2)_{\text{R}}$ R-symmetry would allow to completely determine the $c_\text{R}$, taking advantage of the properties of the $\mathcal{N}=(0,4)$ superconformal algebra which implies $c_\text{R}=3 k_\text{R}$, with $k_{\text{R}}$ the 't Hooft coefficient of the R-symmetry. However, such identification is typically difficult to determine. In \cite{Kim:2019vuc}, $ \text{SU}(2)_u$ is identified with the R-symmetry, but such an assumption cannot hold universally. In fact, both left and right moving fields can transform non-trivially under $\text{SU} (2)_u$, and thus both would contribute to the coefficient of $c_2 (u)$. However, according to $\mathcal{N}=(0,4)$ supersymmetry only right movers should be allowed to transform under the true R-symmetry. As shown in \cite{Angelantonj:2020pyr}, it seems that $\text{SU} (2)_u$ can be identified with the R-symmetry only if a single D1 brane is moved away from the orbifold fixed points, for which the result in \cite{Kim:2019vuc} is thus reproduced. In all other cases, $ c_2 (u)$ also receives contributions from left-moving excitations, and it is not known in literature how to get access to this piece of information purely from anomaly inflow considerations. 

\noindent For BSB orientifolds the situation is even worse. In fact, the D1 and $\text{D}5'$ defects do not, in general, enjoy ${\mathcal N} = (0,4)$ supersymmetry on their world-volume. This is due to the presence of anti-branes, which flip the chirality of the excitations of the strings stretched between the D1 or $\text{D}5'$ and the $\overline{\text{D}5}$ branes, thus explicitly breaking supersymmetry. This means that the notion of R-symmetry is lost, and thus the relation $c_{\text{R}}= 3 k_R$ does not hold.  

\noindent Nevertheless we can take advantage of the fact that when flowing to the IR fixed point, only the massless states survive, so that knowing the light spectrum of the UV degrees of freedom living on the defects allows to determine the central charges of the the left and right moving sectors of corresponding two-dimensional CFT. Such a top-down approach allows then to accumulate \emph{empirical data} on the $2D$ CFT on the D1 and $\text{D}5'$ branes, instrumental for deriving refined consistency conditions in a sought-after bottom-up approach. 

\noindent We can therefore discuss the general features of the defects in our class of vacua with D9 and D5 (anti-)branes. In this case, the D1 branes located on a $\mathbb{Z}_N$ fixed point carry the same gauge group of D9 branes \cite{Dudas:2001wd}, while the gauge group is orthogonal if they are moved in the bulk. The Chan-Paton group of $\text{D}5'$ branes is equal in structure to that of D5 or $\overline{\text{D}5}$ branes sitting on a $\mathbb{Z}_N$ fixed point, although tadpole conditions may force some of the group factors to be absent on the $\overline{\text{D5}}$ branes. Open strings stretched between D1 and D9 branes have ND boundary conditions along the eight transverse directions, and therefore the light excitations involve left-handed fermions which are singlets of the ${\mathcal N} = (0,4)$ superalgebra. Open strings stretched between D1 and D5 branes have DD boundary conditions along the internal space, so that the light excitations now comprise a full right-moving super multiplet (two right-moving scalars and Majorana-Weyl fermions) and a left-moving Majorana-Weyl fermion and a pair of left-moving chiral bosons. Open strings stretched between D1 and $\overline{\text{D}5}$ branes give rise to the same fields but with flipped chirality, which is, indeed, incompatible with ${\mathcal N} = (0,4)$ supersymmetry, as D1 and  $\overline{\text{D}5}$ branes would preserve super-charges of opposite chirality. 

\noindent For the interactions of the $\text{D}5 '$ branes with D9 and D5 branes we have a full super multiplet, together with a left-moving fermion in the $95'$ sector, while the $55'$ sector only involves a left-moving fermion. Trading D5 with $\overline{\text{D}5}$ branes implies a chirality flip for the fermionic singlet in the $\bar 5 5'$ sector. The presence of massless scalars in the $51$ and $95'$ strings suggests that $\text{D}5'$ (D1) branes can be interpreted as (anti-) instantons of D9 (D5 or $\overline{\text{D}5}$) branes. This identification is further supported by the fact that the twisted charges of the putative instanton brane and of the physical brane match. In these cases, the charge vector $Q$ can be expressed in terms of the anomaly vectors $b_i$ associated with the gauge groups of the D9 (D5 or $\overline{\text{D}5}$) branes.

\noindent In general, the defects are charged under the gauge group of the vacuum. In the IR, the theory decouples from the bulk, and the D9 and D5 or $\overline{\text{D}5}$ gauge groups become global symmetries realised on the defect CFT as Ka$\check{\text{c}}$-Moody algebras in both left and right moving sectors\footnote{Actually, this only occurs for the Ka$\check{\text{c}}$-Moody algebra associated with the gauge group admitting the defect as an instanton.}. This is reflected in the sign of the Ka$\check{\text{c}}$-Moody level $k_i $, which can be extracted from the coefficient of $\text{tr}\, F_i^2$ in the anomaly polynomial $I_4$,  and in its contribution to the central charge. 

\noindent For instance, for the gauge group $G_{9,i}$ on the D9 branes with dual Coxeter number $h_i^\vee$, the algebra on the D1 defect is realised at level $ k_i \geq 0$ and thus
\begin{equation}
c_i = \frac{k_i \, \text{dim} \, G_{9,i}}{k_i + h_i^\vee}\,,
\end{equation}
since the Ka$\check{\text{c}}$-Moody algebra is realised by a left-moving fermion, which is a singlet of the ${\mathcal N} = (0,4)$ superalgebra. For the gauge group $G_{\bar 5 ,i}$ or $G_{5,i}$ on the $\overline{\text{D}5}$ or D5 branes, the level of the Ka$\check{\text{c}}$-Moody algebra can be positive or negative. An analogous feature also emerges on the $\text{D}5'$ branes, where the levels of the D5 branes are positive, while those associated with the D9 group can be positive or negative. Similarly, when $\overline{\text{D}5}$ branes replace the D5 ones, the level of the associated Ka$\check{\text{c}}$-Moody algebra is negative.

\noindent Upon removing the contribution of the centre of mass (CM) degrees of freedom (four non-chiral scalars and four right-moving Majorana-Weyl fermions) which decouple in the IR, a natural generalisation of the KSV unitarity constraints is 
\begin{equation}
\sum_{i\,|\, k_i>0} \frac{k_i \, \text{dim} \, G_i}{k_i + h_i^\vee} \le c_\text{L}-4_{\text{CM}} \qquad \text{and} \qquad 
\sum_{i\,|\, k_i<0} \frac{|k_i| \, \text{dim} \, G_i}{|k_i| + h_i^\vee} \le c_\text{R}-6_{\text{CM}}\,,
\label{eq:chiralKSV}
\end{equation}
both for the left and right-moving sectors. However, in the UV, charged (non-chiral) scalar fields are present on the defect, and their role in the realisation of the Ka$\check{\text{c}}$-Moody algebras in the IR is unclear. These scalar fields are non-compact and their dynamics in the IR is difficult to determine. They could be free and generate an abelian algebra or describe an independent interacting sector of the theory. In both cases, they are expected to contribute to the left and right central charges and, if this happens, the inequalities \eqref{eq:chiralKSV} cannot be saturated. Indeed, this is the case in all string constructions where the defects admit an instanton interpretation.

\noindent As a final comment, notice that it is always possible to introduce D1 branes in the bulk of the compactification orbifold, which only couple to the two-forms $C_2^{0}$ and $C_2^1$ from the untwisted sector. In the decompactification limit, one is expected to recover the ten-dimensional type I superstring with a defect coupled to the (non-chiral) R-R two-form $C_2$ existing in $D=10$. Since in $D=6$ $C_2^{0}$ and $C_2^1$ are the two chiral components of $C_2$, the D1 brane in the bulk must carry the same charge, $Q^0 = \pm Q^1$, thus satisfying the \emph{null-charge} conjecture, $Q\cdot Q =0$ \cite{Angelantonj:2020pyr}. We expect that this is the case for any six-dimensional vacuum with at least one tensor multiplet. The existence of \emph{null-charge} branes allowed to put in the swampland various models which pass the KSV constraints, but do not admit a string or F-theory construction \cite{Angelantonj:2020pyr}.

\noindent We can then now move to the discussion of the defects introduced in the examples analysed in this Chapter, namely the supersymmetric and BSB orientifolds of $T^4/\mathbb{Z}_6$.

\subsection{Examples on \texorpdfstring{$T^4 /\mathbb{Z}_6$}{T4/Z6}}

The spectrum characterising the world-sheet theory of the D1 and $\text{D}5'$ defects can be deduced, following \cite{Dudas:2001wd}, by performing an orbifold projection of the toroidal construction, as done for instance in \cite{Angelantonj:2024iwi}. We will present only the light spectrum and refer to the Appendix of \cite{Angelantonj:2024iwi} for the explicit construction of the amplitudes.

\subsubsection{D1 branes in the \texorpdfstring{$T^4 /\mathbb{Z}_6$}{T4/Z6} BSB orientifold}

We start our analysis by considering D1 branes sitting on the $\mathbb{Z}_6$ fixed point of the $T^4 /\mathbb{Z}_6$ orientifold discussed in Section \ref{SSec:BSBZ6}. Such D1 branes support the Chan-Paton gauge group $\text{SO}(r_1)  \times \text{U}(r_2)  \times \text{U}(r_3) \times \text{SO}(r_4) $, reproducing the structure of that of the D9 branes in the same vacuum. The light excitations are summarised in table \ref{tab:D1BSBZ6}, where the representations of the various fields with respect to the $\text{SO} (4) \sim \text{SU} (2)_u \times \text{SU} (2)_v$ group, transverse to the D1 world-volume, are given.  The second line in the table includes the contribution of the centre of mass degrees of freedom which are expected to decouple in the IR. From the microscopic data in the table \ref{tab:D1BSBZ6}, we can compute the anomaly polynomial,
 \begin{equation}
     \begin{aligned}\label{eq:I4D1Z6}
		 I_4&=\tfrac12 \left ( -(r_1+  r_4) \, \text{tr}R^2  - \chi(N) \left ((r_1-r_2)^2 + (r_4-r_3)^2+ (r_2-r_3)^2 \right ) \right.
		\\
		& \qquad + \tfrac{r_1}{2} \, \text{tr} F_{9,1}^2 + \tfrac{r_4}{2} \, \text{tr} F_{9,4}^2 + r_2 \, \text{tr} F_{9,2}^2 + r_3 \, \text{tr} F_{9,3}^2
		\\
		& \qquad  + (r_1-r_2) \, \text{tr} F_{\bar 5,1}^2 + (r_4-r_3) \, \text{tr} F_{\bar 5,4}^2 
  \\
		&\left. \qquad - (r_1+r_3-2r_2) \, \text{tr} F_{\bar 5,2}^2 - (r_4+r_2-2 r_3) \, \text{tr} F_{\bar 5,3}^2 \right ) \, ,
	\end{aligned} 
   \end{equation}
   which cancels the bulk inflow if the charge vector is
   \begin{equation}
       \begin{aligned} \label{eq:D1defZ6}
	Q&= r_1 \, b_5 + r_2 \, b_6 + r_3 \, b_7 + r_4 \, b_8
 \\
 &=\left ( \tfrac{   r_1+2r_2 + 2r_3  + r_4}{2 \sqrt{6}},  -\tfrac{   r_1+2r_2 + 2r_3  + r_4}{2 \sqrt{6}};\tfrac{-r_1-r_2+r_3+r_4}{\sqrt{6}}; \right.
	\\
	&\left. \qquad  \tfrac{-r_1+r_2+ r_3- r_4}{\sqrt{2}} ; \boldsymbol{0}_4; \tfrac{-r_1+2 r_2-2r_3 + r_4}{\sqrt{3}}; \boldsymbol{0}_5 \right )\, .
	\end{aligned}
   \end{equation}
Notice that this solution for $Q$ guarantees the positivity of the tension of the defect, $Q\cdot J >0$, with the K\"ahler form $J$ given in eq. \eqref{eq:Jform}. From \eqref{eq:I4D1Z6} we also extract the levels of the Ka$\check{\text{c}}$-Moody algebras in the IR,
\begin{equation}
    \begin{split}
         &  k_1= r_1  \, ,
         \\
         & k_5= r_1 - r_2 \, ,
    \end{split}
    \qquad 
    \begin{split}
        & k_2=r_2 \, , 
        \\
        & k_6=2 r_2 - r_1 - r_3 \, ,
    \end{split}
    \qquad
    \begin{split}
        & k_3=r_3 \, , 
        \\
        & k_7= 2r_3 - r_2 - r_4 \, ,
    \end{split}
    \qquad 
    \begin{split}
    & k_4=r_4 \, ,
    \\
 & k_8=r_4-r_3 \,.
     \end{split}
\end{equation}
Notice that, according to our choice of the Chan-Paton labels, $k_i$ may be positive or negative, implying the associated the Ka$\check{\text{c}}$-Moody algebra to be realised on the left or right-moving sector of the CFT.

\begin{table}
\centering
\begin{tabular}{| c | c| c|}
\hline
 Representation &   $\text{SO}(1,1)\times \text{SU}(2)_u \times \text{SU}(2)_v  $  &  Sector 
 \\
\hline
$\smalltableau{ \null \\ \null \\}_{1_1} + \smalltableau{ \null \\ \null \\}_{1_4} + \left ( \smalltableau{ \null \\} \times \overline{\smalltableau{\null \\}} \right )_{1_2} + \left ( \smalltableau{ \null \\} \times \overline{\smalltableau{\null \\}} \right )_{1_3}  $  & $(0,1,1) + 2 \times ( \tfrac1 2,2,1)_\text{L}$ & D1-D1
\\
$\smalltableau{ \null \& \null \\}_{1_1} + \smalltableau{ \null \& \null \\}_{1_4} + \left ( \smalltableau{ \null \\} \times \overline{\smalltableau{\null \\}} \right )_{1_2} + \left ( \smalltableau{ \null \\} \times \overline{\smalltableau{\null \\}} \right )_{1_3} $  & $ (1,2,2)+2 \times (\tfrac12,1,2)_\text{R}$ & 
\\
$( \,  \smalltableau{\null \\}_{1_1}, \overline{\smalltableau{\null \\}}_{1_2}) + (\smalltableau{\null \\}_{1_4}, \smalltableau{\null \\}_{1_3} ) + ( \,  \smalltableau{\null \\}_{1_2}, \overline{\smalltableau{\null \\}}_{1_3})$  & $4\times (1,1,1)+2 \times (\tfrac12,2,1)_\text{R}$ & 
\\
$( \,  \smalltableau{\null \\}_{1_1}, \overline{\smalltableau{\null \\}}_{1_2}) + (\smalltableau{\null \\}_{1_3}, \smalltableau{\null \\}_{1_4} ) + ( \,  \smalltableau{\null \\}_{1_2}, \overline{\smalltableau{\null \\}}_{1_3}) $  & $ 2 \times (\tfrac12,1,2)_\text{L}$ & 
\\ 
\hline
$( \,  \smalltableau{\null \\}_{1_1}, \smalltableau{\null \\}_{9_1}  ) + (\, \smalltableau{\null \\}_{1_4}, \smalltableau{\null \\}_{9_4}  ) $  & $ (\tfrac12,1,1)_\text{L}$ & D1-D9
\\
$( \,  \smalltableau{\null \\}_{1_2}, \overline{\smalltableau{\null \\}}_{9_2} ) + (\, \smalltableau{\null \\}_{1_3}, \overline{\smalltableau{\null \\}}_{9_3} )$  & $ (\tfrac12,1,1)_\text{L}$ & 
\\
\hline
$( \,  \smalltableau{\null \\}_{1_1}, \smalltableau{\null \\}_{\bar 5_1}  ) + (\, \smalltableau{\null \\}_{1_4}, \smalltableau{\null \\}_{\bar 5_4} )$  & $ (1,1,2) + 2 \times (\tfrac12,1,1)_\text{L}$ & D1-$\overline{\text{D}5}$ 
\\
$( \,  \smalltableau{\null \\}_{1_2}, \overline{\smalltableau{\null \\}}_{\bar 5_2}  ) + (\, \smalltableau{\null \\}_{1_3}, \overline{\smalltableau{\null \\}}_{\bar 5_3}  )$  & $ 2 \times (1,1,2) + 4 \times (\tfrac12,1,1)_\text{L}$ & 
\\
$(\, \smalltableau{\null \\}_{1_1}, \overline{\smalltableau{\null \\}}_{\bar 5_2} )+ (\, \smalltableau{\null \\}_{1_2}, \overline{\smalltableau{\null \\}}_{\bar 5_3} ) + (\overline{\smalltableau{\null \\}}_{1_2} , \smalltableau{\null \\}_{\bar 5_1} )$  & $  2 \times (\tfrac12,1,1)_\text{R}$ & 
\\
$(\, \smalltableau{\null \\}_{1_3}, \smalltableau{\null \\}_{\bar 5_4} )+ (\, \smalltableau{\null \\}_{1_4}, \smalltableau{\null \\}_{\bar 5_3}  ) + (\overline{\smalltableau{\null \\}}_{1_3} , \smalltableau{\null \\}_{\bar 5_2} )$  & $  2 \times (\tfrac12,1,1)_\text{R}$ & 
\\
\hline
\end{tabular}\\
\caption{The light spectrum for a probe $\text{SO}(r_1)\times \text{U}(r_2) \times \text{U}(r_3)\times \text{SO}(r_4) $ D1 brane at a $\mathbb{Z}_6$ fixed point in the BSB $T^4/\mathbb{Z}_6$ vacuum with D9/$\overline{\text{D5}}$ branes.}
\label{tab:D1BSBZ6}
\end{table}

\noindent From the solution \eqref{eq:D1defZ6}, we read that the D1 branes can be interpreted as instantons of the gauge theories on the $\overline{\text{D}5}$ branes, since $Q \propto b_{5,6,7,8}$. This observation is supported by the presence of moduli in the bi-fundamental representations $( \,  \smalltableau{\null \\}_{1_1}, \smalltableau{\null \\}_{\bar 5_1} ) + (\, \smalltableau{\null \\}_{1_2}, \overline{\smalltableau{\null \\}}_{\bar 5_2} )+ ( \,  \smalltableau{\null \\}_{1_3}, \overline{\smalltableau{\null \\}}_{\bar 5_3}  ) + (\, \smalltableau{\null \\}_{1_4}, \smalltableau{\null \\}_{\bar 5_4}  )$. 

\noindent Finally, one can check that the constraints \eqref{eq:chiralKSV} are satisfied with $c_\text{L}$ and $c_\text{R}$ computed from table \ref{tab:D1BSBZ6}, once the degrees of freedom of the centre of mass are removed. As expected, this implies that the two-dimensional CFT is unitary and this string construction is consistent. As an example, in the $a=0$ vacuum, where the $\overline{\text{D}5}$ support a $\text{USp} (8) \times \text{U}(4) \times \text{U}(4) \times \text{USp}(8)$ gauge group and D9 branes a $\text{SO} (8) \times \text{U}(4) \times \text{U}(4) \times \text{SO}(8)$ gauge group on the fixed point, the simple choice $r_1=1$ and $r_2=r_3=r_4=0$ gives
\begin{equation}
    \begin{split}
    &\sum_{i \, | \, k_i \geq 0} \frac{k_i \ \text{dim} G_i }{k_i + h_i^{\vee}}=4 + 6< c_\text{L} -4_\text{CM} = 28\,, 
    \\
    &\sum_{i \, | \, k_i <0} \frac{|k_i| \ \text{dim} G_i }{|k_i| + h_i^{\vee}} = 4 < c_\text{R} -6_\text{CM} = 20  \,,
\end{split}
\end{equation}
where the centre of mass degrees of freedom have been removed from the counting of $c_\text{L,R}$. 

\begin{table}
\centering
\begin{tabular}{| c | c | c |}
\hline
	 { Representation} &  $\text{SO}(1,1) \times \text{SU}(2)_l \times \text{SU} (2)_R$ & { Sector} \\
	\hline
	$ \smalltableau{ \null \\ \null \\}   $  & $\left (0, 1, 1 \right ) + 2 \times \left (\tfrac{1}{2}, 2, 1 \right )_\text{L}  $  & D1-D1
	\\
	$ \smalltableau{ \null \& \null \\}   $ & $\left (1, 2, 2 \right ) + 2 \times \left (\tfrac{1}{2}, 1, 2 \right )_\text{R}$ & \ \ 
	\\
	$ \smalltableau{ \null \& \null \\}  $  & $4 \times \left (1, 1, 1 \right ) + 2 \times \left (\frac{1}{2}, 2, 1 \right )_\text{R} $ &  \ \ 
 \\
	 $  \smalltableau{ \null \\ \null \\} $  & $ 2 \times \left (\frac{1}{2},  1,2 \right )_\text{L} $ &  \ \
	\\ 
 \hline
	$  (\, \smalltableau{\null \\}_{1}, \smalltableau{\null \\}_{9_1}+\smalltableau{\null \\}_{9_2} ) $  & $  \left (\frac{1}{2},  1, 1 \right )_\text{L} $ & D1-D9
 \\
 $  (\, \smalltableau{\null \\}_{1}, \smalltableau{\null \\}_{9_3}+\overline{\smalltableau{\null \\}}_{9_4} ) $  & $  2 \times \left (\frac{1}{2},  1, 1 \right )_\text{L} $ 
 \\
 \hline
\end{tabular}
\caption{Spectrum for probe D1 branes  in the bulk of the $T^4/\mathbb{Z}_4$ orbifold, supporting an $\text{SO}(r)$ gauge group.}
\label{tab:D1bulkZ4}
\end{table}

\noindent One has the option of moving the D1 branes away from fixed points. As in \cite{Angelantonj:2020pyr}, this turns out to be the configuration which satisfies the \emph{minimal} constraints of \cite{Kim:2019vuc}. The gauge group on the defect is $\text{SO} (r)$ and the microscopic degrees of freedom are listed in table \ref{tab:D1bulkZ4}, while the D1-$\overline{\text{D}5}$ sector is absent, since the open strings with these boundary conditions are now massive. Also in this case, the second line in the table contains the centre of mass degrees of freedom, associated with the singlet in the decomposition of the twofold symmetric representation of orthogonal groups. The anomaly polynomial now reads
\begin{equation}
    I_4= \tfrac{1}{2} \Bigg ( -2 r \, \text{tr} R^2 + \tfrac{r}{2} \text{tr} F_{9,1}^2 + r \text{tr} F_{9,2}^2 + r \, \text{tr} F_{9,3}^2 + \tfrac{r}{2} \text{tr} F_{9,4}^2  \Bigg ) \, ,
\end{equation}
which cancels the inflow of the bulk theory if the charge vector is
\begin{equation}
    Q = \left ( r,-r ; \boldsymbol{0}_{12} \right ) \, .
\end{equation}
As expected, the D1 branes couple only to the two-forms from the untwisted sector, and therefore satisfy the \emph{null-charge} condition \cite{Angelantonj:2020pyr}. For $r=1$, \emph{i.e.} for a single D1 brane, the left and right central charges
\begin{equation}
c_\text{L}= 20 +4_{\text{CM}} \, ,
\qquad
c_\text{R}=  6 +6_{\text{CM}} \, ,
\end{equation}
agree with the those given in \cite{Kim:2019vuc}. We would like to stress, though, that for the case of an arbitrary stack of D1 branes, $r>1$, the expressions of $c_\text{L,R}$ in \cite{Kim:2019vuc} seem no-longer to be correct. 

\noindent Upon reading the coefficients $k_1=k_2=k_3=k_4=1$ from $I_4$, and taking into account that the Ka$\check{\text{c}}$-Moody algebra is realised only in the left-moving sector, the unitarity constraint reads
\begin{equation}
\sum_{i \, | \, k_i \geq 0} \frac{k_i \ \text{dim} G_i }{k_i + h_i^{\vee}}= 16 < c_\text{L}-4_\text{CM} =20 \, ,
\end{equation}
and is clearly satisfied.

\subsubsection{\texorpdfstring{$\text{D}5'$}{D5'} branes in the \texorpdfstring{$T^4/\mathbb{Z}_6$}{T4/Z6} BSB orientifold}
\label{sssec:D5pZ6BSB}

\begin{table}
\centering
\begin{tabular}{| c | c | c |}
\hline
	 { Representation} &  $\text{SO}(1,1) \times \text{SU}(2)_u \times \text{SU}(2)_v $ & { Sector} 
  \\
	\hline
$\smalltableau{ \null \& \null \\}_{1_1} + \smalltableau{ \null \& \null \\}_{1_4} + \left ( \smalltableau{ \null \\} \times \overline{\smalltableau{\null \\}} \right )_{1_2} + \left ( \smalltableau{ \null \\} \times \overline{\smalltableau{\null \\}} \right )_{1_3}  $  & $(0,1,1) + 2 \times ( \tfrac1 2,2,1)_\text{L}$ & D5$'$-D5$'$
\\
$\smalltableau{ \null \\ \null \\}_{1_1} + \smalltableau{ \null \\ \null \\}_{1_4} + \left ( \smalltableau{ \null \\} \times \overline{\smalltableau{\null \\}} \right )_{1_2} + \left ( \smalltableau{ \null \\} \times \overline{\smalltableau{\null \\}} \right )_{1_3} $  & $ (1,2,2)+2 \times (\tfrac12,1,2)_\text{R}$ & 
\\
$( \,  \smalltableau{\null \\}_{1_1}, \overline{\smalltableau{\null \\}}_{1_2}) + (\smalltableau{\null \\}_{1_4}, \smalltableau{\null \\}_{1_3} ) + ( \,  \smalltableau{\null \\}_{1_2}, \overline{\smalltableau{\null \\}}_{1_3})$  & $4\times (1,1,1)+2 \times (\tfrac12,2,1)_\text{R}$ & 
\\
$( \,  \smalltableau{\null \\}_{1_1}, \overline{\smalltableau{\null \\}}_{1_2}) + (\smalltableau{\null \\}_{1_4}, \smalltableau{\null \\}_{1_3} ) + ( \,  \smalltableau{\null \\}_{1_2}, \overline{\smalltableau{\null \\}}_{1_3}) $  & $ 2 \times (\tfrac12,1,2)_\text{L}$ & 
\\ 
\hline
$( \,  \smalltableau{\null \\}_{1_1}, \smalltableau{\null \\}_{9_1} ) + (\, \smalltableau{\null \\}_{1_4}, \smalltableau{\null \\}_{9_4}  )$  & $ (1,2,1) + 2 \times (\tfrac12,1,1)_\text{R}$ & D5$'$-D9
\\
$( \,  \smalltableau{\null \\}_{1_3}, \overline{\smalltableau{\null \\}}_{9_3}  ) + (\, \smalltableau{\null \\}_{1_2}, \overline{\smalltableau{\null \\}}_{9_2}  )$  & $ 2 \times (1,2,1) + 4 \times (\tfrac12,1,1)_\text{R}$ & 
\\
$(\, \smalltableau{\null \\}_{1_1}, \overline{\smalltableau{\null \\}}_{9_2} )+ (\, \smalltableau{\null \\}_{1_2}, \overline{\smalltableau{\null \\}}_{9_3} ) + (\overline{\smalltableau{\null \\}}_{1_2} , \smalltableau{\null \\}_{9_1} )$  & $  2 \times (\tfrac12,1,1)_\text{L}$ & 
\\
$(\, \smalltableau{\null \\}_{1_3}, \overline{\smalltableau{\null \\}}_{9_4} )+ (\, \smalltableau{\null \\}_{1_3}, \overline{\smalltableau{\null \\}}_{9_2}  ) + (\overline{\smalltableau{\null \\}}_{1_4} , \smalltableau{\null \\}_{9_3} )$  & $  2 \times (\tfrac12,1,1)_\text{L}$ & 
 \\ 
 \hline
	$( \,  \smalltableau{\null \\}_{1_1}, \smalltableau{\null \\}_{\bar 5_1} ) + (\, \smalltableau{\null \\}_{1_4}, \smalltableau{\null \\}_{\bar 5_4}  ) $  & $ (\tfrac12,1,1)_\text{R}$  & $\text{D}5'\text{-}\overline{\text{D}5}$ 
\\
$( \,  \smalltableau{\null \\}_{1_2}, \overline{\smalltableau{\null \\}}_{\bar 5_2}  ) + (\, \smalltableau{\null \\}_{1_3}, \overline{\smalltableau{\null \\}}_{\bar 5_3}  )$  & $ (\tfrac12,1,1)_\text{R}$ & 
	 \\ 
  \hline
\end{tabular}
\caption{Spectrum for probe $\text{D}5'$ branes wrapping the entire internal space for the BSB $T^4/\mathbb{Z}_6$ orientifold. The associated Chan-Paton group is $\text{USp}(r_1) \times \text{U}(r_2) \times \text{U}(r_3) \times \text{USp}(r_4)$.}
\label{tab:D5'BSBZ6}
\end{table}

We can now consider the other kind of defects for this vacuum given by $\text{D}5'$ branes wrapping the compactification space, whose associated Chan-Paton gauge group is  $\text{USp}(r_1) \times \text{U}(r_2)  \times \text{U}(r_3) \times \text{Usp}(r_4) $. The spectrum of light excitations is summarised in table \ref{tab:D5'BSBZ6}. Clearly, in general, the $\text{D}5'$ branes probe all $\mathbb{Z}_N$ fixed points, which include the origin where  $\overline{\text{D}5}$ branes are placed. The second line contains the contribution of the centre of mass degrees of freedom of the defects, associated to the various singlets of the Chan-Paton gauge group. The anomaly polynomial is 
   \begin{equation}
       \begin{aligned}
		 I_4&= \tfrac12 \left ( (r_1+r_4) \text{tr}R^2  - \chi(N) \left ((r_1-r_2)^2 + (r_4-r_3)^2+ (r_2-r_3)^2 \right ) \right.
		\\
		&\qquad  -  \tfrac{r_1}{2} \, \text{tr} F_{\bar 5,1}^2 - \tfrac{r_4}{2} \, \text{tr} F_{\bar 5,4}^2 - r_2 \, \text{tr} F_{\bar 5,2}^2 - r_3 \, \text{tr} F_{\bar 5,3}^2
		\\
		&  \qquad  + (r_2-r_1) \, \text{tr} F_{9,1}^2 + (r_3-r_4) \, \text{tr} F_{9,4}^2 
  \\
		&\left. \qquad + (r_1+r_3-2r_2) \, \text{tr} F_{9,2}^2 + (r_4+r_2-2r_3) \, \text{tr} F_{9,3}^2 \right) \, ,
	\end{aligned} \label{eq:I4D5'Z6}
   \end{equation}
   which cancels the anomaly inflow if
   \begin{equation}
       	\begin{aligned} \label{eq:D5'defZ6}
	Q &=  \left ( \tfrac{r_1+2r_2+2r_3+r_4}{2\sqrt{6}},\tfrac{r_1+2r_2+2r_3+r_4}{2\sqrt{6}}; \tfrac{r_1-r_4+r_2-r_3}{\sqrt{6}} ; \tfrac{r_1-r_2-r_3+r_4}{3\sqrt{2}} ; \tfrac{r_1-r_2-r_3+r_4}{3} \boldsymbol{1}_4;  \right.
	\\
 &\left. \qquad \tfrac{r_1-2 r_2+2 r_3-r_4}{4 \sqrt{3}} ; \tfrac{r_1-2 r_2+2 r_3-r_4}{4} \boldsymbol{1}_5 \right ) \, .
	\end{aligned} 
   \end{equation}
From this expression, we can see that $Q \propto b_{1,2,3,4}$, implying that the $\text{D}5'$ branes are naturally interpreted as instantons of the D9 ones. Indeed, the light spectrum involves scalars in the corresponding bi-fundamental representations. 

\noindent The solution \eqref{eq:D5'defZ6}, together with the expression \eqref{eq:Jform} for the K\"ahler $J$-form, guarantees that the $\text{D}5'$ branes have positive tension as well, $Q\cdot J >0$. Moreover, from eq. \eqref{eq:I4D5'Z6} we read
\begin{equation}
\begin{split}
    &k_1= 2(r_2-r_1) \, ,
     \\
    &   k_5= -r_1 \, ,
\end{split}
     \qquad 
     \begin{split}
     & k_2 = r_1 + r_3 -2 r_2 \, , 
     \\
     & k_6 = -r_2 \, , 
     \end{split}
     \qquad
     \begin{split}
     & k_3=r_4+r_2-2r_3 \, , 
     \\
    & k_7=-r_3 \, ,
    \end{split}
    \qquad
    \begin{split}
    & k_4=2(r_3-r_4) \, , 
    \\
    & k_8=- r_4  \, .  
    \label{eq:KacMoodylevelsZ4D5'}
\end{split}
\end{equation}
The unitarity constraints are satisfied both in the left and right-moving CFTs. For instance, still for $a=0$, $r_1=1$ and $r_2=r_3=r_4=0$ and one finds
\begin{equation}
    \begin{split}
    &\sum_{i \, | \, k_i \geq 0} \frac{k_i \ \text{dim} G_i }{k_i + h_i^{\vee}}=  4 < c_\text{L}-4_\text{CM} =20 \, ,
    \\
    &\sum_{i \, | \, k_i <0}  \frac{|k_i| \ \text{dim} G_i }{|k_i| + h_i^{\vee}}=  7 + 4 = 11 < c_\text{R} - 6_\text{CM}= 28 \, .
\end{split}
\end{equation}

\subsubsection{D1 branes in the \texorpdfstring{$T^4 /\mathbb{Z}_6$}{T4/Z6} supersymmetric orientifold}

Analogous considerations can be made when supersymmetry is present. In such a case, the spectrum is organised according to the $\mathcal{N}=(0,4)$ superalgebra representations described in the Section \ref{Sec:defects}. We can start by analysing the D1 brane defects.  In such a case the gauge group is given by $\text{U}(r_1)\times \text{U}(r_2)\times \text{U}(r_3)$ and gives rise to the light spectrum summarised in table \ref{tab:D1SUSYZ6}. The anomaly polynomial reads
    \begin{equation} \label{eq:susy6pol4}
	\begin{aligned}
		 I_4= \tfrac12 & \left (-(r_1+  r_3 ) \text{tr}R^2   - \chi(N) \left ((r_1-r_2)^2 + (r_3-r_2)^2 \right ) \right.
		\\
		& + r_1 \, \text{tr} F_{9,1}^2 + r_2 \, \text{tr} F_{9,2}^2 + r_3 \, \text{tr} F_{9,3}^2
		\\
		&\left.  - (r_1-r_2) \, \text{tr} F_{5,1}^2 - (2 r_2 -r_1-r_3) \, \text{tr} F_{5,2}^2 - (r_3-r_2) \, \text{tr} F_{5,3}^2 \right ) \, ,
	\end{aligned}
\end{equation}
which cancels the bulk inflow if the charge vector is
    \begin{equation}
        Q= \left ( \tfrac{r_1+r_2+r_3}{\sqrt{6}},- \tfrac{r_1+r_2+r_3}{\sqrt{6}};\tfrac{r_3-r_1}{\sqrt{2}}; \tfrac{-r_1-2 r_2+r_3}{\sqrt{2}}; \boldsymbol{0}_4 \right )  \, .
    \end{equation}
From the anomaly polynomial, it is possible to read the level of the Ka$\check{\text{c}}$-Moody algebras
\begin{equation}
\begin{split}
    & k_1= r_1 \, ,
    \\
    & k_4= r_2-r_1 \, , \qquad 
    \end{split}
    \qquad
    \begin{split}
    & k_2= r_2 \, , 
    \\
    & k_5= r_1+ r_3 - 2 r_2 \, , 
    \end{split}
    \qquad 
    \begin{split}
    & k_3= r_3 \, , 
   \\
   &  k_6= r_2-r_3 \, .
\end{split}
\end{equation}
The minimal choice $r_1=1, \, r_2=r_3=0$ induces a realisation of the algebra both on the left and right moving sectors, which can be shown to satisfy the unitarity constraints,
\begin{equation}
    \begin{split}
    &\sum_{i \, | \, k_i \geq 0} \frac{k_i \ \text{dim} G_i }{k_i + h_i^{\vee}}=4 + 8=12  < c_{\text{L}}-4_{\text{CM}}=32  \, ,
    \\
    &\sum_{i \, | \, k_i <0} \frac{|k_i| \ \text{dim} G_i }{|k_i| + h_i^{\vee}}=4  < c_{\text{R}}-6_{\text{CM}}=24 \, ,
\end{split}
\end{equation}
where we have taken $a=0$ for simplicity.
    \begin{table}
\centering
\begin{tabular}{| c | c| c|}
\hline
 Representation &   $\text{SO}(1,1)\times \text{SU}(2)_u \times \text{SU}(2)_v  $  &  Sector 
 \\
\hline
	$\left (\smalltableau{ \null \\} \times \overline{\smalltableau{ \null \\}}\right )_{1_1} + \left ( \smalltableau{ \null \\} \times \overline{\smalltableau{ \null \\}} \right )_{1_2} + \left ( \smalltableau{ \null \\} \times \overline{\smalltableau{ \null \\}} \right )_{1_3} $  & $(0,1,1) + 2 \times ( \tfrac1 2,2,1)_\text{L}$ & D1-D1
\\
$\left (\smalltableau{ \null \\} \times \overline{\smalltableau{ \null \\}} \right )_{1_1} + \left ( \smalltableau{ \null \\} \times \overline{\smalltableau{ \null \\}} \right )_{1_2} +\left ( \smalltableau{ \null \\} \times \overline{\smalltableau{ \null \\}} \right )_{1_3}$  & $ (1,2,2)+2 \times (\tfrac12,1,2)_\text{R}$ &  
\\
$  \overline{\smalltableau{\null \& \null \\}}_{1_1}+ (\smalltableau{\null \\}_{1_1},\overline{\smalltableau{\null \\}}_{1_2} )+ (\smalltableau{\null \\}_{1_2},\overline{\smalltableau{\null \\}}_{1_3}) +  \smalltableau{\null \& \null \\}_{1_3} $  & $4\times (1,1,1)+ 2 \times (\tfrac12,2,1)_\text{R}$ & 
\\ 
$\overline{\smalltableau{\null \\ \null \\}}_{1_1}+ ( \, \smalltableau{\null \\}_{1_1},\overline{\smalltableau{\null \\}}_{1_2} )+ ( \, \smalltableau{\null \\}_{1_2},\overline{\smalltableau{\null \\}}_{1_3}  ) +  \smalltableau{\null \\ \null \\}_{1_3}$  & $2 \times (\tfrac12,1,2)_\text{L}$ & 
\\
\hline
$  (\, \smalltableau{\null \\}_{1_1},  \overline{\smalltableau{\null \\}}_{9_1} ) + (\, \smalltableau{\null \\}_{1_2}, \overline{ \smalltableau{\null \\}}_{9_2}  )+ (\,\smalltableau{\null \\}_{1_3}\, , \, \overline{\smalltableau{\null \\}}_{9_3}  )$  & $ 2 \times (\tfrac12,1,1)_\text{L}$ & D1-D9
\\
\hline
$ ( \,  \smalltableau{\null \\}_{1_1},  \overline{\smalltableau{\null \\}}_{5_1} ) + (\, \smalltableau{\null \\}_{1_2} , \smalltableau{\null \\}_{5_2} )+ (\, \overline{\smalltableau{\null \\}}_{1_3}, \overline{\smalltableau{\null \\}}_{5_3}  )$  & $ 2 \times (1,2,1) + 4 \times (\tfrac12,1,1)_\text{R}$ & D1-D5
\\
$ ( \,  \smalltableau{\null \\}_{1_1},  \smalltableau{\null \\}_{5_1}   ) + ( \,  \overline{\smalltableau{\null \\}}_{1_1} , \smalltableau{\null \\}_{5_2} ) + (\, \smalltableau{\null \\}_{1_2}, \, \overline{ \smalltableau{\null \\}}_{5_1} ) $ & $ 2 \times (\tfrac12,1,1)_\text{L}$ & \ \
 \\
$ (\,  \overline{\smalltableau{\null \\}}_{1_2} ,  \smalltableau{\null \\}_{5_3}   ) + (\,  \smalltableau{\null \\}_{1_3} ,  \overline{\smalltableau{\null \\}}_{5_2}  ) + (\,  \overline{\smalltableau{\null \\}}_{1_3} , \overline{\smalltableau{\null \\}}_{5_3}  )$  & $ 2 \times (\tfrac12,1,1)_\text{L}$ & \ \
\\
\hline
\end{tabular}\\
\caption{The light spectrum for a probe $\text{U}(r_1)\times \text{U}(r_2)\times \text{U}(r_3)$ D1 branes at a $\mathbb{Z}_6$ fixed point in the supersymmetric $T^4/\mathbb{Z}_6$ orientifold with D9 and D5 branes. }
\label{tab:D1SUSYZ6}
\end{table}

\subsubsection{\texorpdfstring{$\text{D}5'$}{D5'} branes in the \texorpdfstring{$T^4/\mathbb{Z}_6$}{T4/Z6} supersymmetric orientifold}
\label{sssec:D5pZ6SUSY}

Finally, we can discuss D5$'$ branes for such vacuum, which also support a $\text{U}(r_1)\times \text{U}(r_2)\times \text{U}(r_3)$ gauge group. The light spectrum is summarised in table \ref{tab:D5'SUSYZ6} \begin{table}
\centering
\begin{tabular}{| c | c| c|}
\hline
 Representation &   $\text{SO}(1,1)\times \text{SU}(2)_u \times \text{SU}(2)_v  $  &  Sector 
 \\
\hline
	$\left (\smalltableau{ \null \\} \times \overline{\smalltableau{ \null \\}} \right)_{5'_1} + \left ( \smalltableau{ \null \\} \times \overline{\smalltableau{ \null \\}} \right )_{5'_2} +\left (\smalltableau{ \null \\} \times \overline{\smalltableau{ \null \\}} \right )_{5'_3} $  & $(0,1,1) + 2 \times ( \tfrac1 2,2,1)_\text{L}$ & D1-D1
\\
$\left ( \smalltableau{ \null \\} \times \overline{\smalltableau{ \null \\}}\right )_{5'_1} + \left ( \smalltableau{ \null \\} \times \overline{\smalltableau{ \null \\}} \right )_{5'_2} + \left ( \smalltableau{ \null \\} \times \overline{\smalltableau{ \null \\}} \right )_{5'_3}$  & $ (1,2,2)+2 \times (\tfrac12,1,2)_\text{R}$ &  
\\
$  \overline{\smalltableau{\null \& \null \\}}_{5'_1}+ (\smalltableau{\null \\}_{1_1},\overline{\smalltableau{\null \\}}_{5'_2} )+ (\smalltableau{\null \\}_{5'_2},\overline{\smalltableau{\null \\}}_{5'_3}) +  \smalltableau{\null \& \null \\}_{5'_3} $  & $4\times (1,1,1)+ 2 \times (\tfrac12,2,1)_\text{R}$ & 
\\ 
$\overline{\smalltableau{\null \\ \null \\}}_{5'_1}+ ( \, \smalltableau{\null \\}_{5'_1},\overline{\smalltableau{\null \\}}_{5'_2}  )+ ( \, \smalltableau{\null \\}_{5'_2},\overline{\smalltableau{\null \\}}_{5'_3}  ) +  \smalltableau{\null \\ \null \\}_{5'_3}$  & $2 \times (\tfrac12,1,2)_\text{L}$ & 
\\
\hline
$ ( \,  \smalltableau{\null \\}_{5'_1},  \overline{\smalltableau{\null \\}}_{9_1} ) + (\, \smalltableau{\null \\}_{5'_2} , \smalltableau{\null \\}_{9_2}  )+ (\, \overline{\smalltableau{\null \\}}_{5'_3}, \overline{\smalltableau{\null \\}}_{9_3} )$  & $ 2 \times (1,2,1) + 4 \times (\tfrac12,1,1)_\text{R}$ & D5$'$-D9
\\
$ ( \,  \smalltableau{\null \\}_{5'_1},  \smalltableau{\null \\}_{9_1}   ) + ( \,  \overline{\smalltableau{\null \\}}_{5'_1} , \smalltableau{\null \\}_{9_2} ) + (\, \smalltableau{\null \\}_{5'_2}, \, \overline{ \smalltableau{\null \\}}_{9_1} ) $ & $ 2 \times (\tfrac12,1,1)_\text{L}$ & \ \
 \\
$ (\,  \overline{\smalltableau{\null \\}}_{5'_2} ,  \smalltableau{\null \\}_{9_3}   ) + (\,  \smalltableau{\null \\}_{5'_3} ,  \overline{\smalltableau{\null \\}}_{9_2}  ) + (\,  \overline{\smalltableau{\null \\}}_{5'_3} , \overline{\smalltableau{\null \\}}_{9_3} )$  & $ 2 \times (\tfrac12,1,1)_\text{L}$ & \ \
\\
\hline
$  (\, \smalltableau{\null \\}_{5'_1},  \overline{\smalltableau{\null \\}}_{5_1}  ) + (\, \smalltableau{\null \\}_{5'_2}, \overline{ \smalltableau{\null \\}}_{5_2}  )+ (\,\smalltableau{\null \\}_{5'_3}\, , \, \overline{\smalltableau{\null \\}}_{5_3} )$  & $ 2 \times (\tfrac12,1,1)_\text{L}$ & D5$'$-D5
\\
\hline
\end{tabular}\\
\caption{The light spectrum for a probe $\text{U}(r_1)\times \text{U}(r_2)\times \text{U}(r_3)$ D5$'$ branes in the supersymmetric $T^4/\mathbb{Z}_6$ orientifold with D9 and D5 branes. }
\label{tab:D5'SUSYZ6}
\end{table}
and gives the anomaly polynomial 
    \begin{equation}
    \begin{aligned}
        I_4= \tfrac12 &\left ( -(r_1+r_3) \, \text{tr}R^2  - \chi(N) \left ((r_1-r_2)^2 + (r_2-r_3)^2 \right ) \right.
		\\
		& + (r_2-r_1) \, \text{tr} F_{9,1}^2 + (r_1+r_3-2 r_2) \, \text{tr} F_{9,2}^2 + (r_2-r_3) \, \text{tr} F_{9,3}^2  
		\\
		&\left. + r_1 \, \text{tr} F_{5,1}^2 + r_2 \, \text{tr} F_{5,2}^2 + r_3 \ \text{tr} F_{5,3}^2  \right ) \, ,
	\end{aligned}
\end{equation}
which is cancelled by the bulk inflow if the charge vector is
\begin{equation}
	\begin{aligned}
		Q &= \left ( \tfrac{r_1 + r_2 + r_3}{\sqrt{6}}, \tfrac{r_1 + r_2 + r_3}{\sqrt{6}}; \tfrac{r_1-r_3}{\sqrt{2}} ; \tfrac{r_1  - 2 r_2+ r_3}{3\sqrt{2}} ; \tfrac{-r_1 + 2 r_2-r_3}{3} \boldsymbol{1}_4 \right ) \, .
	\end{aligned}
\end{equation}
The structure of the anomaly polynomial entails the levels of the Ka$\check{\text{c}}$-Moody algebras to be
\begin{equation}
\begin{split}
     & k_1= r_2-r_1 \, , 
    \\
    & k_4= r_1 \, , 
\end{split}
\qquad  
\begin{split}
& k_2= r_1+ r_3 - 2 r_2 \, ,
\\
& k_5= r_2 \, ,
\end{split}
\qquad 
\begin{split} 
& k_3= r_2-r_3 \, ,
\\
& k_6= r_3 \, .
\end{split}
\end{equation}
As before, we can thus evaluate the anomaly constraints again for the choice $a=0$ and $r_1=1 \, , r_2=r_3=0$
\begin{equation}
    \begin{split}
    &\sum_{i \, | \, k_i \geq 0} \frac{k_i \ \text{dim} G_i }{k_i + h_i^{\vee}}=4  < c_{\text{L}}-4_{\text{CM}}=32  \, ,
    \\
    &\sum_{i \, | \, k_i <0} \frac{|k_i| \ \text{dim} G_i }{|k_i| + h_i^{\vee}}=4 + 8=12  < c_{\text{R}}-6_{\text{CM}}=24 \, .
\end{split}
\end{equation}

\chapter{Unitarity: Dai-Freed Anomalies}\label{ChUnitarity}

\newpage

\section{A modern view on anomalies}\label{sec:introduction}

The discussion in the previous Chapter makes it clear that the implications of unitarity have not been fully exploited, yet despite the intense activity that has been devoted to the subject since the '70s \cite{Adler:1969gk, Bell:1969ts, Alvarez-Gaume:1983ihn, Alvarez-Gaume:1983ict, Alvarez-Gaume:1984zlq, Alvarez-Gaume:1984zst,  Fujikawa:2004cx} (see \emph{e.g.} \cite{Bilal:2008qx, Alvarez-Gaume:2022aak, Harvey:2005it,10.1093/acprof:oso/9780198507628.001.0001} for reviews). Indeed, the analysis of the previous Chapter shows how admitting the possibility that a one-dimensional defect be present in the spectrum entails further constraints on the low-energy effective theory, beyond the simple requirement of anomaly cancellation of the bulk theory. The introduction of this kind of defect, which is natural from a string theory perspective, since it simply reflects the coupling to the ubiquitous $p$-form fields, has been recently linked to the conjectured existence of a unique theory of quantum gravity \cite{McNamara:2019rup}, resulting from the interconnected web of dualities among the known string theory constructions \cite{Witten:1995ex}. This extended notion of duality requires the knowledge of all processes allowed in quantum gravity (even at the non-perturbative level) and of all possible defects, as suggested by the so-called {\em completeness hypothesis} \cite{Polchinski:2003bq, Banks:2010zn}. Combining the latter with unitarity leads to non-trivial requirements, which have been nevertheless shown to hold in F-theory vacua \cite{Kim:2019vuc}, in $D=10$ \cite{Kim:2019vuc} and in $D=6$ perturbative string theory constructions \cite{Angelantonj:2020pyr, Angelantonj:2024iwi}. One can thus approach the issue of describing the shape of the string landscape by adopting a complementary bottom-up view as part of the {\em swampland program} \cite{Vafa:2005ui}, in which unitarity combined with the completeness hypothesis implies further consistency conditions to be satisfied by low-energy theories admitting a UV completion in quantum gravity, along the lines of \cite{Kim:2019vuc,   Lee:2019skh, Kim:2019ths,  Montero:2020icj, Hamada:2021bbz, Katz:2020ewz, Angelantonj:2020pyr, Tarazi:2021duw, Bedroya:2021fbu, Martucci:2022krl, Baykara:2023plc, Hayashi:2023hqa}. In this perspective, one can combine unitarity with other general properties of the string landscape to deduce additional constraints on the string landscape. Requiring that topology change be part of the processes admitted by a theory of quantum gravity entails additional constraints \cite{Garcia-Etxebarria:2018ajm, Monnier:2018nfs,Debray:2021vob, Lee:2022spd, Basile:2023knk, Basile:2023zng}. This picture emerges naturally in the current understanding of anomalies pioneered by the theorems on the Dirac operator by Dai and Freed \cite{Dai:1994kq, Freed:1986hv}. 

\noindent Traditionally, anomalies can be described as the violation of gauge invariance of the one-loop effective action under gauge transformations that are either connected (local anomalies) or disconnected (global anomalies) to the identity \cite{Witten:1982fp, Witten:1985xe, Witten:1985mj, Witten:2019bou}. Both can be ultimately traced back to the ambiguities occurring in the definitions of the partition function of chiral fermionic and bosonic fields. The first can be described by the descent procedure and are encoded in the anomaly density $I_{D+2}$, which is a $D+2$-dimensional polynomial in the curvature of the tangent and gauge bundle \cite{Alvarez-Gaume:1983ict, Alvarez-Gaume:1983ihn,Alvarez-Gaume:1984zlq}, while the others are described by building a mapping torus $X \times [0,1]/ \sim$ where the equivalence relation identifies $(x,0)$ with $(g(x),1)$, with $g$ the considered gauge transformation. The mapping torus thus depends on the homotopy classes of the gauge group \cite{MR1867354} whose associated anomaly vanishes if the topological phase $\phi_T$ of the kinetic operator $O_T$ on the mapping torus \cite{Alvarez-Gaume:1984zst} does. Recently, it has been shown \cite{Witten:2019bou, Yonekura:2016wuc, Hsieh:2020jpj} that these ambiguities can be clarified by describing the chiral fields on a given spacetime $X$ as a boundary mode of a gapped Dirac spinor or a non-chiral $p+1$-form living in a $D+1$-dimensional space $Y$ such that $\partial Y=X$ with suitable elliptic boundary conditions\footnote{These results have been obtained working in Euclidean signature. The case of Minkowski's signature has been covered in \cite{Fukaya:2019qlf}.} \cite{Witten:2019bou, Yonekura:2016wuc, Hsieh:2020jpj, Witten:2018lgb}, required in order for $O_T$ to be self-adjoint on $X$. Schematically, we can write the partition function of these chiral fields living on $X=\partial Y$ as 
\begin{equation} \label{eq:genpartfun}
    \mathcal{Z}(L, Y)= \left | \mathcal{Z}_T(X) \right | e^{-2 \pi i \phi_T(Y)}
\end{equation}
where $|\mathcal{Z}_T(X)|$ corresponds to the absolute value of the partition function of the chiral field living on the boundary once we have imposed the elliptic boundary conditions $L$. Note that, contrary to what happens in standard treatments of anomalies, the definition \eqref{eq:genpartfun} has a well-defined phase, at the cost however of a non-trivial dependence {\em a priori} on the extension $Y$, that should preserve the structure characterising $X$.
The space $Y$ has no physical meaning, and therefore for a physically meaningful theory, the partition function should not depend on the choice of the extension. The dependence on the latter is all contained in the phase $\phi_T(Y)$ and can be quantified by taking the ratio of the partition function on two different extensions \begin{equation}
    \mathcal{A}_T(Y \cup \bar{Y}')= \frac{1}{2 \pi i} \text{log} \frac{\mathcal{Z}(L, Y')}{\mathcal{Z}(L, Y)}
\end{equation}
where $Y \cup \bar{Y}' \equiv Y_{\text{cl}}$ is a closed manifold obtained by glueing $Y$ and $Y'$ with the opposite orientation along the common boundary. The resulting partition function $\mathcal{A}_T(Y_{\text{cl}})$ describes an invertible topological field theory\footnote{An invertible topological field theory is quantum field theory whose Hilbert space on a closed manifold is one-dimensional \cite{Freed:2014iua}.} \cite{Hopkins:2002rd, Yonekura:2016wuc}, known in literature as the {\em anomaly theory}. This nomenclature is justified by the fact that the usual treatment of anomalies is reproduced by specific choices of the closed manifold $Y_{\text{cl}}$. Indeed, whenever it is a boundary of a $D+2$ dimensional space $Y_{\text{cl}}= \partial Z$ the Atiyah-Patodi-Singer (APS) \cite{Atiyah:1975jf} theorem implies
\begin{eqaed}\label{eq:local_anomaly_reproduced}
    \mathcal{A}_{T}(\partial Z) = \text{Index}(O_T) - \int_Z I_{D+2}\, ,
\end{eqaed}
where $\text{Index}(O_T) $ is the index of the operator $O_T$ and it is an integer number. Similarly, the usual global anomalies are reproduced if we choose $Y_{\text{cl}}$ to be a mapping torus. However, these two examples do not exhaust {\em a priori} all the possible $D+1$ dimensional manifolds and therefore different choices of the manifolds $Y_{\text{cl}}$ may be new sources of anomalies, the so-called {\em Dai-Freed} anomalies \cite{Garcia-Etxebarria:2018ajm}. It is then crucial to determine the set of manifolds that are to be considered. For a generic quantum field theory, constraints can restrict the choice of the closed manifolds $Y_{\text{cl}}$ \cite{Witten:2015aba}, but for a theory of quantum gravity, topology is allowed to change and thus {\em all} the manifolds should be considered. Luckily, our task can be simplified further, since it can be shown that if two closed manifolds are equal up to a boundary the difference between the anomaly theories on the two spaces is encoded in the local anomaly. Thus if the latter vanishes, the anomaly depends on the equivalence classes of spaces defined up to a boundary. This space actually forms an abelian group under the disjoint union known as the {\em bordism group} $\Omega_{D+1}^{X_{\text{str}}}$. Therefore to cancel anomalies it is enough to evaluate $\mathcal{A}_T$ on the generator of this group.

\noindent In $6D$ (super)gravity, local anomalies are cancelled by employing the celebrated Green-Schwarz-Sagnotti mechanism, introduced in the previous Chapter, which requires the Bianchi identity  
\begin{equation} \label{twistedstringcond}
	\begin{aligned}
		d H_3^\alpha = X_4^\alpha =  \frac{a^\alpha}{4} \, p_1(R) - \sum_i \frac{b_i^\alpha}{\lambda_i} \, c_2(F_i) \, ,
	\end{aligned}
\end{equation} 
to be satisfied where we have used the definition the characteristic classes\footnote{For the abelian case studied in \ref{sec:anomaly_U1}, the second Chern class should be replaced by $-c_1^2/2$.} $ \frac{1}{4} p_1(R)= -\frac{1}{2}\text{tr} R^2$, $c_2(F)= \frac{1}{2}  \text{tr} F^2 $ in \eqref{eq:Bianchi}. This means that, in the procedure of extending manifolds, one must preserve the structure inherited by the Bianchi identity, the so-called {\em twisted-string structure} \cite{Basile:2023knk, Basile:2023zng, Tachikawa:2024ucm} (or string-$G$, for a given gauge group $G$), in addition to the spin structure with a principal bundle\footnote{In certain settings the spin structure can be modified to a pin$^\pm$, spin$^c$ or spin$^{\mathbb{Z}_4}$ structure. This is crucial to find the correct consistency conditions, for instance when reducing M-theory over non-orientable manifolds where I-fold defects emerge \cite{Montero:2020icj} and certain topological symmetries are broken \cite{McNamara:2019rup, McNamaraThesis}.} valid for the case of gauge theories with fermions. Turning off any gauge contribution reduces to the ordinary string structure, yielding the bordism group $\Omega_7^\text{string} = 0$. This shows that there are no purely gravitational Dai-Freed anomalies in six dimensions when the Green-Schwarz-Sagnotti mechanism cancels the local anomaly. In the presence of gauge fields, twisted string structures can instead generate novel anomalies \cite{Dierigl:2022zll}. If such an anomaly is found it may still be possible to cancel it with a ``topological'' version of the Green-Schwarz-Sagnotti mechanism \cite{Debray:2021vob, Dierigl:2022zll}, but we will not consider this subtle possibility in the present discussion.

\noindent The right-hand side of \eqref{twistedstringcond}, which we dub the Bianchi class, is thus trivial in de Rham cohomology. However, one should require the Bianchi class to be trivial in integral cohomology, which can include torsional classes undetected by differential forms. This was originally argued in \cite{Witten:1985mj}, and (at least for perturbative heterotic strings) follows from Dai-Freed anomaly cancellation on the world-sheet \cite{Basile:2023knk}.  However, specifically for theories in $D=6$, there is an additional subtlety to be taken into account. These settings allow for the presence of many chiral $2$-forms which lead to additional Bianchi identities. As discussed in \cite{Basile:2023zng}, it seems that the correct requirement to impose on admissible anomaly backgrounds is to satisfy at least one identity (at the level of integral cohomology). This turns out to eliminate all anomalies in string theory examples, and can produce non-trivial bordism groups. Imposing multiple identities simultaneously generically kills all independent characteristic classes. Therefore in the following examples, we shall focus on backgrounds satisfying at least one Bianchi identity but spaces trivialising all the available ones have been discussed in \cite{Basile:2023zng}.

\noindent Nevertheless, even imposing one Bianchi identity, the computation of these bordism groups is a complicated mathematical task that has not been fully addressed yet. Therefore, we cannot push until the end the proof of the cancellation of Dai-Freed anomalies for string theory vacua but, we can analyse backgrounds satisfying the Bianchi identity on which the anomaly theory can be computed. These backgrounds are Lens spaces, given by the quotient of the seven-sphere with a discrete group, $L_p^7=S^7/\mathbb{Z}_p$ \cite{Debray:2023yrs}. On the one hand, this allows to verify, without knowing the exact shape of the $\Omega_7^{\text{string-G}}$, the cancellation of anomalies for (non-)supersymmetric heterotic theories and for the Gepner orientifold with no tensor multiplets. On the other hand, it allows to seek constraints for low-energy theories from a bottom-up perspective.

\noindent For our discussion it is therefore crucial to specify which is the anomaly theory for the fields of interest. In the case of $6D$ non-supersymmetric heterotic theories the fields contributing to the anomaly are simply Majorana-Weyl fermions and the $2$-form fields, while if supersymmetry is present there is an additional contribution coming from the gravitini. In the following, we are going to specify the structure of $\mathcal{A}_T$ for each of these fields and their values on the Lens spaces that are needed for the rest of the Chapter.

\subsection{Anomaly: fermions}\label{sec:anomaly_fermions}

Let us start by discussing the contribution of fermionic fields to the anomaly theory. This is the context in which the Dai-Freed theorems were originally formulated \cite{Freed:1986hv,  Dai:1994kq}. The partition function for a chiral fermion is obtained from that of a massive Dirac spinor defined on the extended space $Y$ by imposing the set $L$ of boundary conditions which localises a chiral fermion of a given chirality on the boundary $X$ \cite{Witten:2019bou,Yonekura:2016wuc} 
	\begin{equation}\label{ferpart}
		\mathcal{Z}_{\frac{1}{2},+} \big (L, Y \big )= \left | \text{Pf} \left ( D_{\text{D}}^{+}(X) \right ) \right | e^{-2 i \pi \eta_\text{D} (Y)}
	\end{equation}
where the Pfaffian of the Dirac operator is the regularised product of the positive eigenvalues\footnote{When the Dirac operator in the bulk $Y$ has zero modes, the boundary conditions should be properly modified (see for instance \cite{Yonekura:2016wuc,Dai:1994kq} for details). However for the analysis of anomalies such subtleties play no role and we shall skip them.}. The phase given by $\eta (Y)$ is the so-called {\em eta invariant} \cite{Dai:1994kq,Freed:1986hv} and corresponds to the regularised sum of the signs of the eigenvalues of the Dirac operator. From the general considerations above, because of the glueing properties of the $\eta$-invariant, the anomaly theory is nothing but the phase of \eqref{ferpart} 
\begin{equation}
    \mathcal{A}_{D}(Y_{\text{cl}})= \eta_D(Y_{\text{cl}}) \, ,
\end{equation}
that, in the case $Y_{\text{cl}}=\partial Z$, reduces to \cite{Atiyah:1975jf}
	\begin{equation}
		\text{Index}^{D}= \eta_D(\partial Z) + \int_Z \left ( \hat{A}(R) \, \text{ch}_\text{R}(F) \right )\, .
	\end{equation} 
Here $\text{ch}_\text{R}(F)=\text{tr}_\text{R} \, e^{i F}$ is the Chern character for the associated gauge bundle in the representation $\text{R}$.

\noindent In a similar fashion one can write the partition function of chiral gravitini. However, one must take into account that a massive Rarita-Schwinger field in $D+1$ dimensions localises a chiral Rarita-Schwinger field and a Weyl fermion of opposite chirality on the $D$-dimensional boundary\footnote{We are assuming $D$ to be even in these arguments.} \cite{Debray:2023yrs, Debray:2021vob}, and that the $D$-dimensional Rarita-Schwinger operator identifies the degrees of freedom of the propagating gravitino and a Weyl fermion. As a result, the structure of the partition function for the $D$-dimensional gravitino field is dictated by
	\begin{equation}\label{gravitinopart}
		\mathcal{Z}_{\text{grav},+} \big (L, Y \big )= \frac{ \left | \text{Pf} \left ( D_{\text{RS}}^{+}(X) \right ) \right | }{\left | \text{Pf} \left ( D_{\text{D}}^{+}(X) \right ) \right |^2 } \, e^{-2 i \pi \eta_{\text{RS}} (Y)+ 4 i \pi \eta_{\text{D}}(Y) } \,  ,
	\end{equation}
where the operator $ D_{\text{RS}}^{+}(X)$ is the Rarita-Schwinger operator defined on the product between the spin and the tangent bundle.  The anomaly is thus described by the combination of eta invariants of the Dirac and Rarita-Schwinger operators \cite{Garcia-Etxebarria:2018ajm, Hsieh:2020jpj}
\begin{equation}\label{eq:etagrav}
    \eta_{\text{grav}}(Y_{\text{cl}})= \eta_{\text{RS}}(Y_{\text{cl}}) - 2 \, \eta_D(Y_{\text{cl}}) \, ,
\end{equation}
where the phase $\eta_{\text{RS}}(Y_{\text{cl}})$ is related to the density index via the Atiyah-Patodi-Singer (APS) index theorem \cite{Atiyah:1975jf,Debray:2023yrs, Debray:2021vob} 
	\begin{equation}
		\text{Index}^{\text{RS}}= \eta_{\text{RS}}(\partial Z) + \int_Z  \left ( \hat{A}(R) \, ( \text{ch}(2 R) - 1) \right ) \, ,
	\end{equation} 
with $\text{ch}(2 R)= \text{tr} \, e^{ 2 i R}$.

\noindent On Lens spaces the values of the $\eta$-invariants are known and we will use their expression on $L_p^7$ to compute explicitly the anomaly theory for fermions. To do so, we can embed the $\mathbb{Z}_p$ line bundle into the gauge group bundle, implying the decomposition of a given representation in terms of the $\mathbb{Z}_p$ charge $q$. As a result, the contribution to the anomaly theory of fermions transforming in a given representation of the gauge group comes from $\eta$-invariants carrying non-trivial $\mathbb{Z}_p$ charges associated to the representation itself \cite{Debray:2021vob, Debray:2023yrs} 
\begin{equation} \begin{aligned} \label{eq:etas}
    {\eta_\text{D}}_q(L^{7}_p) & = - \, \frac{p^4 + 10p^2 - 11 - 30 p^2 q^2 - 60 pq + 60 pq^3 - 30 q^4 + 60 q^2}{720p} \, , \\
    {\eta_\text{RS}}_0(L^{7}_p) & = - \, \frac{7p^4 - 170p^2 + 163}{720p}  \, ,
\end{aligned} \end{equation} 
where for our case we only need the uncharged $\eta$-invariant for the Rarita-Schwinger field\footnote{When discussing the consistency of type IIB supergravity under the full S-duality group \cite{Debray:2021vob, Dierigl:2022reg, Debray:2023yrs, Dierigl:2023jdp, Yonekura:2024bvh} charged $\eta$-invariants for the Rarita-Schwinger operator are relevant, but it will not be of interest for us.}.

\noindent The discussion presented so far describes the contribution to the anomaly theory of the fermionic fields. This, however, is not the end of the story, since string theory and (super)gravity include additional relevant fields in their spectra: the (chiral) $p$-form fields.

\subsection{Anomaly: bosons} \label{sec:anomaly_bosons}

In string theory, $p$-form fields are ubiquitous and may give a non-trivial contribution to the anomaly theory. However, although the action and partition function of non-chiral forms are well-understood and under control, there are difficulties in the definition of similar actions for chiral forms, which prevents one from writing down consistent partition functions straightforwardly. Furthermore, the standard techniques that describe abelian gauge fields (whether chiral or not) in terms of differential forms miss all restrictions arising from a consistent coupling to matter in topologically non-trivial backgrounds. The formalism that naturally incorporates the holonomy of the $\text{U}(1)$ gauge bundle encoding a non-trivial topology describes $p$-form fields as classes in a (generalised) differential cohomology $\check{H}^{*}(X)$, known as Cheeger-Simons \cite{10.1007/BFb0075216} or, equivalently, Deligne cohomology \cite{Deligne}. Following \cite{Hsieh:2020jpj, Freed:2006yc, Freed:2006ya, Freed:2000ta, Hopkins:2002rd}, we can translate such a piece of information in terms of cochains, according to which an abelian $p$-form gauge fields is given by
\begin{equation}
    \check{\mathbf{A}}= ( A, N, F) \, \in  \, \check{H}^{p}(X)\, ,
\end{equation}
where $A \in C^{p}(X, \mathbb{R})$ (the space of real $p$-cochains) is the gauge connection, $N \in C^{p+1}(X, \mathbb{Z})$ defines the {\em characteristic class} and $F \in C^{p+1}(X, \mathbb{R})$ is the associated field strength. The non-trivial topological data are encoded in the characteristic integer cohomology element $[N] \in H^{p+1}(X,\mathbb{Z})$, which carries torsional information invisible to de Rham cohomology curvature classes $[F] \in H_{\text{dR}}^{p+1}(X)$. The two are connected via to the co-boundary of $A$
\begin{equation} \label{charclass}
     	N= F - \delta A \, ,
     \end{equation}
which however hides a gauge redundancy \cite{Hsieh:2020jpj}
      	\begin{equation} \label{gaugered}
      	A \to A + \delta a + n \, , \qquad 	N \to N - \delta n \, ,
      \end{equation}
with $ a \in C^{p-1}(X,\mathbb{R}) $ and $ n \in C^{p}(X, \mathbb{Z})$. This constrains the structure of the action and of the partition function, which should make sense at the level of differential cohomology, and, thus must be invariant under the transformations in \eqref{gaugered}. 

\noindent The situation however is substantially different depending on whether the $p$-form uplifted in differential cohomology is or is not chiral. For non-chiral forms, requiring that the action be invariant under the gauge transformations \eqref{gaugered} when $\check{\mathbf{A}}$ is electrically coupled to $\check{\mathbf{B}} \in \check{H}^{p+1}(X)$ and magnetically to $\check{\mathbf{C}} \in \check{H}^{D-p+1}(X)$ implies the presence of an additional SPT phase \cite{ Baez:1995xq, Freed:2007vy, Lurie:2009keu,  Freed:2016rqq, Schommer-Pries:2017sdd, Yonekura:2018ufj} (and local counter-terms) defined on $Y$.  As a result, one can write the partition function for the non-chiral $p$-form fields as
 \begin{equation}\label{eq:nonchiral_partition_function} 
 	\mathcal{Z}_{\check{\mathbf{A}}} (Y,L)= e^{2 \pi i (-1)^{D-p} \big ( \check{\mathbf{C}}, \check{\mathbf{B}} \big )_Y } \int \big [ D \check{\mathbf{A}} \big ] e^{-S[\check{\mathbf{A}},X]} \, ,
 \end{equation}
where we have introduced the cohomology pairing \cite{Hsieh:2020jpj}
	\begin{equation} \label{eq:cohomology_pairing}
		\big ( \check{\mathbf{A}}_1, \check{\mathbf{A}}_2 \big )_Y= \int_Y A_{\check{\mathbf{A}}_1 *\check{\mathbf{A}}_2} \, ,
	\end{equation} 
dictated by the product between differential characters\footnote{The present definition uses the notion of the {\em cup product} \cite{MR1867354} and the homotopy cochain $q(\omega_1,\omega_2)$ acting on differential forms $\omega_1$ and $\omega_2$ according to	
	\begin{equation}
	\omega_1 \wedge \omega_2 - \omega_1 \cup \omega_2 = q(\delta \omega_1, \omega_2)+ (-1)^{p_1} q(\omega_1, \delta \omega_2) + \delta q( \omega_1,  \omega_2) \, .
	\end{equation}}
\begin{equation}
	\begin{aligned}
	   &N_{\check{\mathbf{A}}_1 *\check{\mathbf{A}}_2}= N_{\check{\mathbf{A}}_1} \cup N_{\check{\mathbf{A}}_2} \, ,
	   \\
	   & F_{\check{\mathbf{A}}_1 *\check{\mathbf{A}}_2}= F_{\check{\mathbf{A}}_1} \wedge F_{\check{\mathbf{A}}_2} \, ,
	   \\
	   &A_{\check{\mathbf{A}}_1 *\check{\mathbf{A}}_2}= A_{\check{\mathbf{A}}_1} \cup F_{\check{\mathbf{A}}_2} + (-1)^{p_1+1} F_{\check{\mathbf{A}}_1} \cup A_{\check{\mathbf{A}}_2} + q(F_{\check{\mathbf{A}}_1}, F_{\check{\mathbf{A}}_2} ) \, .
	\end{aligned}
\end{equation}
The definition in \eqref{eq:nonchiral_partition_function} holds only when the electrically coupled field $\check{\mathbf{B}}$ is topologically trivial on $X$. Although such requirement seems rather restrictive, it actually corresponds to the unique available choice, since whenever $\check{\mathbf{B}}$ is non-trivial the partition function vanishes (see \cite{Hsieh:2020jpj} for details). Furthermore, in writing the expressions above we have extended the definitions of the background fields to $\check{\mathbf{C}} \in \check{H}^{D-p+1}(Y)$ and $\check{\mathbf{B}} \in \check{H}^{D-p+1}(Y)$, for which the topological triviality must hold only on the boundary $X$.
 
 \noindent The anomaly is thus given by
\begin{equation} \label{eq:anomaly_nonchiral}
    \mathcal{A}_{\check{\mathbf{A}}}(Y_{\text{cl}})= (-1)^{D-p} \big ( \check{\mathbf{C}}, \check{\mathbf{B}} \big )_{Y_{\text{cl}}} \, .
\end{equation}
In our context, the background fields are the combination of the first Pontyagin and the second Chern classes entering eq. \eqref{eq:Bianchi}. On Lens spaces, the fields are flat and thus the cohomology pairing depends only on the integer flux. However, requiring the twisted string structure to hold implies that at least one of the two fields is also topologically trivial, thus yielding a vanishing pairing.  

\noindent If the $p$-forms are instead chiral, we can describe the partition function of the $p$-form field as a boundary mode of a massive $p+1$-form field, similarly to what has been done for chiral fermions. The resulting expression is quite cumbersome and can be found explicitly in \cite{Hsieh:2020jpj} wich also provides a detailed derivation. For our purpose it is enough to notice that the expression for the anomaly theory of chiral $p$-forms coupled to a background field $\mathbf{\check{C}} \in \check{H}^{p+1}(Y)$ is given by 
\begin{equation}
    \mathcal{A}_{\check{\mathbf{A}}}(Y_{\text{cl}})= \widetilde{\mathcal{Q}}_{Y_{\text{cl}}}(\mathbf{\check{C}}) \equiv \mathcal{Q}_{Y_{\text{cl}}}(\mathbf{\check{C}}) - \mathcal{Q}_{Y_{\text{cl}}}(0) \, ,
\end{equation}
where we have introduced the (inhomogeneous) {\em quadratic refinement} $\mathcal{Q}$ \cite{Hopkins:2002rd} of the cohomology pairing, defined via the characteristic equation (see \emph{e.g.} \cite{Dierigl:2022zll} for an application in a physical context)
\begin{equation} \label{eq:quadrefcharacteristiceqinh}
    \mathcal{Q}_W( \mathbf{\check{A}}_1+ \mathbf{\check{A}}_2 ) -  \mathcal{Q}_W( \mathbf{\check{A}}_1 ) -  \mathcal{Q}_W(  \mathbf{\check{A}}_2 ) + \mathcal{Q}_W ( 0 )= \big ( \mathbf{\check{A}}_1,  \mathbf{\check{A}}_2 \big )_{W} \, ,
\end{equation} 
on a probe manifold $W$. In general, the choice of the quadratic refinement is dictated by a general solution of the characteristic equation, subject to the constraint that whenever $Y_{\text{cl}}= \partial Z$
\begin{equation}\label{eq:QW}
    Q_{\partial Z}(\mathbf{\check{C}})= \int_Z \left \{ \frac{1}{2} \left ( w + F_{\mathbf{\check{C}}} \right )^2 - \frac{1}{8} \hat{L} \left (R \right )\right \} \, ,
\end{equation}
where $\hat{L} \left (R \right )$ is the Hirzebruch polynomial that has already appeared in eq. \eqref{eq:OpactionWZ}. In eq. \eqref{eq:QW}, we have introduced an integral lift $w$ of the {\em Wu class} defined as the characteristic element of the cup product pairing \cite{Monnier:2018nfs}, corresponding in six dimensions to $w=- \frac{p_1}{4}$. In our case the background field $\mathbf{\check{C}}$ is given by the Chern character $\mathbf{\check{c}}$, whose curvature and characteristic classes are simply given by the second Chern class $c_2(F)$ as an element of de Rham and integer cohomology, respectively \cite{Hsieh:2020jpj}. In general, these requirements do not admit a unique solution and up to now it is unclear how to choose the right quadratic refinement systematically. Moreover, for the settings relevant for our discussion, chiral fields also take values in a lattice of signature $(1,n_T)$, depending on the number $n_T$ of tensor multiplets. Hence, the previous considerations have to be slightly modified. Specifically, chiral forms arise as boundary modes of $\check{H}^4(Y) \otimes \Lambda$ \cite{Michelangelo}, where $\Lambda$ is the $SO(1,n_T)$ lattice determined by the structure of local anomalies \eqref{factor-1}. In this scenario, the lattice-valued quadratic refinement on $Y_{\text{cl}}= \partial Z$ reads
\begin{equation}\label{eq:lattice_quadratic_refinement}
    Q_{\partial Z}^{\Lambda}(\mathbf{\check{c}})= \int_Z \left \{ \frac{1}{2} \left ( a_0 w +  b_0 F_{\mathbf{\check{c}}} \right )^2 -  \frac{1}{2} \sum_{i=1}^{n_T} \left ( a_i w +  b_i F_{\mathbf{\check{c}}} \right )^2 + \frac{n_T-1}{8} \, \hat{L} \left (R \right ) \right \} \, .
\end{equation}
Notice that \eqref{eq:lattice_quadratic_refinement} can be equivalently interpreted as the contribution of one antiself-dual and $n_T$ self-dual fields, as befits the traditional additive view on anomalies for chiral $p$-form fields.

\noindent Determining the quadratic refinement on a generic manifold $Y$ can be cumbersome and is uniquely fixed only for null-bordant manifolds, as described in \cite{Hsieh:2020jpj} for Spin-bordism. In the settings at stake, we do not know if the twisted string bordism group vanishes, nor if the relevant backgrounds actually lie in the trivial class, so $\mathcal{Q}_Y$ could {\em a priori} be different. Nevertheless, we can exploit the fact that in the purely gravitational case $\Omega_7^{\text{string}}=0$ to constrain $\mathcal{Q}_Y(0)$. Applying the APS index theorem to \eqref{eq:lattice_quadratic_refinement}, this gravitational contribution can be cast in the form\footnote{Strictly speaking, since the Bianchi class is either $\tfrac{p_1}{4}$ or $\tfrac{3}{4} p_1$, it is not obvious that the expression of $\mathcal{Q}(0)$ is unique as in the case of the standard string structure trivialising $\tfrac{p_1}{2}$.}
\begin{equation}\label{eq:quadratic_refinement_0}
    \mathcal{Q}_{\partial Z} ( 0)= \int_Z \left ( \frac{1}{2} \left ( \frac{a}{4} p_1  \right )^2-\, \frac{1}{8} \hat{L} \left (R \right ) \right ) = - \, \frac{7(35 a^2-3) }{8} \, \eta_D(\partial Z) +\frac{(a^2-1) }{8} \, \eta_{\text{grav}}(\partial Z) \, ,
\end{equation}
where $a$ is the relevant coefficient in the Bianchi classes in \eqref{twistedstringcond} and $\eta_{\text{grav}}$ is defined in \eqref{eq:etagrav}. Then, one can use the formulae in \eqref{eq:etas} to evaluate this expression on Lens spaces. However, turning on the gauge fields spoils the preceding observations and one can only consider a generic solution of the characteristic equation. On Lens  spaces, a general solution to the characteristic equation can be explicitly computed and can be parametrised by an integer $m=0, \ldots, 2p-1$ as \cite{Hsieh:2020jpj}
\begin{equation}\label{eq:tildeQ_lens}
    \widetilde{\mathcal{Q}}({\mathbf{\check{A}}} )= - \, \frac{a(a+m)}{2p} \, , 
\end{equation}
since the cohomology pairing on Lens spaces is
	\begin{equation}
		\big ( \mathbf{\check{A}}, \mathbf{\check{B}} \big )_{L_p^7}= - \, \frac{a b}{p} \, ,
	\end{equation}  
where $\left [ N_{\mathbf{\check{A}}} \right ]= a y$ and $\left [ N_{\mathbf{\check{B}}} \right ]= b y$, with $y$ a generator of the integral cohomology group.

\noindent In top-down examples, the choice of the quadratic refinement may not be arbitrary. For instance, in \cite{Dierigl:2022zll} it has been argued that F-theory selects a single possibility for $\mathbb{Z}_3$ gauge groups, although a general rule to determine $\widetilde{\mathcal{Q}}$ is not known. A complete analysis that describes the combination of quadratic refinements allowing to cancel the anomaly can be achieved in the simple setting of $\text{SU}(2)$ models, but in general one would need some restrictions on the possible quadratic refinements.

\noindent Now that we have specified all the ingredients of our anomaly theory, we can proceed to evaluate Dai-Freed anomalies for supergravity and string theory vacua. In \cite{Basile:2023zng}, the analysis has been carried out for supergravity with at most one tensor multiplet and simply-laced gauge groups. In the following, we shall only discuss the simple setting for the $\text{SU}(2)$ with $n_T=1$ and the abelian $\text{U}(1)$ gauge group without tensor multiplets, since the cases for all the other gauge groups have only increasing technical difficulties, without any additional conceptual novelty. We shall then discuss the compactification of the non-supersymmetric $\text{SO}(16) \times \text{SO}(16)$ heterotic theory on the $T^4/\mathbb{Z}_6$ orbifold and the Gepner orientifold with no tensor multiples, following the analysis in \cite{Basile:2023zng}.  

\section{Dai-Freed anomalies for \texorpdfstring{$\text{SU}(2)$}{SU(2)} in \texorpdfstring{$6D$}{6D} supergravity}\label{sec:anomaly_SU2}

In this Section we shall focus on six-dimensional supergravity with one tensor multiplet and $\text{SU}(2)$ as a gauge group. In such a setting, the contribution to the anomaly theory involves $3$ (the adjoint representation) vector multiplets, $d_\text{T}$\footnote{In the following the multiplicities of the various representations R will be denoted $d_{\text{R}}$.}  ($d_{\text{F}}$) hyper multiplets transforming into the trivial (fundamental) representation, the gravity multiplet and the neutral tensor multiplet. The latter two comprise, aside from the usual contribution coming from chiral fermions, $2$-form fields, which, as described in the previous Section, play a non-trivial role in the structure of the anomaly theory. The fermionic contribution is already highly constrained by local anomalies. Indeed, the cancellation of the irreducible purely gravitational piece of the anomaly polynomial implies that the number of trivial and fundamental hyper multiplets satisfy   
\begin{equation} \label{irrgravsu2}
	d_{\text{T}} +  2 d_{\text{F}} =247 \, . 
\end{equation}     
Notice that here we are looking at constraints for {\em would-be } UV complete theories with $\text{SU}(2)$ as the full gauge group. If $\text{SU}(2)$ is embedded into a larger gauge group, eq. \eqref{irrgravsu2} still holds by defining $d_T$ as a signed integer, where the negative contribution arises from the vector multiplets transforming into the adjoint representation of the turned-off gauge groups. Such a case would imply a much larger family to consider, and the control over these theories is lost. For the purpose of uncovering the role played by Dai-Freed anomalies in sharpening the string landscape or testing particular (top-down) examples, this consideration can be ignored. Therefore, taking into account the cancellation of the irreducible gravitational anomalies, the reducible anomaly polynomial takes the form
\begin{equation} \label{anpolsu2}
    I_8 = \left ( \text{tr} R^2 \right )^2- \frac{1}{24} \text{tr} R^2 \, \text{tr} F^2 \left ( d_{\text{F}}-4\right ) + \frac{1}{48} \left ( \text{tr} F^2 \right )^2  \left ( d_{\text{F}} - 16 \right ) , 
\end{equation}
where we used the decomposition $ \text{tr} \ F^4= \frac{1}{2} \left ( \text{tr} \ F^2 \right )^2$ valid for $\text{SU}(2)$. In order to cancel the reducible piece of the anomaly, we shall impose the factorisation \eqref{factor-1}, where however the implementation of $p$-form fields as elements of differential cohomology requires that the 't Hooft coefficients be integer-valued and embedded into a self-dual lattice \cite{Seiberg:2011dr}. With one tensor multiplet, such embedding is possible in two ways, depending on whether the metric $\Omega_{\alpha \beta}$ is either diagonal or off-diagonal. This translates into constraints on the allowed values of $d_T$ and $d_F$. For instance, the off-diagonal case,
	\begin{equation}
		\Omega=	\begin{pmatrix}
			0 & 1 \\
			1 & 0
		\end{pmatrix} \, ,
	\end{equation}
allows a factorisation of the anomaly polynomial if $d_{\text{F}}=4 + 12 s$, which yields
	\begin{equation}\label{poloffsu2}
		I_8 = \frac{1}{4} \left ( 2 \text{tr} R^2 -  \text{tr} F^2  \right )  \left ( 2 \text{tr} R^2 +  (1 - s)  \text{tr} F^2 \right  ) .	
	\end{equation} 
This factorisation determines the structure of the twisted string bordism group implied by the Bianchi identities
	\begin{equation} \label{bianchioffsu2}
		\begin{aligned}
			& d  H_1= \tfrac{1}{2} p_1 + c_2
			\, , \\
			& d  H_2= \tfrac{1}{2} p_1 +(s-1) c_2 \, .
		\end{aligned}
	\end{equation}  
\noindent We now build $\text{SU}(2)$ bundles on $L_p^7$ backgrounds by including the defining line bundle of $\mathbb{Z}_p$ into $\text{SU}(2)$. Following \cite{Debray:2023yrs}, there are two possibilities to describe such inclusion. Indeed, we can embed $\mathbb{Z}_p$ as a subgroup of $\text{U}(1)$ into $k$ diagonal blocks in the fundamental representation of $\text{SU}(2)$, which however only allows two values for $k$, namely $k=0,1$. For $k=0$ the gauge bundle is trivial and one only probes the gravitational anomaly via the uncharged $\eta$ invariants \eqref{eq:etas} 
\begin{eqaed}\label{eq:grav_anomaly}
    \mathcal{A}(L_p^7) = - \, \frac{p^4 + 11p^2 - 12}{3p} \, .
\end{eqaed}
Notice that in this case there is no quadratic refinement in general, but the only allowed backgrounds are ordinary string manifolds with $\tfrac{p_1}{2} = 0$. For Lens spaces this requires $p=2$, and $L_2^7 = \mathbb{R}P^7$ is the real projective 7-space. The resulting gravitational anomaly
\begin{eqaed}\label{eq:grav_anomaly_vanishes_off-diag}
    \mathcal{A}(\mathbb{R}P^7) &= -8 
    \\
    &= 0 \ \ \text{mod} \, 1
    \\
    &\equiv_1 0
\end{eqaed}
consistently with the fact that it is a bordism invariant of $\Omega_7^\text{string} = 0$. Therefore, in the following analysis of $n_T=1$ models, we will not consider trivial gauge bundles. Here and in the following we shall use the short-hand notation $\equiv_p$ to indicate the equivalence modulo $p$.

\noindent For $k=1$, we are embedding the $\mathbb{Z}_p$ subgroup into the Cartan subalgebra of $\text{SU}(2)$, and we can use the eigenvalues for the Cartan generator to deduce how the contribution of charged fermions decomposes into suitable combinations of $\eta_q$. The eta invariants $\eta_q$ are known and thus the anomaly for $k=1$ is given by
	\begin{equation} \label{ansu2Lp7off}
		\begin{aligned}
			\mathcal{A} \big (L_p^7 \big) &= d_{\text{F}} \left ( \eta_1 + \eta_{-1} \right )- \left ( \eta_2 + \eta_{-2}+\eta_0 \right ) + (d_T+1) \eta_0- \eta_{\text{grav}} 
             \\
             &\equiv_1 \frac{d_{\text{F}} (p^2-1) - 4p^4 + 40}{12p} \, .
		\end{aligned}
	\end{equation}

\noindent The Bianchi classes are now encoded in \eqref{bianchioffsu2}, and one must evaluate them on Lens spaces by using the known fact that $p_1/4=y$ generates the integer cohomology group $H^4(L_p^7, \mathbb{Z}) = \mathbb{Z}_p$, and the second Chern class is given by $c_2(L_p^7)=-k y$. This means that eq.\eqref{bianchioffsu2} evaluates to $(2-k)y$ and $(2-k(s-1))y$. For $k=1$, The only Bianchi class that can be trivialised on a Lens space is $(3-s)y$. Thus, setting $s = 3 + m p$ for some integer $m$ the Lens background has the appropriate twisted string structure, and one finds
\begin{eqaed} \label{eq:anomalyoff}
    \mathcal{A}(L^7_p) \equiv_1 - \, \frac{(p-1) \, p \, (p + 1)}{3} \, ,
\end{eqaed}
which also vanishes because the numerator is always divisible by three.

\noindent For the diagonal lattice, in which the bilinear form reads
	\begin{equation}
		\Omega=	\begin{pmatrix}
			1 & 0 \\
			0 & -1
		\end{pmatrix} \, ,
	\end{equation}
the number of fundamentals allowing an integral factorisation  of the anomaly polynomial has to be of the form
	\begin{equation} \label{su2famdiag}
		d_{\text{F}}=10 + 12 s \, .
	\end{equation}
For these values it is easy to see that the polynomial factorises according to 
	\begin{equation} \label{poldiagsu2}
		I_8 = \frac{1}{8} \left \{   \left ( 3 \text{tr} R^2 - s \text{tr} F^2  \right )^2 -  \left ( \text{tr} R^2 +  (1 - s)  \text{tr} F^2 \right  )^2 \right \}.	
	\end{equation} 
 
\noindent A complete study of Dai-Freed anomalies of such theories would require knowing the proper $\text{SU}(2)$-twisted string bordism group, where the twisted string structure is spelt out by the Bianchi identities
	\begin{equation} \label{bianchidiagsu2} 
		\begin{aligned}
			& d  H_1= \tfrac{3}{4} p_1 + s \ c_2 \, ,
			\\
			& d  H_2= \tfrac{1}{4} p_1 + (s-1) c_2 \, .
		\end{aligned}
	\end{equation} 
Following the analysis above, we restrict to Lens spaces $L_p^7$ which provide particularly simple families of candidate backgrounds to work with, even though we do not know if these exhaust all the nontrivial bordism representatives. 
	
\noindent We build $\text{SU}(2)$ bundles on $L_p^7$ backgrounds by including the defining line bundle of $\mathbb{Z}_p$ into $\text{SU}(2)$ according to the possible embeddings described before for $k=0,1$. For $k=0$, the anomaly theory reduces to the purely vanishing gravitational anomaly
\begin{eqaed}\label{eq:grav_anomaly_vanishes}
    \mathcal{A}(L_p^7) = - \, \frac{p^4 + 11p^2 - 12}{3p} + \mathcal{Q}_+(0) - \mathcal{Q}_-(0) = 0 \, .
\end{eqaed}
For $k=1$, charged fermions decompose into representations with different $\mathbb{Z}_p$ charges, for which the eta invariants $\eta_q$ are easily calculated. The anomaly for $k=1$ is thus given by
	\begin{equation} \label{ansu2Lp7diag1}
		\begin{aligned}
			\mathcal{A} \big (L_p^7 \big) \equiv_1 \frac{d_{\text{F}} (p^2-1) - 4p^4 + 40}{12p} + {\mathcal{Q}}_+ - {\mathcal{Q}}_- \, ,
		\end{aligned}
	\end{equation}
where we have added to \eqref{ansu2Lp7off} the contribution of the quadratic refinement for the two chiral $2$-forms that on Lens spaces is given by \eqref{eq:tildeQ_lens}.

\noindent We can now proceed to find the Lens spaces which satisfy the twisted string structure on which we are going to compute the relevant $\eta$-invariants and the quadratic refinement. For $k=1$ the Bianchi classes are $(3-s)y$ and $(2-s)y$ and one can show that for any Lens space trivialising one of the Bianchi classes, the fermionic anomaly is always an integer multiple of $1/p$. This means that there always exists an appropriate choice of $m_\pm$ in \eqref{eq:tildeQ_lens} that trivialises the anomaly theory. This is crucial, since the heterotic and F-theory landscapes include models in this family.  However, the choice of the quadratic refinement constrains the dynamics of chiral $p$-form fields in a very specific way, making other realisations of chiral forms inconsistent. In particular, if one uses the suitable combination of $\eta$ invariants reproducing locally the term \eqref{eq:lattice_quadratic_refinement} along the lines of \cite{Hsieh:2020jpj}, the anomaly cannot be cancelled.

\noindent Finally, there is a family of models that admits both the diagonal and off-diagonal embedding given by $d_{\text{F}}=16+24s$. For this family, we can evaluate the anomaly theory in the same way as it has been done for the off-diagonal model and indeed the anomaly theory can be shown to be given by \eqref{eq:anomalyoff}.

\noindent To summarise, $6D$ supergravity models with fermions in both trivial and fundamental representations give rise either to a vanishing anomaly if the $2$-form fields are non-chiral in generalised differential cohomology, or to an anomaly that a specific choice of the quadratic refinement can cancel. As shown in \cite{Basile:2023zng}, the discussion follows a similar pattern and leads to the same conclusions if the set-up is modified by considering $\text{U}(1)$, $\text{SU}(n)$, $\text{Spin}(2n)$, $E_7$ and $E_8$ gauge groups with hyper multiplets in fundamental, adjoint, symmetric and anti-symmetric representations, still under the assumption of the presence of one tensor multiplet. We can perform a similar analysis even for the case of no tensor multiplets that can be studied by considering the sample of models given by the infinite families of \cite{Taylor:2018khc}.

\section{Dai-Freed anomalies for \texorpdfstring{$\text{U}(1)$}{U(1)} in \texorpdfstring{$6D$}{6D} supergravity}\label{sec:anomaly_U1}

In this Section we are going to focus on $\text{U}(1)$ models with no tensor multiplets, that are known to give rise to different infinite families satisfying all the known swampland criteria but for which no string or F-theory realisation is known \cite{Taylor:2018khc}. Without tensor multiplets there is only one possible embedding of the anomaly which unavoidably entails chiral $2$-forms. This means that the choice of the quadratic refinement is crucial in order to address the consistency of these theories, and indeed we are going to argue that the anomaly theory is eventually non-trivial for large charges, thus excluding all but a finite number of these models. Actually, it is always possible to choose a quadratic refinement that cancels the anomaly but, to date, we are still lacking a top-down mathematical principle selecting a suitable $\mathcal{Q}$\footnote{It can be also argued that the presence of families with arbitrary large charge increase the effective gauge coupling $g_\text{eff} = gQ$, which in $D=6$ has dimensions of length, and it can be observed to violate perturbatively the (upper bound to the) quantum gravity cutoff $\Lambda_\text{QG}$ determined by the magnetic weak gravity conjecture \cite{Arkani-Hamed:2006emk, Harlow:2022ich}. Intuitively, this may suggest that large charges are obstructed in weakly coupled gauge theories when quantum gravity is involved.}.

\noindent As we mentioned, without tensor multiplets there is only one possible lattice embedding with $a=3$, and the factorised anomaly polynomial is
\begin{eqaed} \label{0tensorI8}
	I_8&= \tfrac{1}{2} \left ( \tfrac{3}{4} p_1 + \tfrac{b}{2} c_1^2  \right )^2 \, .
\end{eqaed}    
In \cite{Taylor:2018khc} a family of charges $q,r,q+r$ admitting such factorisation was found, with fixed multiplicities
\begin{equation}\label{eq:u1_no_tensor_fam}
	\begin{aligned}
		d_q= 54 \, , \qquad d_r= 54 \, , \qquad d_{q+r}= 54 \, .
	\end{aligned}
\end{equation}
Here $q,r$ are integers determining the anomaly coefficient $b=- 6( q^2 +r q + r^2)$.

\noindent The anomaly theory for these models is described by
\begin{equation} \label{an0T}
\begin{aligned}
	\mathcal{A} \big (Y \big ) =& 27 ( \eta_q + \eta_{-q}- 2 \eta_0) +  27 ( \eta_r + \eta_{-r}- 2 \eta_0)
 \\
 &+  27 ( \eta_{q+r} + \eta_{-q-r}- 2 \eta_0) -\frac{b(b+2m)}{8p} \, ,
 \end{aligned}
\end{equation}
where we have considered the general form of the quadratic refinement in terms of an integer parameter $m$\footnote{Notice that the gravitational anomaly is cancels out because of the definition of the quadratic refinement and the triviality of the corresponding string bordism group.}. Using the expressions contained in \eqref{eq:etas}, we can compute the value of \eqref{an0T} for Lens spaces $L_p^7$ with $p$ a divisor of $| 3-3( q^2 +r q + r^2)|$, which trivialises the Bianchi class. For general $m$ and $p$, one has
\begin{eqaed}\label{eq:anom_abelian_no_tensors}
    \mathcal{A}(L_p^7) \equiv_1 \frac{\beta (p^2+3m-18)}{2p} \, ,
\end{eqaed}
where $\beta \equiv -b/6 = q^2+q r+r^2$ is the only combination of charges that appears. This is also true for the Bianchi class, which is $3(1-\beta)y$. For $p=3$ one has 
\begin{eqaed}
    \mathcal{A}(L_3^7) \equiv_1 \frac{\beta (m-3)}{2} \, ,
\end{eqaed}
which vanishes either for even $\beta$ ($q$ and $r$ even) or $m = 3$ (mod $2$). For general $\beta$, $p=\beta-1$ remains a valid choice, and we now study this case for large charge. We restrict to choices of quadratic refinement such that $m$ is bounded (mod $2p$), and show that all but finitely many theories are excluded with this assumption unless $m=6$. For this family of models the choice of \cite{Hsieh:2020jpj} amounts to $m = p^2-2$, which for $p$ even is bounded mod $2p$ for $p = \beta-1 \gg 1$. For $p$ odd, $m \equiv_{2p} p-2 = \beta - 3$ with $\beta$ even, so for this choice $m$ is not bounded and a separate analysis is needed.

\noindent To begin with, notice that if $(q,r) \to \infty$ in $\mathbb{R}^2$ then $\beta \to +\infty.$ Indeed,
\begin{eqaed}
    2\beta - |(q,r)|^2 = (q+r)^2 \geq 0 \; \Longrightarrow \; \beta \geq \frac{1}{2} \, |(q,r)|^2 \to +\infty \, .
\end{eqaed}
For $m=6$ the anomaly always vanishes. Assuming that $m \neq 6$ and bounded (taken between $0$ and $2p-1$), letting $p = \beta-1$ one has
\begin{eqaed}
    \mathcal{A}(L_{\beta-1}^7) & \overset{\beta \gg 1}{\sim} \frac{\beta \, \left(\beta^2-\beta+3(m-6)\right) + 3(m-6)}{2\beta} + \frac{3(m-6)}{2\beta^2} \\
    & \equiv \frac{\beta \, N(\beta) + N_0}{2\beta} + \frac{N_0}{2\beta^2} \, ,
\end{eqaed}
where for our purposes we only need use that $N(\beta) \in \mathbb{Z}[\beta]$ is a polynomial with integer coefficients and that $N_0 \in \mathbb{Z}$ is an integer. The first term cannot be integer for large enough $\beta$, since the numerator cannot be a multiple of $\beta$, for $\beta > N_0$. The second term is subleading for large $\beta$, and thus cannot compensate the fractional part since the series is asymptotic. Furthermore, having packaged the divergent part in the first term, all other terms resum to a finite result, namely the limit of \eqref{eq:anom_abelian_no_tensors} (as a rational number rather than mod 1) with its divergent contribution subtracted. All in all, in order to cancel the anomaly on $L_{\beta-1}^7$ one necessarily needs $m=6$, whereas for $p=3$ only even charges survive unless $m=3$ mod 2. If $m$ does not depend on the choice of background, one can save at most even charges, if any.

\noindent In order to conclude our analysis for the choice of quadratic refinement of \cite{Hsieh:2020jpj}, for even $\beta \equiv 2z$ and $p=\beta-1$, the parameter $m = p^2-2 \equiv_{2p} p-2 = 2z-3$ is not bounded for $z \gg 1$. The anomaly simplifies to
\begin{eqaed}
    \mathcal{A}(L_{2z-1}^7) \equiv_1 \frac{2z^2+z-13}{2z-1} \equiv_1 - \, \frac{12}{2z-1} \, ,
\end{eqaed}
which once more shows that the anomaly is eventually non-vanishing.

\noindent The result just described is confirmed by a numerical scan. Indeed, fixing the quadratic refinement to the choice $m=p^2-2$, which corresponds to the definition in \cite{Hsieh:2020jpj}, only $6$ models out of $441$ are anomaly-free\footnote{There are additional models for which the analysis cannot be performed, since they would formally yield $p=0$ as valid backgrounds. These correspond to $(q,r)=\{ (-1,0), (0,-1), (1,0), (0,1), (1,-1), (-1,1) \}$.}. The charges are allowed to vary with $|q|\leq 10$ and $|r|\leq 10$, but the non-anomalous models are bounded by $|q|=2$ and $|r|=2$, as listed in table \ref{u1family0TY}.
 \begin{table}[] \centering
 	\begin{tabular}{ |p{1.4 cm}|} 
 		\hline
 		$(r,s)$
 		\\
 		\hline
        $(-2, 1)$
        \\
        $(-1, -1)$
        \\
        $(-1, 2)$
        \\
        $(1, -2)$
        \\
        $(2, -1)$
        \\
        $(1,1)$
 		\\
 		\hline
 	\end{tabular}
 	\caption{Anomaly-free families of the $\text{U}(1)$ models in \eqref{eq:u1_no_tensor_fam} with no tensor multiplets in terms of $q,r$ with $n_T=0$ and $|q| \, , \, |r| \leq 10$, choosing $m=p^2-2$.}
 	\label{u1family0TY}
 \end{table}
 
\noindent To summarise our findings, without a mechanism indicating which specific quadratic refinements are selected in string or F-theory, we cannot place these theories in the swampland with certainty. N\"aively one can expect to exclude most of them, since only a very specific choice for each background cancels all the anomalies that we have investigated. It would be interesting to further explore this issue, trying to deduce a top-down criterion to select the allowed quadratic refinement(s). Nonetheless, we now know that, unless $m = 6$, no choice with $m$ bounded can cancel the anomaly on $L_{\beta-1}^7$ for infinitely many charges, while the anomaly on $L_3^7$ only saves even charges unless $m=3$ mod 2.

\section{The Gepner orientifold with no tensor multiplets}\label{sec:gepner}

The importance and role played by the quadratic refinement can be further investigated from the allowed top-down string theory realisations. Gepner \cite{Gepner:1987qi} has shown that consistent superstring world-sheet theories in a background with $\text{SU}(n)$ holonomy can be realised in terms of tensor products of $\mathcal{N}=2$ superconformal minimal models, corresponding to special points of the moduli space of a compactification on Calabi-Yau manifolds $\text{CY}_n$. Focusing on type IIB strings compactified in $D=6$ on $\text{CY}_2$, the chiral spectrum is unique and comprises the gravity multiplet and twenty-one tensor multiplets with $\mathcal{N}=(2,0)$ spacetime supersymmetry. The orientifold algorithm \cite{Sagnotti:1987tw, Pradisi:1988xd, Horava:1989vt, Bianchi:1990yu, Bianchi:1990tb, Bianchi:1991eu, Dudas:2000bn, Angelantonj:2002ct} then allows to halve the number of supercharges to $\mathcal{N}=(1,0)$, but the light spectra depend on the point of the moduli space in which the model is constructed. In the following, we will focus on the orientifold projection of the Gepner model with $81$ characters worked out in \cite{Angelantonj:1996mw}, comprising ten hyper multiplets in the antisymmetric representation of the Chan-Paton gauge group $\text{SO}(8)$ from the open-string sector and twenty-one neutral hyper multiplets from the closed-string sector. Our interest in this model is motivated by the absence of tensor multiplets, which highlights the single chiral $2$-form contained in the gravity multiplet.

\noindent The full anomaly theory for a generic closed manifold $Y$ contains a single quadratic refinement, and takes the form
\begin{equation} \label{angepnerY}
		\mathcal{A} \big ( Y \big )= (10-1) \, \eta_{\mathbf{28}} (Y) + 21 \, \eta_{\mathbf{0}} (Y) - \eta_{\text{grav}} (Y) + \mathcal{Q}(Y) \, .
\end{equation}
On boundaries $Y=\partial Z$, \eqref{angepnerY} reproduces the anomaly polynomial
	\begin{equation} \label{angepner}
		\begin{aligned}
			\mathcal{A} \big ( \partial Z \big ) &\equiv_1 - \int_Z I_8 = - \frac{9}{8} \int_Z \left ( \text{tr} R^2 - \text{tr} F^2\right )^2 \, ,
			\end{aligned}
	\end{equation}
from which one reads the Bianchi identity
\begin{equation} \label{bianchigepner}
		d H= 3 \left ( \tfrac{1}{4} p_1 + c_2 \right ) = X_4 \, .
	\end{equation}
The Green-Schwarz-Sagnotti term cancelling \eqref{angepner} arises from the quadratic refinement $ \mathcal{Q}_{Y} ( 3 \mathbf{\check{c}})=\widetilde{\mathcal{Q}}_Y( 3 \mathbf{\check{c}}) + \mathcal{Q}_Y(0) $, contributing to the gravitational anomaly as
\begin{equation}
    \mathcal{Q}_{Y} ( 0)= \int_Z \left \{ \frac{1}{2} \left ( \frac{a}{4} p_1  \right )^2-\, \frac{1}{8} L \right \} = - \, \frac{7(35 a^2-3) }{8} \, \eta_D(Y) +\frac{(a^2-1) }{8} \, \eta_{\text{grav}}(Y) \, ,
\end{equation}
where $a=3$ in the present case. Thus the only unknown contribution is $\widetilde{\mathcal{Q}}_Y( 3 \mathbf{\check{c}})$, which solves \eqref{eq:quadrefcharacteristiceqinh} and can be parametrised as in \eqref{eq:tildeQ_lens} on Lens spaces. To evaluate the Bianchi identity, we need to specify the second Chern class for $\text{Spin}(2n)$ bundles, which corresponds to $c_2(L_p^7)=- 2k y$ \cite{Basile:2023zng}, with $k=0,1,2$ parametrising the allowed bundles. Eq. \eqref{bianchigepner} singles out Lens spaces $L_p^7$ with  $p=3$, $p = 1 - 2 k$ or $p = 3(1 - 2k)$. Thus, $p=3 \, , 9$ are allowed, depending on the value of $k$. In these cases, the anomaly in \eqref{angepnerY} evaluates to 
\begin{equation}\label{eq:an_gepner_lens}
\begin{aligned}
    \mathcal{A} (L_p^7) &= 9 (k (2 k - 1)(\eta_1+ \eta_{-2}) + 4 k (4 - 2 k)( \eta_1+ \eta_{-1})  + \widetilde{\mathcal{Q}}_Y( 3 \mathbf{\check{c}})
    \\
    & \qquad + ( 18 (k (2 k - 1) + 4 k (4 - 2 k))  28 \cdot 9 + 21 - 7 \cdot 39 )\eta_0
    \\
    & = \frac{3k}{p} \left( m + 3 (p^2 - 2kp + p - 2) \right) 
    \\
    &\equiv_1 \frac{3k}{p} \, (m - 6) \, ,
\end{aligned}
\end{equation}
where the quadratic refinement is parametrised as in \eqref{eq:tildeQ_lens},
\begin{equation}\label{eq:tildeQ_gepner}\widetilde{\mathcal{Q}}_Y( 3 \mathbf{\check{c}})= - \frac{-6 k(-6k+m)}{2p} \, , \qquad \text{with} \ \ m=0,\ldots, 2p-1 \ \ \text{mod} \ \ 2p \, .
\end{equation}
Choosing $p=3$, valid for all $k$, manifestly trivialises \eqref{eq:an_gepner_lens}, while for $k=2$ and $p=9$ one finds
\begin{equation}
   \mathcal{A} (L_9^7) \equiv _1 \frac{2}{3} \, m \, ,
\end{equation}
which cancels only for $m=0$ mod 3. This is not the value of $m$ one finds for the choice in \cite{Hsieh:2020jpj}, corresponding to $m=2$. All in all, we have shown that the anomaly vanishes in this model for a specific choice of quadratic refinement for all the allowed $k$ and $p$, although the rationale behind this choice remains obscure.

\section{Non-supersymmetric heterotic models}\label{sec:non-susy_models}
	
In the previous Sections we have discussed the role played by Dai-Freed anomalies as a consistency swampland criterion for six-dimensional $\mathcal{N}=(1,0)$ supergravity theories with simply laced and abelian gauge groups. The reason why one could call this a swampland condition lies in the origin of Dai-Freed anomalies, which require spacetime topology change. It is natural to think of this as an intrinsically quantum-gravitational effect, whereby cancellation of such anomalies may be unnecessary if the theory is not coupled to gravity. 

\noindent The analysis performed in \cite{Basile:2023zng} of supergravity theories in $D=6$ involve $K3$ compactification in the orbifold limit of the heterotic $\text{SO}(32)$ \cite{Honecker:2006qz} and $E_8 \times E_8$ \cite{Walton, Honecker:2006qz} theories, for which it has been shown that anomalies are cancelled or there exists a specific choice of the quadratic refinement to guarantee it. This result is neither new nor surprising, since \cite{Tachikawa:2021mby} has shown that anomalies of this kind are always absent in heterotic constructions when supersymmetry is present. 

\noindent However, an analogous general result is not available for non-supersymmetric vacua\footnote{To our knowledge, neither for supersymmetric and non-supersymmetric orientifold vacua a similar result is available. Hence, also the possibility to cancel anomalies for the Gepner orientifold was not guaranteed.} and thus there is no \emph{a priori} guarantee that Dai-Freed anomalies would cancel for these theories as well. Nevertheless, the cancellation of Dai-freed anomalies is connected to the consistency under topology change and thus makes no reference to supersymmetry. Hence, it is important to check their absence in non-supersymmetric settings. This program has been initiated in \cite{Basile:2023knk}, where ten-dimensional tachyon-free models have been shown to be free of Dai-Freed anomalies. This also implies that any smooth geometric compactification thereof is anomaly-free, but, in principle, lower-dimensional vacua may arise from different constructions. In light of these considerations, the purpose of this section is to verify that Dai-Freed anomalies on Lens spaces cancel for certain non-supersymmetric heterotic orbifolds in six dimensions. For concreteness, we shall discuss $\text{SU}(2)$ anomalies on Lens spaces for the $\text{SO}(16)\times \text{SO}(16)$ heterotic model\footnote{The global form of the gauge group is not in fact $\text{SO}(16) \times \text{SO}(16)$, but this subtlety will be immaterial for our considerations.} compactified on $K3$ in its orbifold limits. For simplicity we shall study orbifolds \cite{Dixon:1985jw,Dixon:1986jc} whose point group $P$ is a discrete subgroup of the $\text{SU}(2)$ holonomy of $K3$, and particular emphasis will be put on the $T^4/\mathbb{Z}_6$ orbifold realisation\footnote{The complete analysis can be found in \cite{Basile:2023zng}.}. In addition, in the following we shall also restrict our attention to the standard embedding, in which the gauge connection on one of the $\text{SO}(16)$ factors is identified with the spin connection, so that the orbifold action on the gauge world-sheet fermions and on those on space-time are identical. Since in the settings at stake supersymmetry is absent, these restrictions are not imposed by general principles. Rather, they are chosen for convenience and it is easy to relax these requirements to investigate other corners of the landscape. With this setup, the point group action on the world-sheet fermions is dictated by the shift vectors
\begin{equation} \label{vectshifts}
    v_{\text{st}}=\frac{1}{6} (1,-1) \, , \qquad	v_{\text{gauge}}= \frac{1}{6} (0^6,1,-1) \otimes (0^8) \, ,
\end{equation}        
for spacetime and gauge degrees of freedom respectively, acting on world-sheet fermions and bosons via
\begin{equation} \label{pointgroupaction}
	g \cdot \psi^i_R= e^{2 \pi i v_{\text{gauge}}^i} \psi^i_R \, , \quad g \cdot \psi^i_L = e^{2 \pi i v_{\text{st}}^i} \psi^i_L \, ,  \qquad \text{and} \qquad 	g \cdot z^i= e^{2 \pi i v_{\text{st}}^i} z^i \, , 
\end{equation}
where $\psi^i_{L,R}$ denotes the complex combinations of Majorana spinors determined by the complex structure on the $i$-th torus $T^2$ parametrised by $z_i$. With this choice of shift vectors, the point group yields a complex action on world-sheet fermions and bosons, preventing the gauge group enhancement $\text{U}(1) \to \text{SU}(2)$ from taking place, as in the case of $\mathbb{Z}_2$. This means that the breaking of the gauge group is maximal and reflects a simple $K3$ compactification \cite{Walton}, 
\begin{equation} \label{gaugegroupzN}
	\text{SO}(16) \times \text{SO}(16) \to \text{SO}(12) \times \text{SU}(2) \times \text{U}(1) \times \text{SO}(16) \, .
\end{equation} 
Furthermore, the $\text{SO}(4)$ characters are no longer eigenvectors of the orbifold action, and indeed the corresponding partition function can only be expressed in terms of $\text{SU}(2)$ and $\text{U}(1)$ characters. However, there is no ``algorithmic'' procedure to determine the level of the affine $\text{U}(1)$ algebra, and thus the characters appearing in the partition function, which means that we can only determine the $\text{SU}(2)$ representations of the string states from the the $q$-expansion of the modular blocks, leaving the $\text{U}(1)$ charges undetermined. This difficulty can be ignored if our aim is simply to study Dai-Freed anomalies for the $\text{SU}(2)$ gauge group, in which the $\text{U}(1)$ factor in \eqref{gaugegroupzN} is turned off. In any case, a more complete discussion would require computing the relevant twisted string bordism group.

\noindent Following the partition function reported in \cite{Basile:2023zng}, it is possible to the deduce the light spectrum, which gives the gauge boson in the adjoint of $\text{SO}(12) \times \text{SU}(2) \times \text{SO}(16) \times \text{U}(1)$, the graviton, the non-chiral Kalb-Ramond field and the dilaton in the singlet of the gauge group, eight scalars in the representation $(\mathbf{1},\mathbf{1},\mathbf{1})$, four scalars in the representation $(\mathbf{12},\mathbf{2},\mathbf{1})$, a doublet of left-handed fermions in the representation  $(\mathbf{32}_s,\mathbf{2},\mathbf{1}) \oplus (\mathbf{1},\mathbf{1},\mathbf{128}_s) \oplus (\mathbf{1},\mathbf{2},\mathbf{16}) $ and a doublet of right-handed fermions in the representation $(\mathbf{32}_c,\mathbf{1},\mathbf{1}) \oplus (\mathbf{12},\mathbf{1},\mathbf{16})$. From the twisted sectors we have instead two hundred and sixteen scalars in the representation  $(\mathbf{1},\mathbf{1},\mathbf{1})$, thirty-six scalars in the representation  $(\mathbf{12},\mathbf{2},\mathbf{1})$, nine doublets of left-handed fermions in the representation  $(\mathbf{1},\mathbf{2},\mathbf{16})$ and nine doublets of right-handed fermions in the representation $(\mathbf{32}_c,\mathbf{1},\mathbf{1})$.

\noindent Form the light spectrum the anomaly theory can be straightforwardly deduced, yielding 
\begin{equation} \label{feranso16xso16Z6}
	\begin{aligned}
		\mathcal{A}_{\text{fermions}} \big ( Y \big )&= \eta^D_{(\mathbf{32}_c,\mathbf{1},\mathbf{1})} \big ( Y \big )+ \eta^D_{(\mathbf{12},\mathbf{1},\mathbf{16})} \big ( Y \big ) + 9 \, \eta^D_{(\mathbf{32}_c,\mathbf{1},\mathbf{1})} \big ( Y \big )
		\\
		&- \eta^D_{(\mathbf{1},\mathbf{1},\mathbf{128}_s)} \big ( Y \big )- \eta^D_{(\mathbf{32}_s,\mathbf{2},\mathbf{1})} \big ( Y \big )- \eta^D_{(\mathbf{1},\mathbf{2},\mathbf{16})} \big ( Y \big )- 9 \, \eta^D_{(\mathbf{1},\mathbf{2},\mathbf{16})} \big ( Y \big ) \, ,
	\end{aligned}
\end{equation} 
which locally reproduces the corresponding factorised anomaly polynomial \begin{equation} \label{polso16xso16z3}
\begin{aligned}
	I_8 &= \frac{1}{2}\left ( 2 \ \text{tr} R^2  -  \tfrac{1}{2} \  \text{tr} F_1^2  -  \text{tr} F_2^2 - \tfrac{1}{2} \ \text{tr} F_3^2   \right ) \left ( \text{tr} F_3^2 - 2 \  \text{tr} F_1^2  + 8 \  \text{tr} F_2^2  \right )
 \\
 &= \frac12 \, X_4^1 \wedge X_4^2\, ,
\end{aligned}
\end{equation}
as required by the Green-Schwarz-Sagnotti mechanism. The above factorisation thus tells us that the bosonic piece of the anomaly arises from a non-chiral $2$-form field and it is encoded in the cohomology pairing \eqref{eq:anomaly_nonchiral} 
\begin{equation} \label{bfieldsan}
	\mathcal{A}_{\text{B-fields}} \big ( Y \big )= ( \check{\mathbf{A}}^1, \check{\mathbf{A}}^2)_Y \, ,
\end{equation} 
where $\check{\mathbf{A}}^i=(  N^i, A^i, X_4^i )$ are the Cheeger-Simons characters with $ [N^i]= [ X_4^i ]_{\mathbb{Z}}$ and $ [X_4^i]= [ X_4^i ]_{\text{dR}}$ \cite{Hsieh:2020jpj,Freed:2000ta,Freed:2006ya,Freed:2006yc}. Thus, the complete expression for the anomaly theory on a general background is given by
\begin{equation} \label{anomaly}
	\mathcal{A} \big (Y \big )= \mathcal{A}_{\text{fermions}} \big ( Y \big ) -  ( \check{\mathbf{A}}^1, \check{\mathbf{A}}^2)_Y \, .
\end{equation}
If the $i=1$ character is topologically trivial, {\em i.e.} $ [X_4^1]_{\mathbb{Z}} =0$, the coupling reduces to the known Chern-Simons coupling $A^1 \wedge X_4^2$, and thus, when $Y=\partial Z$, this contribution reproduces the Green-Schwarz-sagnotti term $X_4^1 \wedge X_4^2$. Requiring that the differential characters $\check{\mathbf{A}}^i$ be topologically trivial entails a twisted string structure on $Y$, since at the level of de Rham cohomology one obtains the Bianchi identities
\begin{equation} \label{z3pol}
	\begin{aligned}
		& X_4^1= \tfrac{1}{2} p_1 + \tfrac{1}{2} c_2(F_1)+ c_2(F_2) + \tfrac{1}{2} c_2(F_3) \, ,
		\\
		&X_4^2= 4 \,  c_2(F_1) - 16 \, c_2(F_2)  -2 \, c_2(F_3) \, .
	\end{aligned}
\end{equation}
One can choose to trivialise either class at the integral level, but of course the consistency of the model should not depend on this choice. The corresponding twisted string bordism groups ought to classify equivalent anomaly backgrounds.

\noindent We now study anomalies on Lens spaces, turning off all the groups aside from the first $\text{SU}(2)$\footnote{Turning on only the second $\text{SU}(2)$ makes the second Bianchi identity trivial. However, this forbids the Lens space as a valid background for any $p$, as will be explained in the following.}. In this case $L_p^7$ trivialises $X_4^2$ for $p=2,4,8,16$, since the Pontryagin and Chern classes are $ \frac{1}{2} p_1= 2y$ and $c_2=-y$ with $y$ a generator of degree-four cohomology. On these backgrounds, the contribution from \eqref{bfieldsan} vanishes, as in the off-diagonal supersymmetric models, and thus the anomaly is simply given in terms of the net number of fundamentals for the first $\text{SU}(2)$, $d_{\text{F}}=12 \cdot 16$ as in \eqref{ansu2Lp7off}. As a result,
\begin{equation} \label{anlensz2}
	\begin{aligned}
		\mathcal{A} \big ( L_p^7 \big ) = 12 \cdot 16 \, \tilde \eta_1  = 12 \cdot 16 \,  \frac{p^2-1}{12 p} \equiv_1 0 \, ,
	\end{aligned}
\end{equation}
for our specific values of $p$.
It is worth noting that the available backgrounds are the only ones for which the anomaly vanishes, while it would not have been the case for any other value of $p$.

\chapter{Conclusion} \label{conclusion}

\newpage 

\section{Summary and outlook} \label{sec:conclusions}

\noindent In the present dissertation we have analysed various aspects of the string landscape. Particular emphasis has been devoted to the study of tachyonic instabilities and the dynamics of (non-)supersymmetric string vacua, along with unitarity conditions induced by the completeness hypothesis and the consistency under topology change. This latter issue has been addressed both by looking at the consistency of the low-energy supergravity models and by studying explicit examples of string theory vacua, even without supersymmetry.

\noindent In Chapter \ref{ChMisSUSY}, we have addressed the absence of tachyons, of non-supersymmetric string vacua in various dimensions in connection with the large-mass behaviour of the net number of degrees of freedom encoded in the sector-averaged sums associated with one-loop amplitudes. When restricting to the closed-string case, we have shown that the latter depends on the mass of the lightest state, independently of whether it is a tachyon or massless particle, and is strictly smaller than $C_\text{tot}$ in the presence of fermions. This result allowed us to show that the necessary and sufficient condition for classical stability is $C_\text{eff} =0$ (and $\langle d (n) \rangle ({\mathcal T})$), as conjectured in \cite{Dienes:1994np}. 

\noindent The analysis performed in this context is based on the analytic continuation of the degrees of freedom obtained crucially continuing the physical discrete masses to real values. This approach is different from those used in  \cite{Kutasov:1990sv} and \cite{Angelantonj:2010ic} which, using number theoretic methods, allowed to connect the behaviour of the signed sum over the physical spectrum with the vacuum energy, assumed to be finite. It would be interesting to connect misaligned supersymmetry and the latter approach.

\noindent For orientifold vacua, the situation changes dramatically. In this scenario, the exponential growth of the number of degrees of freedom of the one-loop partition functions is related to infrared divergences of the tree-level amplitudes, and {\em vice versa}. This is not surprising, since this UV/IR interplay descends from the dual interpretation of the Klein bottle, annulus and M\"obius strip amplitudes as loop or tree-level diagrams. This allowed us to interpret misaligned supersymmetry associated with the loop channel as the decoupling of closed-string tachyons from D-branes and O-planes. Hence, it provides a {\em necessary}, but not sufficient, condition for their absence from the physical spectrum. Similarly, the vanishing of the sector-averaged sums associated with both the torus and transverse annulus amplitudes is a {\em sufficient}, but not necessary, condition for the absence of tachyons, since the Klein bottle and M\"obius strip amplitudes only implement the orientifold projection. However, such a condition is not necessary, since tachyons might be removed through a non-trivial orientifold projection. In these cases, misaligned supersymmetry would require non-trivial cancellations among amplitudes with different topologies. However, the sector-averaged sums for the torus and the tree-level $\Tilde{\mathcal K}$, $\Tilde{\mathcal A}$ and $\Tilde{\mathcal M}$ amplitudes are built on CFTs on different Riemann surfaces, and thus carry different information. As a result, it appears difficult, if not impossible, to compare them. 

\noindent In this sense, {\em misaligned supersymmetry} cannot uncover the reason behind the IR finiteness of the vacuum energy for general orientifold vacua, and the question is still open. New ideas and technologies need to be developed to fully address this issue.

\noindent In Chapter \ref{ChBSB}, we have focused our attention on a class of classically stable vacua in six dimensions, in which supersymmetry is realised non-linearly through the insertion of orientifold planes of positive tension and charge. In particular, for K3 orientifolds, the orientifold action is dressed with an involution that introduces O5$_+$ planes, and thus requires  $\overline{\text{D5}}$ branes for tadpole cancellation. In this set-up, known as  Brane Supersymmetry Breaking, we have explicitly presented the orientifold of $T^4/\mathbb{Z}_6$ orbifold admitting a non-trivial solution of the tadpole conditions, which forbids further recombination of branes. The situation is actually different in the supersymmetric case, where, regardless of the choice of the specific solution to R-R tadpole conditions, it is always possible to deform the model by moving brane into the bulk, as shown for instance for the $T^4/\mathbb{Z}_6$ example.

\noindent Furthermore, we have discussed the implications of the introduction of string defects coupled to the R-R $2$-forms of the bulk super multiplets. Such defects couple to a combination of tensor-multiplet scalars, denoted as $J$, in such a way that the tension of the defect and the contributions to the energy-momentum tensor from the kinetic terms of scalars and gauge fields are positive, thus avoiding the presence of ghosts.
Although for the supersymmetric case the existence of such $J$-form has been shown since the '90s \cite{Romans:1986er,Sagnotti:1992qw, Sagnotti:1996qj, Riccioni:1998th}, doubts on the possibility of extending its definition to BSB vacua were raised in the literature \cite{Angelantonj:2020pyr}. In the present thesis, we have uncovered its stringy origin, determining its correct expression in the case of BSB vacua. Once the $J$ form has been defined, it is possible to couple defects with positive tension that are also required to support a microscopic unitary theory. This translates into non-trivial conditions for the central charges of the Ka$\check{\text{c}}$-Moody algebras realising the D-branes gauge groups. In general, the algebra is realised on either the left or right-moving sector, but whenever the defects are instantons of a gauge group of the bulk theory, the algebra can be realised on both, implying additional constraints. The bottom-up analysis known up to now in the literature \cite{Kim:2019vuc} only covers the first case when one defect is considered, and there is no model-independent study describing instanton strings or situations in which more coincident defects are involved. However, since for orientifold vacua the microscopic theory living on string defects is known, it has been possible to evaluate these conditions explicitly, and we have found a correlation between the presence of scalars in the string/five brane-amplitudes and the interpretation of the defects as the corresponding gauge instantons. However, up to now, there is no knowledge on how to extend such results from a bottom-up perspective, which would provide {\em sharper} unitarity conditions on string probes. 

\noindent As shown in \cite{Angelantonj:2024iwi}, non-trivial "rigid" solutions of the tadpole conditions, implying the presence of fewer open-string moduli that forbid brane recombination are also present in the $4D$ $\mathbb{Z}_2 \times \mathbb{Z}_2$ orientifold with discrete torsion \cite{Vafa:1994rv, Angelantonj:1999ms}. Therefore a straightforward extension would be seeking similar solutions for the other orientifolds in $D=4$, for which it would be interesting to study the consistency of probe defects along the lines of \cite{Martucci:2022krl}.

\noindent In the last part of the dissertation, Chapter \ref{ChUnitarity}, we have continued to explore the implications of unitarity by addressing the cancellation of generalised global anomalies arising from topology change, known as {\em Dai-Freed anomalies}. We have adopted a twofold approach to address this issue based on the interplay between a bottom-up study of the consistency of $6D$ supergravity theories with a minimal amount of supersymmetry and a top-down analysis of known string models. In particular, we have explicitly presented the case of supergravity theories with an $\text{SU}(2)$ gauge group and one tensor multiplet and the case of a $\text{U}(1)$ gauge group with no tensor multiples. In the former case, we have found that the anomaly theory changes according to the nature of the $2$-form fields. If the latter is trivial the anomaly theory is trivialised on all the tested backgrounds satisfying the twisted-string structure dictated by the Bianchi identity associated with the Green-Schwarz-Sagnotti mechanism. If the $2$-forms are chiral the anomaly theory depends on the choice of the quadratic refinement describing the partition function of such fields, which cannot be determined {\em a priori}. A similar conclusion is reached for the abelian case with no tensor multiples, for which we have seen that Dai-Freed anomalies exclude infinite families of vacua, up to a specific choice of the quadratic refinement. Extending this analysis to theories with more tensor multiplets, and understanding the details of the twisted string structure, {\em i.e.} whether the bordism groups depend on the choice of Bianchi identity to trivialise, are interesting challenges to be pursued in the future. On the other hand, we have studied how such anomalies are cancelled for string vacua, including the non-supersymmetric heterotic theory with an $\text{SO}(16) \times \text{SO}(16)$ gauge group compactified on the $T^4/\mathbb{Z}_6$ orbifold and a Gepner orientifold with no tensor multiplets that goes beyond the validity of the result of \cite{Tachikawa:2021mvw, Tachikawa:2021mby}, which guarantees their vanishing for heterotic supersymmetric theories.

\noindent A natural development of the discussion in the last Chapter would be an investigation of the anomaly inflow on defects at the Dai-Freed level. Indeed, even in a bottom-up approach, exploiting the completeness principle \cite{Polchinski:2003bq, Banks:2010zn, Heidenreich:2020pkc, Heidenreich:2021xpr} arising from cobordism triviality \cite{McNamara:2019rup, McNamara:2021cuo} and holography \cite{Harlow:2018tng, McNamaraThesis} can predict the existence of novel non-perturbative defects in the theory (see \emph{e.g.} \cite{Debray:2021vob, Debray:2023yrs} for a detailed analysis in type IIB supergravity). In turn, the consistency of anomaly inflow on these defects can restrict the theory even with lower amounts of supersymmetry, as exemplified in \cite{Martucci:2022krl}. Another promising avenue is the study of equivariant topological modular forms \cite{douglas2014topological, hopkins, Gukov:2018iiq}, a generalised cohomology theory believed to encode deformation classes of $2D$ SQFTs (and thus of heterotic world-sheets) \cite{Gaiotto:2019asa} according to the Stolz-Teichner conjecture \cite{Stolz:2011zj}. As already mentioned, this approach was fruitfully employed in \cite{Tachikawa:2021mvw, Tachikawa:2021mby} to exclude all anomalies in supersymmetric heterotic strings, but an equivariant version \cite{Chua_2022, gepner2023equivariant} could apply to non-supersymmetric settings and to refined invariants taking into account gauge charges. This approach could exclude further models from the perturbative heterotic landscape.





\backmatter
\phantomsection
\addcontentsline{toc}{chapter}{\numberline{}Bibliography}


\clearpage
\pagestyle{empty}
\null\newpage
\clearpage

\end{document}